\documentclass[useAMS,usenatbib]{mn2e}
\usepackage{amsmath,color}
\usepackage{graphicx}
\usepackage{marvosym}
\usepackage{mathtools}
\usepackage{cancel}
\usepackage{bm}
\usepackage[T1]{fontenc}
\usepackage{aecompl} 
\usepackage{arydshln}
\usepackage{subfigure,multirow}
\usepackage{mathrsfs}
\usepackage{algorithm}
\usepackage[noend]{algpseudocode}

\newcommand{\conj}[1]{\overline{#1}}
\newcounter{NameOfTheNewCounter}
\newcommand{\Gcal}{\bmath{\mathcal{G}}}
\newcommand{\Ycal}{\bmath{\mathcal{Y}}}
\newcommand{\Rcal}{\bmath{\mathcal{R}}}
\newcommand{\Mcal}{\bmath{\mathcal{M}}}

\newcommand{\Acal}{\bmath{\mathcal{A}}}

\newcommand{\Ccal}{\bmath{\mathcal{C}}}

\newcommand{\Pcal}{\bmath{\mathcal{P}}}
\newcommand{\Xcal}{\bmath{\mathcal{X}}}
\newcommand{\GG}{\bmath{G}}
\newcommand{\PP}{\bmath{P}}
\newcommand{\bal}{\bmath{b}}
\newcommand{\bs}{\bmath{s}}

\newcommand{\bb}{\bmath{b}}
\newcommand{\ba}{\bmath{a}}
\newcommand{\bg}{\bmath{g}}

\newcommand{\bR}{\bmath{R}}

\newcommand{\bA}{\bmath{A}}
\newcommand{\bB}{\bmath{B}}
\newcommand{\bC}{\bmath{C}}
\newcommand{\bD}{\bmath{D}}
\newcommand{\bT}{\bmath{T}}
\newcommand{\bG}{\bmath{G}}
\newcommand{\bI}{\bmath{I}}
\newcommand{\bZ}{\bmath{Z}}

\newcommand{\bx}{\bmath{x}}

\newcommand{\bV}{\bmath{V}}

\newcommand{\bLambda}{\bmath{\Lambda}}

\newcommand{\mI}{\mathcal{I}}
\newcommand{\mJ}{\mathcal{J}}
\newcommand{\mT}{\mathcal{T}}
\newcommand{\mS}{\mathcal{S}}

\newcommand{\pqavg}[2]{\left\langle#1\right\rangle_{{#2}}}

\def\bgamma{{\mbox{\boldmath{$\gamma$}}}}

\def\bone{\mathbf{1}}
\def\bonem{\breve{\mathbf{1}}}
\def\bzero{\mathbf{0}}

\newcommand{\vectorize}[1]{\mathrm{vec} \left ( {#1} \right )}

\newcommand{\diag}[1]{\mathrm{diag} \left ( {#1} \right )}

\def\refpinv{\eqref{eq:inv_matrix}}

\numberwithin{equation}{section}

\title[Phase-only calibration and primary beam correction]{Calibration artefacts in radio interferometry.\\III. Phase-only calibration and primary beam correction}
\author[T.~L. Grobler et al.]{T.~L. Grobler$^{1,2}$\thanks{E-mail: trienkog@gmail.com},
A.~J. Stewart$^{3}$,
S.~J. Wijnholds$^{4}$,
J.~S. Kenyon$^{1,2}$
\newauthor
and O.~M. Smirnov$^{1,2}$
\\
$^1$Department of Physics and Electronics, Rhodes University, PO Box 94, Grahamstown, 6140, South Africa\\
$^2$SKA South Africa, 3rd Floor, The Park, Park Road, Pinelands, 7405, South Africa\\
$^3$Department of Astrophysics, University of Oxford, Oxford, OX1 3RH, UK\\
$^4$ASTRON, P.O. Box 2, 7900 AA, Dwingeloo, The Netherlands}
\begin{document}

\date{Accepted 2016 June 14. Received 2016 June 14; in original form 2016 March 11.}

\pagerange{\pageref{firstpage}--\pageref{lastpage}} \pubyear{2016}

\maketitle

\label{firstpage}

\begin{abstract}
This is the third installment in a series of papers in which we investigate calibration artefacts. Calibration artefacts (also known as ghosts or spurious sources) are created when we calibrate with an incomplete model. In the first two papers of this series we developed a mathematical framework which enabled us to study the ghosting mechanism itself. An interesting concomitant of the second paper was that ghosts appear in symmetrical pairs. This could possibly account for spurious symmetrization. Spurious symmetrization refers to the appearance of a spurious source (the anti-ghost) symmetrically opposite an unmodelled source around a modelled source. The analysis in the first two papers indicates that the anti-ghost is usually very faint, in particular when a large number of antennas are used. This suggests that spurious symmetrization will mainly occur at an almost undetectable flux level. In this paper, we show that phase-only calibration produces an anti-ghost that is $N$-times (where $N$ denotes the number of antennas in the array) as bright as the one produced by phase and amplitude calibration and that this already bright ghost can be further amplified by the primary beam correction.   
\end{abstract}


\section{Introduction}
\label{sec:intro}

Interferometric data can be severely degraded by instrumental and environmental errors. These errors need to be removed by \emph{calibration}. Direction independent calibration is normally achieved by finding antenna gains that minimize the difference between observed and modelled visibilities \citep{Rau2009, Wijnholds2010, RRIME1, Yatawatta2012, vanderVeen2013, Salvini2014, Smirnov2015}. However, when we calibrate with an incomplete sky model we inadvertently cause \emph{calibration artefacts}, which appear as spurious positive or negative flux in the image. We refer to these spurious emission features as \emph{ghost sources} or \emph{ghosts}.

This is the third installment in a series of papers on these ghost phenomena. In the first paper \citep{Grobler2014}, which we will refer to as Paper I, we studied ghost formation by a regular east-west array using the WSRT as an example. This analysis was extended to arbitrary array layouts in the second paper \citep{Wijnholds2016}, which we will refer to as Paper II. An interesting result of Paper II was that ghosts appear to form in symmetrical pairs. It could therefore explain the phenomenon of \emph{spurious symmetrization}. In the context of this paper, spurious symmetrization refers to the appearance of spurious emission symmetrically opposite the unmodelled emission around modelled emission in a radio interferometric image \citep{Taylor1999}. Various cases of spurious symmetrization have been reported \citep{Linfield1986, Wilkinson1988, Taylor1999, Stewart2014, Stewart2016, Wijnholds2016}. \citet{Taylor1999} were one of the first to realize that a direct link exists between calibrating with an incomplete model and spurious symmetrization. They also observed that spurious symmetrization is particularly noticeable if an observation is made with a sufficiently small number of antennas; the implication being that the more antennas we use the less likely it is for spurious symmetrization to occur at a detectable flux level.

In Papers I and II, we considered a simple two-source test case with one modelled and one unmodelled source. We found that a negative \emph{suppression ghost} forms on top of the unmodelled source with a flux that is inversely proportional to the number of antennas in the array. In addition to the suppression ghost an anti-ghost also forms. The \emph{anti-ghost} forms symmetrically opposite the unmodelled source around the modelled source. The anti-ghost is positive and has a flux that is inversely proportional to the square of the number of antennas in the array. This instance of spurious symmetization is consistent with the assertion by \citet{Taylor1999} that the more antennas we use, the lower the flux level becomes at which spurious symmetrization will be detectable. Based on the analysis in Paper I and II, the brightness of the anti-ghost should also be independent of the location of the unmodelled source.

\begin{figure*}
\centering
\subfigure[Anti-ghost not visible.]{\includegraphics[width=0.45\textwidth]{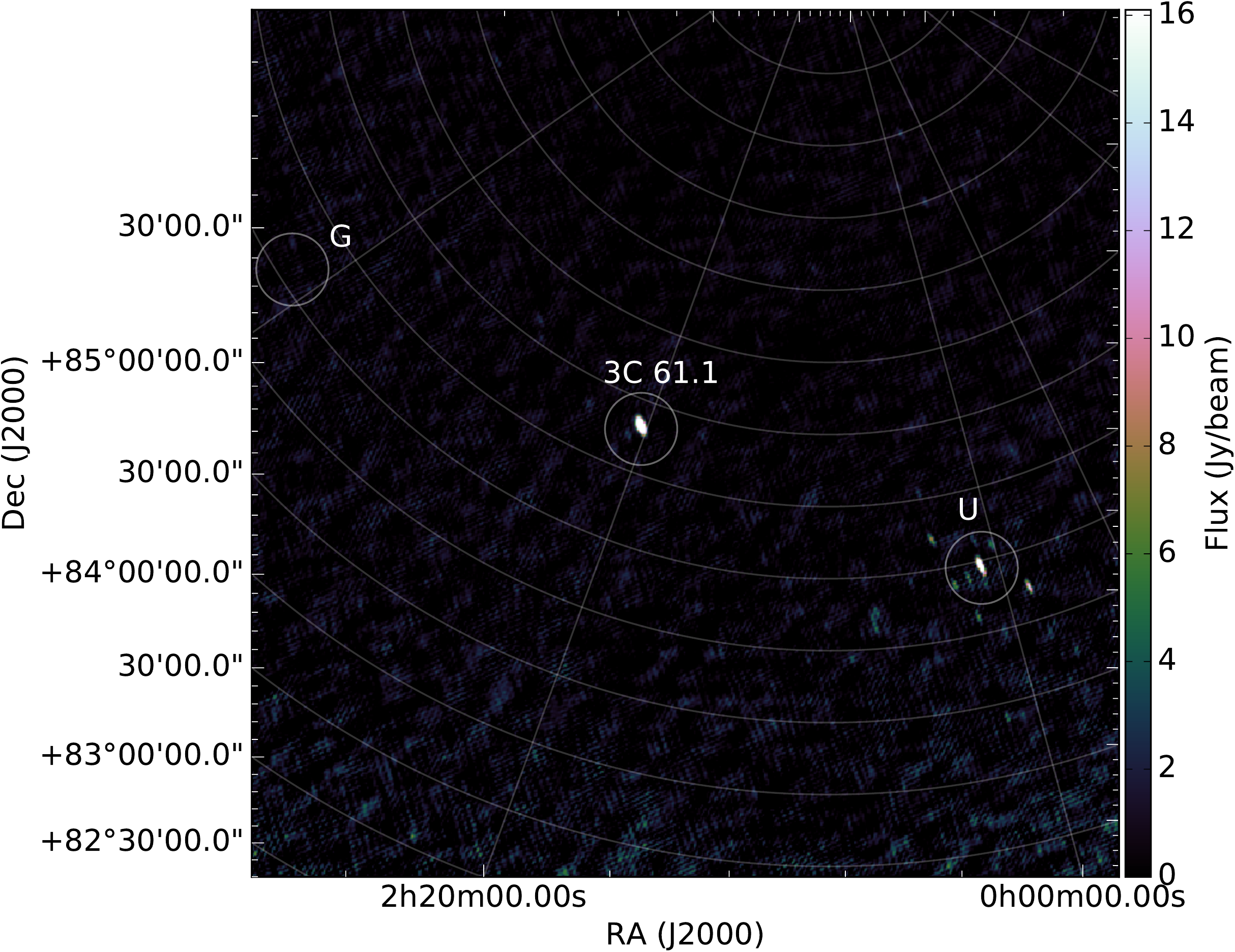}\label{fig:adam1}}
\subfigure[Anti-ghost visible.]{\includegraphics[width=0.45\textwidth]{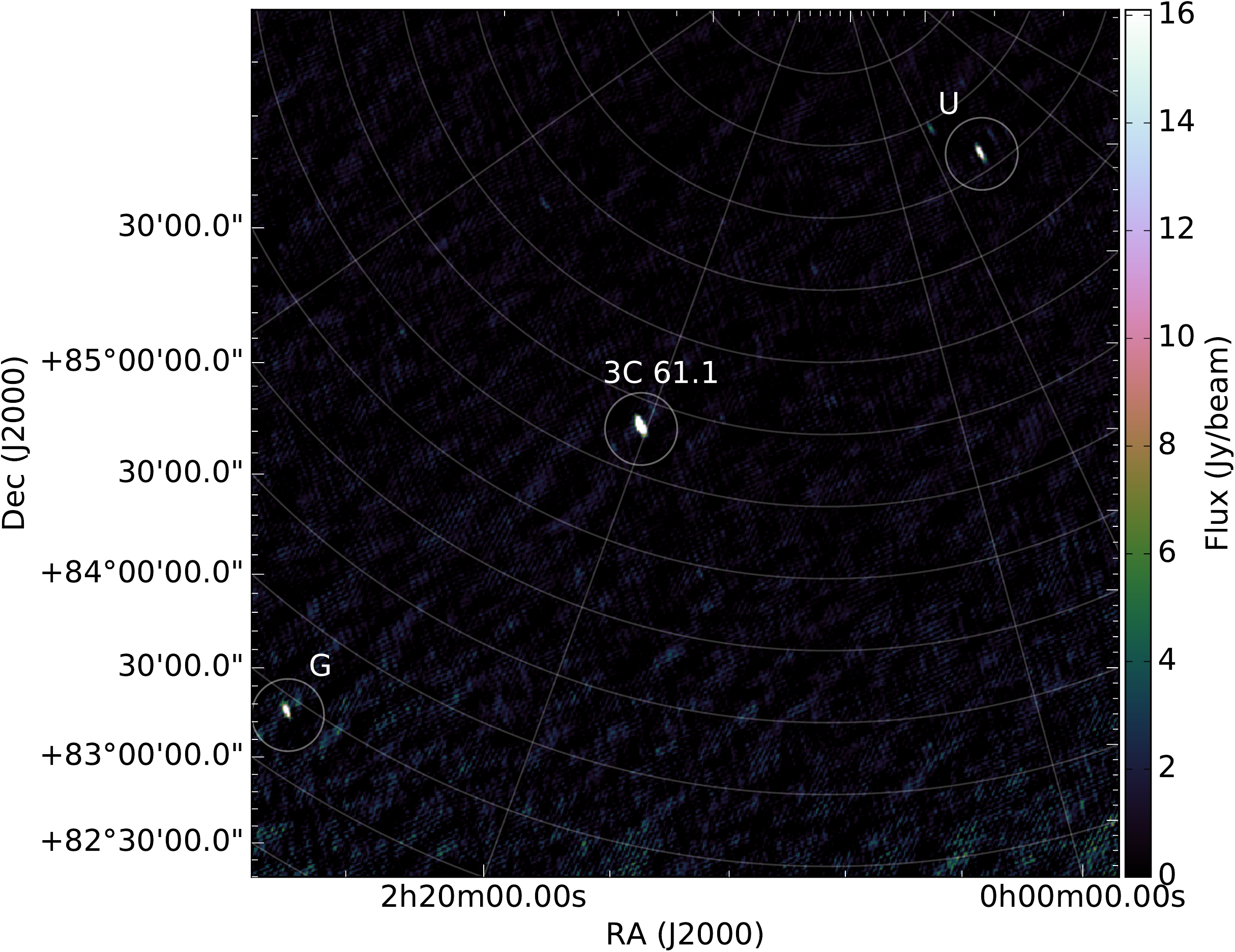}\label{fig:adam2}}
\caption{Two snapshots created during Stewart's experiment. Stewart added a transient to an existing LOFAR dataset which he then ran through the SIP. He found that when he placed the unmodelled transient in some positions (left image) no anti-ghost formed, but when he placed  it at other locations a bright anti-ghost appeared (right image).
\label{fig:adam_exp}} 
\end{figure*}

However, \citet{Stewart2014} recently reported on an intriguing instance of spurious symmetrization in which the anti-ghost was much brighter than expected. Stewart was investigating whether a transient would be detected with LOFAR's Standard Imaging Pipeline (SIP) \citep{Heald2015}. He devised this test to help him make sense of a puzzling transient that was recently detected by the Transients Pipeline (TraP) \citep{Stewart2016, Swinbank2015}. For this test, he added a 60~Jy transient to an existing NCP (North Celestial Pole) dataset. He then determined whether one could detect the synthetic transient after running it through the SIP. The 60~Jy source was not included in the calibration model. He ran the experiment multiple times, varying the location of the transient. The NCP field to which the 60-Jy source was added is dominated by 3C 61.1. The outcome of his test was quite unexpected as he was able to produce a very bright anti-ghost when the transient was placed in some locations and no anti-ghost at all when placed at others. In some instances the anti-ghost was even brighter than the transient itself, which could lead to it being incorrectly identified as a source. Two snapshots of Stewart's experiment are presented in Fig.~\ref{fig:adam_exp}.

Stewart's result is surprising for two reasons:
\begin{enumerate}
\item According to \citet{Taylor1999} and Paper II, the anti-ghost should be extremely faint when we use an array like LOFAR due to its large number of elements.
\item Paper II established that the brightness of the anti-ghost was independent of the location of the unmodelled source.
\end{enumerate}
Obviously, we want to be able to explain the outcome of Stewart's experiment quantitatively, since understanding Stewart's experiment will deepen our understanding of ghost phenomena and spurious symmetrization in particular. There are two major differences between Stewart's experimental setup and the analysis in Paper II:
\begin{enumerate}
\item Phase-only calibration was used in Stewart's case while full-complex calibration was used in Paper II. Phase-only calibration can produce more stable solutions since we have fewer unknowns. Moreover, it is generally accepted that at the imaging stage the amplitude error will be small and constant. The phase-error, however, can still be large and varying. Furthermore, as phase-only calibration conserves phase closures it is also accepted to be quite safe\footnote{http://veraserver.mtk.nao.ac.jp/VERA/kurayama/\\ WinterSchool/difmap6.htm}.
\item The primary beam was included in Stewart's simulations, but was not taken into account in Paper II.
\end{enumerate}
In this paper we will therefore study the impact of using phase-only calibration instead of full-complex calibration on ghost formation as well as the impact of primary beam correction. We will show that these two factors can indeed quantitatively explain the surprising instance of spurious symmetrization found in Stewart's experiment.

Studying the impact of phase-only calibration and primary beam correction further deepens our understanding of ghost phenomena. As mentioned earlier in this series, this is becoming crucial in light of the improved sensitivity (and, consequently, increased susceptibility to artefacts) of current and upcoming telescopes like the Low Frequency Array (LOFAR) \citep{Haarlem2013}, the Karoo Array Telescope (MeerKAT) \citep{Jonas2009} and the Square Kilometre Array (SKA) \citep{Dewdney2009}. Another important motivation, as demonstrated by Stewart's experiment, is that transient pipelines run the risk of mistakingly identifying the anti-ghost as a transient source. An improved understanding of ghost phenomena may help to avoid such errors.

In the next two sections, we formulate the full-complex as well as the phase-only calibration problem and briefly review some of the ghost analysis techniques we developed in the first two papers. In Sec.~\ref{sec:g_pat_phase} we present a theoretical analysis of ghost formation with phase-only calibration. We compare the results with the results for full-complex calibration described in Paper II in Sec.~\ref{sec:comparison}. The effect of primary beam correction is investigated in Sec.~\ref{sec:p_beam} before we conclude our paper with a summary of our findings. The images in this paper use the cubehelix color scheme \citep{Green2011}.

\section{Calibration}
\label{sec:cal_mainsec}

\subsection{Full-Complex Calibration}
\label{ssec:full_complex}

In its most basic form, calibration boils down to minimizing the difference between observed and predicted visibilities by estimating the complex instrumental gain response. Mathematically speaking, unpolarized calibration is realized by minimizing the following optimization problem
\begin{equation}
\label{eq:cal}
\min_{\GG} \left \| \Rcal - \GG\Mcal\GG^H \right \| = \min_{\Gcal} \left \| \Rcal - \Gcal\odot\Mcal \right \| ,
\end{equation}
where:
\begin{enumerate}
 \item The operator $(\cdot)^H$ denotes the Hermitian transpose. Moreover, $\odot$ denotes the Hadamard product and $\|\cdot\|$ denotes the appropriate matrix or vector norm.
 \item $\Rcal$ is the \emph{observed visibility matrix}. Each entry of $\Rcal$, which we denote by $r_{pq}$, represents the visibility that was observed by the baseline formed by antennas $p$ and $q$.
 \item $\Mcal$ is the \emph{model visibility matrix} and we denote an entry of $\Mcal$ with $m_{pq}$. The entries of $\Mcal$ contain model visibilities, i.e. synthetic visibilities that were created from the calibration sky model.
 \item $\bG$ is the \emph{antenna gain matrix} and is equal to $\diag{\bg}$, with $\bg = [g_{1},g_{2},\cdots,g_{N}]^T$ being the complex instrumental responses of the antennas. The operator $\diag{\cdot}$ creates a matrix by placing the input vector on the main diagonal of an otherwise zero matrix, while $(\cdot)^T$ denotes the transpose of its operand. We can decompose the gain of the $p$th element into its amplitude $\gamma_p$ and a phasor $\rho_p = e^{i \varrho_p}$. The amplitudes can be stacked in a vector $\bgamma = \left [ \gamma_1, \cdots, \gamma_N \right ]^T$ and the phasors in a vector $\brho = \left [ \rho_1, \cdots, \rho_N \right ]^T$, such that $\bg = \bgamma \odot \brho$. We will denote the number of antennas in our array by $N$ and the number of baselines by $B$. The number of baselines can be calculated from the number of antennas with $B = (N^2-N) /\ 2$. 
 \item $\Gcal = \bg\bg^H$ is known as the \emph{visibility gain matrix}. An entry of $\Gcal$ is denoted by $g_{pq}$. We can compute $g_{pq}$ with 
 \begin{equation}
   g_{pq}= g_p\conj{g}_q,
 \end{equation}
 where $\conj{(\cdot)}$ denotes complex conjugation. The Hadamard inverse of $\Gcal$ is denoted with $\Gcal^{\odot-1}$ and is known as the \emph{visibility calibration matrix}. We use the visibility calibration matrix to correct our visibilities once we have successfully estimated $\bg$.
 \end{enumerate}


\subsection{Phase-only Calibration}
\label{sec:per_phase_only}

When we perform full-complex calibration we assume that the phase and amplitude of our visibilities were corrupted by our antenna gains. In the case of phase-only calibration we assume that only the phases of our visibilities are affected by the antenna gains. The phase-only equivalent of Eq.~\eqref{eq:cal} is equal to
\begin{equation}
\label{eq:cal_phase}
\min_{\PP} \left \| \Rcal - \PP\Mcal\PP^H \right \| = \min_{\Pcal} \left \| \Rcal - \Pcal\odot\Mcal \right \| ,
\end{equation}
where
\begin{itemize}
 \item $\PP = \diag{\brho}$ is the \emph{antenna phase matrix}.
 \item $\Pcal = \brho\brho^H$ denotes the \emph{visibility phase matrix}. An entry of $\Pcal$ is denoted by $\rho_{pq}$.
\end{itemize}

Once we have estimated $\brho$ we can correct our visibilities, i.e. remove the errors caused by the antenna gains, with
\begin{equation}
\label{eq:Rcal_c}
\Rcal_{\Pcal}^{(c)} = \Pcal^{\odot-1} \odot \Rcal, 
\end{equation}
where 
\begin{enumerate}
 \item $\Rcal_{\Pcal}^{(c)}$ is the \emph{corrected visibility matrix}.
 \item $\Pcal^{\odot-1}$ represents the \emph{visibility phase calibration matrix}. The phase calibration matrix is computed by taking the Hadamard inverse
 of $\Pcal$.
\end{enumerate}


\section{Preliminaries}
\label{sec:prelim}

In this section, we present the remaining building blocks needed for our analysis. In Sec.~\ref{ssec:two_source}, we present the experimental assumptions we adhere to in 
this paper. In Sec.~\ref{ssec:array_geometry}, we define a couple of useful linear transformations. The \emph{extrapolation} algorithm is presented in 
Sec.~\ref{ssec:extrap}. Extrapolation employs the linear transformations in Sec.~\ref{ssec:array_geometry} to transform $\Rcal$ and $\Mcal$ into function-valued matrices. 
We can now use these newly created matrices and standard calibration to create per-baseline artefact maps (a map containing only the calibration systematics). Moreover, these artefact maps contain no sidelobes as extrapolation operates on continuous functions (see Algorithm~\ref{algo:extrap_algo}). 

\subsection{Two-source scenario}
\label{ssec:two_source}
Unless explicitly stated otherwise, all results presented in this paper are subject to the following assumptions:
\begin{enumerate}
 \item We are dealing with a monochromatic observation.
 \item No error was added to our visibilities by the interferometer, i.e. $\bg_t = \bone$. We denote an all one vector of size $N$ with $\bone$. The subscript $t$ of $\bg_t$ is used to indicate that we are referring to the true antenna gain vector which is corrupting the observed visibilities instead of the gain vector estimate which is obtained when we calibrate. 
 \item We assume the $w$-term is negligible, i.e. we are observing in a narrow field-of-view.
 \item We do not consider the effect of noise as O.M. Smirnov has shown that including noise does not alter the positions at which the ghosts form \citep{Smirnov2010ghosts}. 
\end{enumerate}

With regards to the true sky and the calibration sky model we assume the following:
\begin{enumerate}
 \item Our true sky consists of two sources, one in the field center with flux $A_1$ and another off-center with flux $A_2$. We denote the positional vector of the off-center source by $\bmath{s}_0 = (l_0,m_0)$ (a direction-cosine vector). Stated differently, if we were to observe these two sources then antennas $p$ and $q$ would observe
 \begin{equation}
 \label{eq:r_pq}
 r_{pq} = A_1 + A_2e^{-2\pi i \bb_{pq}^T\cdot\bs_0}, 
\end{equation}
where $\bb_{pq} = (u_{pq},v_{pq})^T$ denotes the $uv$-coordinate at which $r_{pq}$ was measured. We use the symbol ``$\cdot$'' to denote the standard dot product which operates on equal length row vectors.
It is important to realize that the coordinates $u_{pq}$ and $v_{pq}$ in Eq.~\eqref{eq:r_pq} are already expressed in wavelengths and not in m (we have already divided by the observational wavelength $\lambda$). In Paper II, we showed that all the $uv$-tracks of an array can be generated from a single circular reference track $\bb_0^\circ(t)$. The circular reference track $\bb_0^\circ(t)$ is depicted in Fig.~\ref{fig:circ_to_ellipse}. 

\item We only include the center source in our calibration model, i.e.
\begin{equation}
\label{eq:m_pq}
m_{pq} = A_1.
\end{equation} 
\item Since Eq.~\eqref{eq:cal_phase} is invariant to amplitude scaling and positional shifts, we may assume that our modelled source is in the center and that  $A_1 = 1$ without any loss of generality, 
\item Unless otherwise stated $A_2 \ll A_1$, i.e. that our calibration model is reasonably accurate. 
\end{enumerate}

\subsection{Linear transformations}
\label{ssec:array_geometry}
All the $uv$-tracks of an interferometer can be mapped onto a single circular reference track $\bb_0^\circ(t)$ by using three linear transformations, namely
\begin{enumerate}
\item a \emph{scaling} described by
\begin{equation}
\phi_{pq}\boldsymbol{D}(\delta_0) =\phi_{pq} 
\begin{bmatrix}
1 & 0 \\
0 & \sin(\delta_0)
\end{bmatrix},
\end{equation}
\item a \emph{rotation} described by
\begin{equation}
\boldsymbol{T}(\theta_{pq}) = 
\begin{bmatrix}
\cos(\theta_{pq}) & -\sin(\theta_{pq})\\
\sin(\theta_{pq}) & \cos(\theta_{pq})
\end{bmatrix},
\end{equation}
\item and a \emph{translation} $\Delta \bb_{pq}$.
\end{enumerate}

\begin{figure*}
 \centering
 \includegraphics[width=1\textwidth]{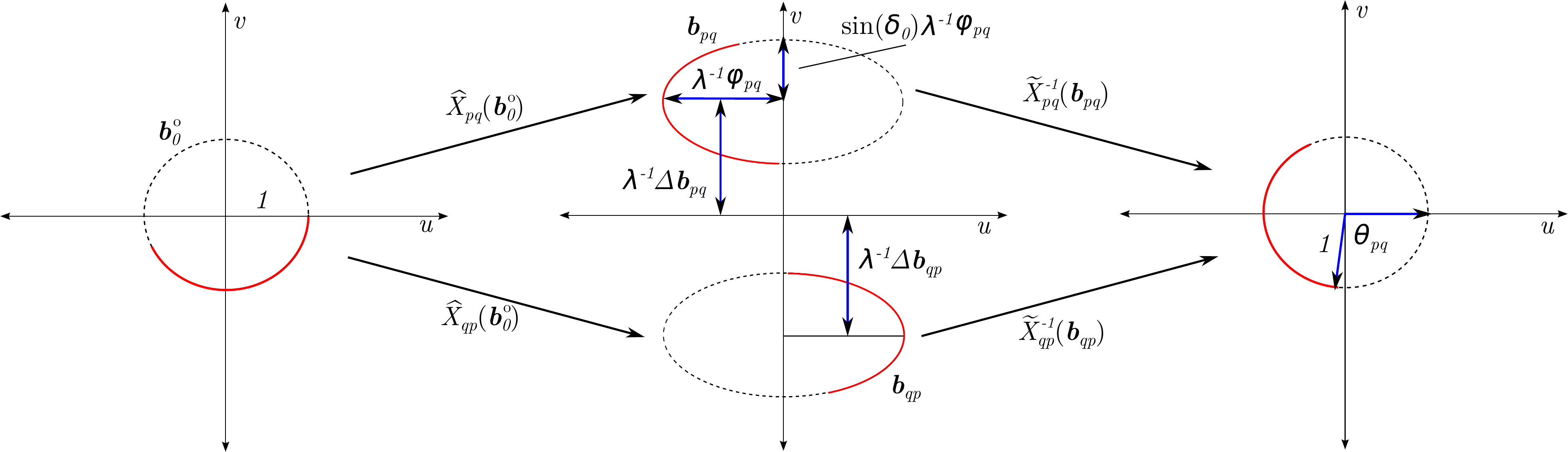}
 \caption{The linear transformations $\widehat{X}_{pq}(\bb_0^{\circ})$ and $\widetilde{X}_{pq}^{-1}(\bb_{pq})$.}
 \label{fig:circ_to_ellipse} 
\end{figure*}

A composition of two or more linear transformations results in another linear transformation. In this paper, we will use the following compositions:
\begin{enumerate}
\item a composition of scaling, rotation and translation
\begin{equation}
\widehat{X}_{pq}(\bb_{0}^{\circ}) = \frac{\phi_{pq}}{\lambda}\bD(\delta_0)\bT(\theta_{pq})\bb_{0}^{\circ} + \frac{\Delta \bb_{pq}}{\lambda}
\end{equation}
\item a composition of scaling and translation
\begin{equation}
\widetilde{X}_{pq}(\bb_{0}^{\circ}) = \frac{\phi_{pq}}{\lambda}\bD(\delta_0)\bb_{0}^{\circ} + \frac{\Delta \bb_{pq}}{\lambda}
\end{equation}
\item a composition of scaling and rotation
\begin{equation}
X_{pq}(\bb_{0}^{\circ}) = \frac{\phi_{pq}}{\lambda}\bD(\delta_0)\bT(\theta_{pq})\bb_{0}^{\circ}
\end{equation}
\end{enumerate}
In the expressions above, $\delta_0$ is the declination of the field center. The quantities $\phi_{pq}$, $\theta_{pq}$ and $\Delta \bb_{pq} = (0,\Delta b_{pq})^T$ are determined by the array geometry and are formally defined in Fig.~\ref{fig:circ_to_ellipse}. The linear transformations $\widehat{X}_{pq}(\bb_0^{\circ})$ and $\widetilde{X}_{pq}^{-1}(\bb_{pq})$ are also depicted in Fig.~\ref{fig:circ_to_ellipse}.

We denote the functional inverse of $\widehat{X}_{pq}$, $\widetilde{X}_{pq}$ and $X_{pq}$ with $\widehat{X}_{pq}^{-1}$, $\widetilde{X}_{pq}^{-1}$ and $X_{pq}^{-1}$ respectively. 

Within this paper we use each of the aforementioned linear tranformations for a specific purpose:
\begin{enumerate}
 \item $\widehat{X}_{pq}$: The circular reference track $\bb_{0}^{\circ}(t)$ is depicted in the left most coordinate system of Fig.~\ref{fig:circ_to_ellipse}.
 We can generate any $uv$-track of an array by applying $\widehat{X}_{pq}$ to $\bb_{0}^{\circ}(t)$. This fact is illustrated by the mapping of $\bb_{0}^{\circ}(t)$ in the left most
 coordinate system to $\bb_{pq}(t)$ in the middle coordinate system of Fig.~\ref{fig:circ_to_ellipse}. The linear transformation $\widehat{X}_{pq}(\bb_{0}^{\circ})$ is exploited by the extrapolation procedure (presented in
 Sec.~\ref{ssec:extrap}) to create per-baseline artefact maps.
 \item $\widetilde{X}_{pq}$: The effect of applying $\widetilde{X}_{pq}^{-1}$ to $\bb_{pq}(t)$ is depicted by the mapping from $\bb_{pq}(t)$ in the middle 
 coordinate system of Fig.~\ref{fig:circ_to_ellipse} to the $u'v'$-track in the right most coordinate system of Fig.~\ref{fig:circ_to_ellipse}. Since there is no rotation transformation 
 present in $\widetilde{X}_{pq}$, the $u'v'$-track in the right most coordinate system of Fig.~\ref{fig:circ_to_ellipse} does not start at  coordinte (1,0). An additional
 derotation by $\theta_{pq}$ is required to achieve this, i.e. $\widetilde{X}_{pq}^{-1}(\bb_{pq})$ allows us to calculate the rotation angle $\theta_{pq}$; which is a quantity that is required by $\widehat{X}_{pq}$ to function properly. 
 \item $X_{pq}$: It turns out that the translation $\Delta \bb_{pq}$ does not affect the positions of the ghosts, only their fluxes as discussed in Paper II. Since $X_{pq}$ does not contain a translation transformation, it can be used to determine the positions at which the ghosts arise (see Eq.~\ref{eq:p_gen}).
\end{enumerate}

\subsection{Extrapolation}
\label{ssec:extrap}

Extrapolation is a technique, which exploits the fact that all $uv$-tracks of an interferometer can be derived from a single imaginary reference track (via three linear transformations) to create clean per-baseline artefact maps (an artefact map untarnished by $uv$-sampling).

Extrapolation requires the following three steps:
\begin{enumerate}
\item Reformulate visibilities as functionals. Rewrite the visibility matrix $\Ycal$ as a function of the imaginary reference track $\bb_0^{\circ}$, i.e. $\Ycal(\bb_0^{\circ})$. The  $\Ycal$ matrix is a proxy matrix and can refer to an observed visibility, model visibility, predicted visibility, visibility gain, visibility calibration, corrected visibility matrix or any one of the distilled visibility matrices. Distillation is discussed in Sec.~\ref{ssec:dist}.
\item Derive the intermediate extrapolated visibility matrix. Replace $\bb_0^{\circ}$ with the free parameter $\bb = (u,v)^T$ to obtain $\Ycal(\bb)$.
\item Calculate the extrapolated visibility matrix. It turns out that the entries of $\Ycal(\bb)$ have no physical meaning if left as is. To make the entries meaningful we substitute $\bb$ with $\widehat{X}_{pq}^{-1}(\bb)$ to obtain $\Ycal(\widehat{X}_{pq}^{-1}(\bb))$. 
\end{enumerate}

If we now image the $pq$th entry of $\Ycal(\widehat{X}_{pq}^{-1}(\bb))$ then the resulting image will be completely devoid of sidelobes as the $pq$th entry of $\Ycal(\widehat{X}_{pq}^{-1}(\bb))$ is defined over the entire $uv$-plane. Extrapolation is a per-baseline operation. 

In Algorithm~\ref{algo:extrap_algo} we use the extrapolation procedure to specifically construct an empirical estimate of $g_{pq} ( \widehat{X}_{pq}^{-1} (\bb) )$. In this algorithm $\mathcal{K}$ denotes the set containing all the index values of the sources that were ``observed'', while $\mathcal{H}$ denotes the set containing the index values of the sources that were included in the calibration model, i.e., $\mathcal{H}\subseteq\mathcal{K}$. The extrapolation procedure can therefore be applied to more complicated sky model scenarios (and is not limited to the two-source test case presented in Sec.~\ref{ssec:two_source}). The artefact map associated with $g_{pq} ( \widehat{X}_{pq}^{-1} (\bb) )-1$ (we subtract one to distill out the calibration artefacts) is obtained by taking the inverse Fast Fourier Transform (FFT) of the empirical estimate of $g_{pq} ( \widehat{X}_{pq}^{-1} (\bb) )-1$, which is now trivial as 
the empirical estimate of $g_{pq} ( \widehat{X}_{pq}^{-1} (\bb) )-1$ is defined on a regular grid. This results in an artefact map without any sidelobes. Moreover, in Algorithm~\ref{algo:extrap_algo},
``$\circ$'' denotes operator composition. Also note that any calibration algorithm may be used to realise Algorithm~\ref{algo:extrap_algo}. 

A more detailed explanation of extrapolation can be found in Paper II.

\begin{algorithm}
\caption{Extrapolation: $g_{pq} ( \widehat{X}_{pq}^{-1} (\bb) )$}\label{algo:extrap_algo}
\begin{algorithmic}[1]
\State $\bmath{u} \gets \text{linspace}(u_{\text{min}},u_{\text{max}},u_{\text{dim}})$
\State $\bmath{v} \gets \text{linspace}(v_{\text{min}},v_{\text{max}},v_{\text{dim}})$ 
\State $g_{pq} \gets \text{zeros}((u_{\text{dim}},v_{\text{dim}}),\text{dtype}=\text{complex})$ \Comment {define your 2D extrapolated regular $uv$-grid}
\State $i \gets 1$
\For{$i\leq u_{\text{dim}}$}  
\State $j \gets 1$
\For{$j\leq v_{\text{dim}}$}  
\State $\bb \gets (\bmath{u}_i,\bmath{v}_j)^T$
\State $\Rcal \gets \left\{ \sum_{k\in\mathcal{K}}A_k e^{-2\pi i [\widehat{X}_{rs} \circ \widehat{X}^{-1}_{pq}(\bb)]^T\cdot\bs_k} \right \}_{rs}$
\State $\Mcal \gets \left\{ \sum_{h\in\mathcal{H}}A_h e^{-2\pi i [\widehat{X}_{rs} \circ \widehat{X}^{-1}_{pq}(\bb)]^T\cdot\bs_h} \right \}_{rs}$
\State $\Gcal \gets \text{calibrate}(\Rcal,\Mcal)$
\State $g_{pq}^{ij} \gets [\Gcal]_{pq}$
\State $j \gets j + 1$
\EndFor
\State $i \gets i + 1$
\EndFor
\end{algorithmic}
\end{algorithm}

\section{Phase-only Ghost Pattern}
\label{sec:g_pat_phase}

In this section, we derive the phase-only ghost pattern that is associated with the two-source test case presented in Sec.~\ref{ssec:two_source}. We do so by following the same approach we took for full-complex calibration in Paper II. We therefore start by developing a general perturbation analysis framework for phase only calibration. 
 We then use this general framework to analyze the two-source test case in Sec.~\ref{ssec:perturbation_phase_only}. We then apply extrapolation in Sec.~\ref{sec:extrap_phase_only} to the result from the perturbation analysis. 
For the sake of completeness we also briefly touch on the topics of distillation and corrected visibilities in Sec.~\ref{ssec:dist} and Sec.~\ref{ssec:corr_vis}. A brief overview of imaging and how it pertains to calibration artefacts is presented in Sec.~\ref{ssec:imaging}.
In the last section we extend the basic per-baseline taxonomy from Paper II. This extended taxonomy makes it easier to compare the ghost patterns that are produced by full-complex and phase-only calibration respectively. 

\subsection{Perturbation Analysis}
\label{ssec:pert1}
Perturbation analysis is based on the assumption that we are using a fairly accurate calibration sky model, implying that the antenna gains are only slightly perturbed due to the incomplete sky model. 
In the perturbation approach that we present here we estimate the amount of perturbation the gains experience, by first linearizing the estimation problem, making it easier to solve. This linearization is only possible
if the perturbation we are trying to estimate is relatively small.

The observed visibility matrix is equal to
\begin{equation}
\label{eq:Rcal}
\Rcal = \bG_t(\Rcal_k + \Rcal_u)\bG_t^H = \bg_t\bg_t^H\odot(\Rcal_k + \Rcal_u), 
\end{equation}
where $\Rcal_k$ contains the visibilities of the sources that we know exist, i.e. $\Rcal_k = \Mcal$, $\Rcal_u$ contains the visibilities of the sources that we are unaware of, $\bg_t$ is the true antenna gain vector and $\bG_t = \diag{\bg_t}$. We can also decompose $\bg_t$ into an amplitdue vector $\bgamma_t=\left [ \gamma_1^t, \cdots, \gamma_N^t \right ]^T$ and phasor vector $\brho_t=\left [ e^{{i\varrho_1^t}}, \cdots, e^{i\varrho_N^t} \right ]^T$, such that $\bg_t = \bgamma_t \odot \brho_t$. Eq.~\eqref{eq:Rcal} allows us to reformulate Eq.~\eqref{eq:cal_phase} as
\begin{equation}
\label{eq:phase_min}
\underset{\brho}{\mathrm{argmin}} \left \| \bg_t \bg_t^H \odot \left ( \bR_k + \bR_u \right ) - \brho \brho^H \odot \bR_k \right \|.
\end{equation}

If we ignore the true amplitude variations $\bgamma_t$ or we neglect the mismatch between the model visibilities (described by $\bR_k$) and the observed visibilities (described by $\bR_k + \bR_u$), then we will induce a phase bias $\Delta \bvarrho$ into $\brho$, i.e.
\begin{equation}
\brho = e^{i \left ( \bvarrho_t + \Delta \bvarrho \right )}
\end{equation}

Without loss of generality, we may assume that $\bvarrho_t = \bzero$. Assuming that we are justified in using phase-only calibration and that we have a reasonably accurate sky model, the biases will be small. We may therefore use the approximation $e^{i\Delta \bvarrho} \approx 1 + i \Delta \bvarrho$. It then follows that
\begin{equation}
\label{eq:Pcal_app}
 \Pcal \approx \left ( \bone + i \Delta \bvarrho \right ) \left ( \bone + i \Delta \bvarrho \right )^H. 
\end{equation}

Eq.~\eqref{eq:Pcal_app} allows us to rewrite Eq.~\eqref{eq:phase_min} to
\begin{eqnarray}
&\underset{\Delta\bvarrho}{\mathrm{argmin}}& \Big \| \bg_t \bg_t^H \odot \left ( \bR_k + \bR_u \right ) \nonumber\\
& & - \left ( \bone + i \Delta \bvarrho \right ) \left ( \bone + i \Delta \bvarrho \right )^H \odot \bR_k \Big \|
\end{eqnarray}
Since $\Delta \bvarrho$ is small, we may ignore the product $\Delta \bvarrho \Delta \bvarrho^H$ and write our minimization problem as
\begin{eqnarray}
 &\underset{\Delta \bvarrho}{\mathrm{argmin}} &\Big \| \bg_t \bg_t^H \odot \bR_k + \bg_t \bg_t^H \odot \bR_u \nonumber\\
& & - \bone\bone^H\odot\bR_k + i \bone \Delta \bvarrho^H \odot \bR_k - i \Delta \bvarrho \bone^H \odot \bR_k \Big \| \nonumber\\
& = & \underset{\Delta \bvarrho}{\mathrm{argmin}} \Big \| \vectorize{\bI \bG_t \bR_k \bG_t^H } - \vectorize{\bI \bI \bR_k \bI^H }\nonumber\\
& & + \vectorize{\bI \bG_t \bR_u \bG_t^H} + i \vectorize{\bR_k \diag{\Delta \bvarrho}^H \bI}\nonumber\\
& & - i \vectorize{\bI \diag{\Delta \bvarrho} \bR_k} \Big \|
\label{eq:phase_only_est1}
\end{eqnarray}
In Eq.~\eqref{eq:phase_only_est1}, we made use of the fact that $\bx\bx^H\odot\Xcal = \diag{\bx}\Xcal\diag{\bx}^H$, where $\bx$ is a vector and $\Xcal$ is matrix of compatible size. In the next step, we will use
\begin{equation}
\label{eq:identity_rao}
\vectorize{\Acal~ \diag{\bb} \Ccal} = \left ( \Ccal^T \bullet \Acal \right ) \bb,
\end{equation}
where $\Acal$ and $\Ccal$ are matrices of compatible size and $\bullet$ denotes the Khatri-Rao or column-wize Kronecker product of two matrices. We swapped around and added identity matrices to Eq.~\ref{eq:phase_only_est1}, denoted by $\bI$, as visual cues to enable the reader to follow how one can apply Eq.~\eqref{eq:identity_rao} to Eq.~\eqref{eq:phase_only_est1} to obtain Eq.~\eqref{eq:normal_eq}. We thus find
\begin{eqnarray}
\lefteqn{\underset{\Delta\bvarrho}{\mathrm{argmin}} \Big \| \left ( \bG_t^H \bR_k^T \bullet \bI \right ) \bg_t -  \left ( \bI \bR_k^T \bullet \bI \right ) \bone} \nonumber\\
& & + \left ( \bG_t^H \bR_u^T \bullet \bI \right ) \bg_t + i \left ( \bI \bullet \bR_k \right ) \Delta \bvarrho - i \left ( \bR_k^T \bullet \bI \right ) \Delta \bvarrho  \Big \| \nonumber\\
& = & \underset{\Delta\bvarrho}{\mathrm{argmin}} \Big \|  \left ( \conj{\bG_t} \conj{\bR_k} \bullet \bI \right ) \bg_t -  \left ( \bI \conj{\bR_k} \bullet \bI \right ) \bone \nonumber\\
& & + \left ( \conj{\bG_t \bR}_u \bullet \bI \right ) \bg_t - i \left ( \left ( \conj{\bR}_k \bullet \bI \right ) - \left ( \bI \bullet \bR_k \right ) \right ) \Delta \bvarrho \Big \| \label{eq:normal_eq}
\end{eqnarray}
Note that the first mentioned $\Delta\bvarrho$ in the first equation of Eq.~\ref{eq:normal_eq} is not conjugated, although strictly speaking it should be. We do not conjugate $\Delta\bvarrho$ in Eq.~\ref{eq:normal_eq}, since $\Delta\bvarrho$ is per definition a real valued vector.

If we apply the normal equation method to Eq.~\eqref{eq:normal_eq} we readily obtain
\begin{equation}
\label{eq:p_only_per_result}
\Delta \bvarrho  \approx  -i \left [\bA^H\bA \right ]^{-1}\bA^H\times \left \{ \bB\bg_t - \bC\bone + \bD\bg_t\right \},
\end{equation}
where
\begin{eqnarray}
\bA & = &  \left ( \conj{\bR}_k \bullet \bI \right ) - \left ( \bI \bullet \bR_k \right ), \nonumber \\
\bB & = & \conj{\bG_t} \conj{\bR_k} \bullet \bI, \nonumber \\
\bC & = & \conj{\bR_k} \bullet \bI, \nonumber \\
\bD & = & \conj{\bG_t} \conj{\bR_u} \bullet \bI. \label{eq:matrices_ABCD}
\end{eqnarray}
There is an inherent ambiguity present in Eq.~\eqref{eq:p_only_per_result}, as we did not explicitly select a reference antenna.  However, this ambiguity has no effect on the outcome of our analysis as it drops out when we calculate $\Pcal$.

\subsection{Two-source Perturbation Results}
\label{ssec:perturbation_phase_only}

In this section we apply the general result derived in the previous section to our two-source test case. It follows from Eq.~\eqref{eq:r_pq} and Eq.~\eqref{eq:m_pq} (if $A_1=1$) that
\begin{eqnarray}
\Rcal_k &=& \bone\bone^H,\\
\Rcal_u &=& A_2\ba\ba^H,
\end{eqnarray}
where $\ba = \left [ e^{-2\pi i \bb_1^T \cdot \bs_0},\cdots, e^{-2\pi i \bb_N^T \cdot \bs_0} \right ]^T$ and $\bb_n$ is the $(x,y,z)$-position of the $n$-th antenna in units of $\lambda$. It also follows from Sec.~\ref{ssec:two_source} that $\bg_t = \bone$. Substitution of these results in Eq.~\eqref{eq:matrices_ABCD} gives
\begin{eqnarray}
\label{eq:matrices_ABCD_2source}
\bA & = & \left ( \bone \bone^H \bullet \bI \right ) - \left ( \bI \bullet \bone \bone^H \right )\nonumber\\
\bB & = & \bone \bone^H \bullet \bI \nonumber\\
\bC & = & \bone \bone^H \bullet \bI ~~=~~ \bB\nonumber\\
\bD & = & A_2 \overline{\left ( \ba\ba^H \right)} \bullet \bI
\end{eqnarray}
Note that the dimensions of the matrices in Eq.~\eqref{eq:matrices_ABCD_2source} are $N^2\times N$.

To find an expression for our two source model using Eq.~\eqref{eq:p_only_per_result}, we need to find the inverse of
\begin{equation}
\label{eq:mat_to_invert}
\bA^H \bA = 2(N\bI - \bone\bone^H).
\end{equation}
Since this matrix is rank deficient, we will resort to the pseudo-inverse, denoted by $()^\dagger$ to emulate the behaviour of most least squares fitting routines. We find that
\begin{equation}
\label{eq:inv_matrix}
 \left [ \bA^H\bA \right ]^{\dagger} = \frac{1}{2N}\bI - \frac{1}{2N^2}\bone\bone^H,
\end{equation}
for which a proof can be found in Appendix~\ref{sec:app:deriv_pinv}.

If we substitute Eqs.~\eqref{eq:matrices_ABCD_2source} and \eqref{eq:inv_matrix} in Eq.~\eqref{eq:p_only_per_result}, we obtain
\begin{equation}
\label{eq:delta_rho}
\Delta \bvarrho \approx \frac{A_2}{N}\mathscr{I} \left \{  \left (\ba^H\bone \right )\ba \right \},
\end{equation}
where $\mathscr{I}\{\cdot\}$ denotes the imaginary operator. Substitution of this result into Eq.~\eqref{eq:Pcal_app} gives
\begin{eqnarray}
\Pcal &\approx& \bone\bone^H - i\bone\Delta \bvarrho^H + i\Delta\bvarrho\bone^H\nonumber\\
&=& \bone\bone^H - c\mathscr{I} \left \{  \left (\ba^H\bone \right )\bone\ba^T \right \}\nonumber\\  
& & + c\mathscr{I}\left \{  \left (\ba^H\bone \right )\ba\bone^H \right \},\label{eq:Pcal_result}
\end{eqnarray}
where $c = A_2i/N$. Concentrating on a single element of the visibility phase matrix, we find
\begin{equation}
\label{eq:g_pq_after_Taylor}
 [\Pcal]_{pq} = \rho_{pq}  \approx 1 + i(\Delta \varrho_p - \Delta \varrho_q). 
\end{equation}
Simplifying Eq.~\eqref{eq:g_pq_after_Taylor} results in
\begin{equation}
\label{eq:sum_sines}
\rho_{pq} \approx 1 - \sum_{s\neq p} c\sin(2\pi \bb_{ps}^T \cdot \bs_0)-\sum_{r\neq q} c\sin(2\pi \bb_{rq}^T \cdot \bs_0),  
\end{equation}
where $\bb_{ps} = \bb_p - \bb_s$ and $\bb_{rq} = \bb_r - \bb_q$ (also see Sec.~\ref{ssec:two_source}).  

The fact that a sine can be written in terms of two complex exponentials allows us to
rewrite Eq.~\eqref{eq:sum_sines} to
\begin{equation}
\label{eq:g_pq_phase}
\rho_{pq} \approx \sigma_{pq,0} + \sum_{rs\in \mI} \sigma_{pq,rs}e^{-2\pi i \bb_{rs}^T\cdot\bs_0},
\end{equation}
where 
\begin{equation}
\sigma_{pq,0} = 1, 
\end{equation}
and
\begin{equation}
   \sigma_{pq,rs} =
   	\begin{cases}
	   \frac{A_2}{N} & \text{if}~rs\in\mI_1\\
       -\frac{A_2}{N} & \text{if}~rs\in\mI_1'\\
       \frac{A_2}{2N} & \text{if}~rs\in\mI_2\\
       -\frac{A_2}{2N} & \text{if}~rs\in\mI_2'\\
       \frac{A_2}{2N} & \text{if}~rs\in\mI_3\\
       -\frac{A_2}{2N} & \text{if}~rs\in\mI_3'\\
       0 & \textrm{otherwise}
     \end{cases},
   \label{eq:c_pq_rs_phase}
\end{equation} 
with
\begin{eqnarray}
\mI_1 &=& \{rs|(r=p)\wedge (s=q) \wedge (r\neq s)\} \nonumber \\
\mI_1' &=& \{rs|(r=q)\wedge (s=p) \wedge (r\neq s)\} \nonumber \\
\mI_2 &=& \{rs|(r\neq p)\wedge (s=q) \wedge (r\neq s)\} \nonumber \\
\mI_2' &=& \{rs|(r\neq q)\wedge (s=p) \wedge (r\neq s)\} \nonumber \\
\mI_3 &=& \{rs|(r = p)\wedge (s\neq q) \wedge (r\neq s)\} \nonumber \\
\mI_3' &=& \{rs|(r = q)\wedge (s\neq p) \wedge (r\neq s)\}, \label{eq:index_sets}
\end{eqnarray}
where ``$\wedge$'' denotes the logical and.  Moreover, we define
\begin{equation}
\label{eq:baseline_indices}
\mI = \bigcup_{k=1}^3 \mI_k\cup\mI_k'.
\end{equation}

\subsection{Extrapolation}
\label{sec:extrap_phase_only}

We will use extrapolation to show that each term in Eq.~\eqref{eq:g_pq_phase} produces a ghost artefact in the image. We can express $\rho_{pq}$ as a function of the reference baseline $\bb_0^{\circ}$, i.e.
\begin{equation}
\label{eq:p_pq1} 
\rho_{pq}(\bb_0^{\circ}) = \sigma_{pq,0} + \sum_{rs\in\mI}\sigma_{pq,rs} e^{-2\pi i [\widehat{X}_{rs}(\bb_0^{\circ})]^T\cdot \bs_0}. 
\end{equation}
To extrapolate, we replace $\bb_0^{\circ}$ with $\bb=(u,v)^T$ to obtain 
\begin{equation}
\label{eq:p_pq4} 
\rho_{pq}(\bb) \approx \sigma_{pq,0} + \sum_{rs\in\mI}\sigma_{pq,rs} e^{-2\pi i [\widehat{X}_{rs}(\bb)]^T\cdot \bs_0}. 
\end{equation}
As discussed in Section~\ref{ssec:extrap}, we can create physically meaningful extrapolated visibilities by replacing $\bb$ in Eq.~\eqref{eq:g_pq_phase} with $\widehat{X}_{pq}^{-1}(\bb)$ to obtain  
\begin{eqnarray}
\rho_{pq}(\widehat{X}_{pq}^{-1}(\bal)) &\approx& \sigma_{pq,0} + \sum_{rs\in\mI} \sigma_{pq,rs} e^{-2\pi i[\widehat{X}_{rs}\circ \widehat{X}_{pq}^{-1}(\bal)]^T\cdot\bmath{s}_0},\nonumber\\
&=& \tau_{pq,0} + \sum_{rs\in\mI} \tau_{pq,rs} e^{-2\pi i[X_{rs}\circ X_{pq}^{-1}(\bal)]^T\cdot\bmath{s}_0},\nonumber\\
\label{eq:p_gen}
&=& \tau_{pq,0} + \sum_{rs\in\mI} \tau_{pq,rs} e^{-2\pi i \bal^T\cdot\bmath{s}_{pq}^{rs}}
\end{eqnarray}
where $\tau_{pq,0}=\sigma_{pq,0}$, $\tau_{pq,rs}=\sigma_{pq,rs}\beta_{pq,rs}$ and
\begin{equation}
\label{eq:beta}
\beta_{pq,rs} = e^{2\pi i\big[\frac{\phi_{rs}}{\lambda\phi_{pq}}\bD(\delta_0)\bT(\theta_{rs}-\theta_{pq})\bD^{-1}(\delta_0)\Delta\bb_{pq} - \frac{\Delta\bb_{rs}}{\lambda}\big]^T\cdot\bs_0}
\end{equation}
and
\begin{equation}
\label{eq:position}
\bmath{s}_{pq}^{rs} = \frac{\phi_{rs}}{\phi_{pq}} \bmath{s}_0 \bZ,
\end{equation}
with
\begin{equation}
\bZ =
\left[ \begin{array}{ll}
\cos(\theta_{rs}-\theta_{pq}) & -\frac{\sin(\theta_{rs}-\theta_{pq})}{\sin(\delta_0)} \\
\sin(\delta_0)\sin(\theta_{rs}-\theta_{pq}) & \cos(\theta_{rs}-\theta_{pq}) \end{array}\right].
\end{equation}

Since the Fourier inverse of a complex exponential is a delta function, each term in the sum of Eq.~\eqref{eq:p_gen} can be interpreted as a spurious point source, i.e. as a calibration artefact or ghost. The amplitude of the spurious point source is equal to $\tau_{pq,rs}\in\mathbfss{C}$ and its position vector is equal to $\bs_{pq}^{rs}$. Furthermore, the amplitudes of the ghosts exhibit the following symmetry 
\begin{equation}
\label{eq:symmetry}
\tau_{pq,rs} = -\conj{\tau_{pq,sr}}.
\end{equation}

In Eq.~\eqref{eq:baseline_indices}, we defined a set of baseline indices to uniquely identify all baselines that produce a ghost in the ghost pattern for baseline $pq$. Since each term in Eq.~\eqref{eq:p_gen} is associated with a specific ghost, we can map our set of baseline indices to a set of ghost locations for baseline $pq$, i.e., we can define the mapping $M_{pq}: \mathbfss{P}(\mJ) \rightarrow \mathbfss{P}(\mathcal{Q}_{pq})$ as
\begin{equation}
\label{eq:mapping_indices}
M_{pq} \left ( \mJ \right ) = \bigcup_{rs \in \mJ} \left \{ \bmath{s}_{pq}^{rs} \right \},
\end{equation}
where $\mJ = \{rs\}_{r\neq s}$, $\mathcal{Q}_{pq} = \{\bs_{pq}^{rs}\}_{r\neq s}$ and $\mathbfss{P}(\cdot)$ denotes the power set of its operand (which has to be a set).
We can use Eq.~\eqref{eq:mapping_indices} to group all per-baseline phase-only ghost positions that are predicted by our perturbation analysis into one set
\begin{equation}
\label{eq:sources_pq_phase}
\mT_{pq} = M_{pq} \left ( \mI \right ).
\end{equation}
We are actually interested in studying the ghost pattern that is generated by the instrument as a whole. It is therefore more appropriate to merge the per-baseline perturbation results into one set $\mT$. We can construct $\mT$ as follows 
\begin{equation}
\label{eq:sources_phase}
\mT = \bigcup_{pq} \mT_{pq}. 
\end{equation}
Note that $\mT_{pq}$ and $\mT$ contain no recurring elements\footnote{Recall that $\{1,2,3\}\cup\{1,2,4\} = \{1,2,3,4\}$}. 

Let $\bzero = (0,0)$. Eq.~\eqref{eq:sources_phase} allows us to reformulate Eq.~\eqref{eq:p_gen} as
\begin{equation}
\label{eq:p_final_irr}
\rho_{pq}(\widehat{X}_{pq}^{-1}(\bal)) \approx \sum_{\tilde{\bmath{s}}\in\widehat{\mT}} \tau_{pq,\tilde{\bs}} e^{-2\pi i \bal^T\cdot \tilde{\bmath{s}}},
\end{equation}
where $\widehat{\mT} = \mT \cup \{\bzero\}$ and  
\begin{equation}
\label{eq:coeff_p_only}
\tau_{pq,\tilde{\bs}} =
\left\{ \begin{array}{ll}
\sum_{rs\in \mI_{pq,\tilde{\bmath{s}}}}\tau_{pq,rs} & \textrm{if}~\mI_{pq,\tilde{\bmath{s}}} \neq \emptyset\\
\tau_{pq,0} & \textrm{if}~\tilde{\bmath{s}} = \bmath{0} \\
0 & \textrm{otherwise} \end{array}\right..
\end{equation}
In Eq.~\eqref{eq:coeff_p_only}, $\mI_{pq,\tilde{\bmath{s}}}=\{rs|(\bmath{s}_{pq}^{rs}\in \mT_{pq}) \wedge (\tilde{\bmath{s}} = \bmath{s}_{pq}^{rs}) \wedge (\tilde{\bmath{s}} \neq \bzero)\}$.

\subsection{Distillation}
\label{ssec:dist}

In our analysis, we assume that our model is reasonably accurate, such that the deviations from the true gain values will be small. As a result, the ghosts will be very faint. We therefore have to distill $\Pcal$ and $\Pcal^{\odot-1}$ to make the ghosts visible. Distillation effectively removes all real emission from the image, such that we can image only the deviations from the true gain values (see Papers I and II). We can distill $\Pcal$ by subtracting $\bonem = \bone \bone^H$ from it, i.e.
\begin{equation}
\label{eq:dis_cal_pq_phase_old}
 [\Pcal(\widehat{X}_{pq}^{-1}(\bb))-\bonem]_{pq} \approx \sum_{\tilde{\bmath{s}} \in \mT} \tau_{pq,\tilde{\bs}} e^{-2\pi i \bal^T\cdot \tilde{\bmath{s}}}.
\end{equation}

The visibility phase calibration matrix can be described by
\begin{equation}
\label{eq:Pcal_inv}
\Pcal^{\odot-1} \approx \bone\bone^H + i\bone\Delta \bvarrho^H - i\Delta\bvarrho\bone^H,
\end{equation}
since $(\rho_{pq})^{-1} = \conj{\rho}_{pq}$. As was the case in Paper II, Eq.~\eqref{eq:Pcal_inv} implies that the brightest ghost positions are preserved under the Hadamard inverse. Although the positions are preserved, the fluxes of the ghosts change sign. Comparing Eqs.~\eqref{eq:Pcal_result} and \eqref{eq:Pcal_inv}, we can easily see that
\begin{equation}
\label{eq:dis_cal_pq_phase}
 [\Pcal^{\odot-1}(\widehat{X}_{pq}^{-1}(\bb))-\bonem]_{pq} \approx \sum_{\tilde{\bmath{s}}\in\mT} \tau_{pq,\tilde{\bs}}^{\odot-1} e^{-2\pi i \bal^T\cdot \tilde{\bmath{s}}}.
\end{equation}
where $\tau_{pq,\tilde{\bs}}^{\odot-1} = - \tau_{pq,\tilde{\bs}}$.
Note that the index set $\mT$ is used in Eq.~\eqref{eq:dis_cal_pq_phase_old} and Eq.~\eqref{eq:dis_cal_pq_phase}, while the index set $\widehat{\mT}$ was used in Eq.~\eqref{eq:p_final_irr}.
The reason being, according to the perturbation approach presented in Sec.~\ref{ssec:perturbation_phase_only}, there is no ghost at $\bzero$. 

\subsection{Corrected Visibilities}
\label{ssec:corr_vis}
The corrected visibility matrix $\Rcal_{\Pcal}^{(c)}$ was defined in Eq.~\eqref{eq:Rcal_c}. As in the case of $\Pcal$, distilling $\Rcal_{\Pcal}^{(c)}$ allows us to study the fainter spurious emission in $\Rcal_{\Pcal}^{(c)}$. The distilled corrected visibility matrix $\Rcal^{\Delta}_{\Pcal}$ (also known as the residual visibility matrix) can be calculated with 
\begin{equation}
\label{eq:cor_der1}
\Rcal^{\Delta}_{\Pcal} = \Rcal_{\Pcal}^{(c)}-\Rcal = (\Pcal^{\odot-1}-\bonem)\odot\Rcal, 
\end{equation}
implying that
\begin{equation}
\label{eq:cor_der2}
[\Rcal^{\Delta}_{\Pcal} (\widehat{X}_{pq}^{-1}(\bb))]_{pq} \approx \sum_{\tilde{\bs}\in\mT} \tau_{pq,\tilde{\bs}}^{\odot-1} e^{-2\pi i \bb^T\cdot \tilde{\bs}} (1 + A_2 e^{-2\pi i \bb^T \cdot \bs_0}).
\end{equation}

Inspecting Eq.~\eqref{eq:cor_der2} reveals that when we correct we actually convolve the atomic ghost pattern in $\Pcal^{\odot-1}-\breve{\bone}$ with the modelled and unmodelled source. Our corrected residuals are therefore made up of two ghost patterns, one around the modelled source and one around the unmodelled source, that were added together coherently (Papers I and II). 

\subsection{Imaging}
\label{ssec:imaging}

Imaging and how it relates to calibration systematics was discussed at length in Paper II.  In this section we will only summarize the most important quantities we obtained in Paper II. We denote the dirty images associated with $\Pcal^{\odot-1}-\breve{\bone}$ and $\Rcal^{\Delta}_{\Pcal}$ with $I_D^{\Pcal^{\odot-1}-\breve{\bone}}$ and $I_D^{\Rcal^{\Delta}_{\Pcal}}$ respectively. As we showed in Sec.~\ref{sec:extrap_phase_only} each baseline has its own unique ghost pattern. When we image, we effectively average all the individual per-baseline ghost patterns together. If a ghost happens to form consistently in each baseline its per-baseline flux will add up coherently. On the other hand, if a ghost is only present in one baseline averaging the different baselines together will result in a dimmer ghost. In mathematical terms, the aforementioned result translates into the following: if we assume natural weighting then we can calculate the flux of the ghost at $\tilde{\bs}$ in $I_D^{\Pcal^{\odot-1}-\breve{\bone}}$ with 
\begin{equation}
\label{eq:average_flux}
\xi_{\tilde{\bs}}^{\odot-1} = \pqavg{\mathscr{R} \left \{ \tau_{pq,\tilde{\bs}}^{\odot -1} \right \}}{p < q},
\end{equation}
where $\pqavg{\cdot}{p<q}$ denotes averaging over all the baselines and $\mathscr{R}\{\cdot\}$ denotes the real operator. In Paper II we showed that only the real flux of a ghost manifests at $\tilde{\bs}$, which explains why we only average the real flux together in Eq.~\eqref{eq:average_flux}.  It is actually more sensible to calculate the maximum possible flux that the ghost at position $\tilde{\bs}$ can contribute to the image, since some of the imaginary flux of a ghost manifests in the vicinity of $\tilde{\bs}$ due to the Ghost Spread Function (GSF) (Papers I and II). We therefore also define
\begin{equation}
\label{eq:total_flux}
\widehat{\xi}_{\tilde{\bs}}^{\odot-1} = \pqavg{ \left | \tau_{pq,\tilde{\bs}}^{\odot -1} \right |}{p<q}.
\end{equation}
Eq.~\eqref{eq:total_flux} represents the maximum flux the ghost at $\tilde{\bs}$ could have contributed to the image. It contributes this maximum amount if it is completely real, and can therefore be interpreted as a type of upper bound. Eq.~\eqref{eq:cor_der2} also allows us to define 
\begin{equation}
 \xi_{\tilde{\bs}}^{\Delta} = \xi_{\tilde{\bs}}^{\odot-1} + A_2\xi_{\tilde{\bs}-\bs_0}^{\odot-1}
\end{equation}
and
\begin{equation}
\label{eq:total_flux2}
 \widehat{\xi}_{\tilde{\bs}}^{\Delta} = \widehat{\xi}_{\tilde{\bs}}^{\odot-1} + A_2\widehat{\xi}_{\tilde{\bs}-\bs_0}^{\odot-1},
\end{equation}
which are the  $I_D^{\Rcal^{\Delta}_{\Pcal}}$ equivalent of Eq.~\eqref{eq:average_flux} and Eq.~\eqref{eq:total_flux}. 

\subsection{Augmented Per-baseline Ghost Taxonomy}
\label{ssec:tax_baseline}
Based on Paper II, the result equivalent of Eq.~\eqref{eq:g_pq_phase} for full-complex calibration is
\begin{equation}
\label{eq:g_pq_full complex}
\left [ \Gcal \right ]_{pq} = g_{pq} = c_{pq,0} + \sum_{rs\in\mJ} c_{pq,rs} e^{-2 \pi i \bb_{rs}^T \cdot \bs_0}, 
\end{equation}
where
\begin{equation}
c_{pq,0} = 1 + \frac{A_2}{N},  \label{eq:c0pq}
\end{equation}
and
\begin{equation}
c_{pq,rs} =
  \begin{cases}
   \frac{2A_2}{N} - \frac{A_2}{N^2} & \text{if } rs \in \mI_1  \\
   \frac{A_2}{N} - \frac{A_2}{N^2} & \text{if } rs \in \mI_2 \\
   \frac{A_2}{N} - \frac{A_2}{N^2} & \text{if }  rs \in \mI_3 \\
   -\frac{A_2}{N^2} & \text{if }  rs \in \mI_4 \\
   0 & \text{otherwise}
  \end{cases},
  \label{eq:cpqrs}
\end{equation}
\label{lastpage}
The index sets $\mI_1$, $\mI_2$ and $\mI_3$ are already defined in Eq.~\eqref{eq:index_sets}. The fourth index set is defined as
\begin{equation}
\label{eq:index_4}
\mI_4 = \{rs|(r\neq p) \wedge (s\neq q) \wedge (r\neq s)\}. 
\end{equation}
The index set $\mJ$ was defined in Sec.~\ref{sec:extrap_phase_only}.

We can make the following observations if we inspect Eq.~\eqref{eq:g_pq_phase} and Eq.~\eqref{eq:g_pq_full complex}: 
\begin{enumerate}
\item Some of the coefficients ($\sigma_{pq,rs}$ and $c_{pq,rs}$) in Eq.~\eqref{eq:g_pq_phase} and  Eq.~\eqref{eq:g_pq_full complex} are inversely proportional to $N$ while others are inversely proportional to $N^2$.
The ghosts that are associated with the coefficients that are inversely proportional to $N$ are called proto-ghosts, and those associated with the coefficients that 
are inversely proportional to $N^2$ are called deutero-ghosts (Paper II).
\item Some coefficients in Eq.~\eqref{eq:g_pq_phase} and  Eq.~\eqref{eq:g_pq_full complex} have a positive sign, while others have a negative sign. If the sign is positive and the ghost flux is completely real, it will manifest itself as a negative ghost, i.e. as a suppressing ghost. If the sign is negative and the ghost flux is completely real, it will manifest itself as a positive ghost, i.e. as an amplifying ghost.
The inverse relation between the sign of the coefficients and whether a ghost manifests as a positive or negative ghost is due to the the fact that $(\rho_{pq})^{-1} = \conj{\rho}_{pq}$ (also see Eq.~\eqref{eq:Pcal_inv}).
\end{enumerate}
We can augment the per-baseline taxonomy of Paper II by using the second observation. The details of this proposed augmentation follows below.

Assigning the false color blue to the coefficients with a positive sign and the false color red to those with a negative sign, enables us to divide $\mI$ into the following mutually exclusive subsets: $\mI_{PB} = \bigcup_{i=1}^3 \mI_i$, $\mI_{PR} = \bigcup_{i=1}^3 \mI_i^\prime$, $\mI_{DB} = \emptyset$ and $\mI_{DR} = \emptyset$.
Similarly, we can divide $\mJ$ into $\mJ_{PB} = \bigcup_{i=1}^3 \mI_i$, $\mJ_{PR} = \emptyset$, $\mJ_{DB} = \emptyset$ and $\mJ_{DR} = \mI_4$. The aforementioned index sets are also presented in Table~\ref{tab:tax_per_baseline_vis_index}.

\begin{table*}
\caption{Comparison between the composite baseline index sets produced by phase-only calibration and full-complex calibration. The entries in the last row  of each subtable were obtained
by taking the union of all the other entries in the column in which they reside. The same applies to the entries in the last column of each subtable.}
\label{tab:tax_per_baseline_vis_index}
\begin{tabular}{|l|c|c|c||c|c|c|}
\cline{2-7}
\multicolumn{1}{c}{} & \multicolumn{3}{|c||}{phase-only} & \multicolumn{3}{|c|}{full-complex}\\
\cline{1-7}
\rule{0pt}{2.5ex} sign & $\frac{1}{N}$ & $\frac{1}{N^2}$ & all & $\frac{1}{N}$ & $\frac{1}{N^2}$ & all\\
\cline{1-7}
\rule{0pt}{2.5ex}\textcolor{blue}{$+$} & $\mI_{PB} = \bigcup_{i=1}^3 \mI_i$ & $\mI_{DB} = \emptyset$ & $\mI_B = \mI_{PB} \cup \mI_{DB}$ & $\mJ_{PB} = \bigcup_{i=1}^3 \mI_i$ & $\mJ_{DB} = \emptyset$ & $\mJ_B = \mJ_{PB} \cup \mJ_{DB}$\\
\rule{0pt}{2.5ex}\textcolor{red}{$-$} & $\mI_{PR} = \bigcup_{i=1}^3 \mI_i^\prime$ & $\mI_{DR} = \emptyset$ & $\mI_R = \mI_{PR} \cup \mI_{DR}$ & $\mJ_{PR} = \emptyset$ & $\mJ_{DR} = \mI_4$ & $\mJ_R = \mJ_{PR} \cup \mJ_{DR}$\\
\cline{1-7}
\multicolumn{1}{c|}{} & $\mI_P = \mI_{PB} \cup \mI_{PR}$ & $\mI_D = \emptyset$ & $\mI$ & $\mJ_P = \mJ_{PB} \cup \mJ_{PR}$ & $\mJ_D = \mJ_{DB} \cup \mJ_{DR}$ & $\mJ$\\
\cline{2-7}
\end{tabular}

\caption{Comparison between the index sets of ghost locations produced by phase-only calibration and full-complex calibration using the augmented taxonomy with blue and red proto-ghosts and blue and red deutero-ghosts.
Since $M_{pq}$ is structure preserving, the entries in the last row of each subtable can also be obtained by taking the union of all the other entries in the column in which they reside. The same applies to the entries in the last column of each subtable. 
Note that in the table we have used the shorthand $\tilde{0}$ to denote $\{\bzero\}$.
}
\label{tab:tax_per_baseline_pos_index}
\begin{tabular}{|l|c|c|c||c|c|c|}
\cline{2-7}
\multicolumn{1}{c}{} & \multicolumn{3}{|c||}{phase-only} & \multicolumn{3}{|c|}{full-complex}\\
\cline{1-7}
color & proto & deutero & all & proto & deutero & all\\
\cline{1-7}
\rule{0pt}{2.5ex}\textcolor{blue}{B} & $\mT_{pq}^{PB} = M_{pq} \left ( \mI_{PB} \right )$ & $\mT_{pq}^{DB} = \emptyset$ & $\mT_{pq}^B = M_{pq}(\mI_B)$ & $\mS_{pq}^{PB} = M_{pq} \left ( \mJ_{PB} \right ) \cup \tilde{0}$ & $\mS_{pq}^{DB} = \emptyset$ & $\mS_{pq}^B =M_{pq}\left ( \mJ_B \right )\cup \tilde{0}$\\
\rule{0pt}{2.5ex}\textcolor{red}{R} & $\mT_{pq}^{PR} = M_{pq} \left ( \mI_{PR} \right )$ & $\mT_{pq}^{DR} = \emptyset$ & $\mT_{pq}^R = M_{pq}(\mI_R)$ & $\mS_{pq}^{PR} = \emptyset$ & $\mS_{pq}^{DR} = M_{pq} \left ( \mJ^{DR} \right )$ & $\mS_{pq}^R = M_{pq}\left ( \mJ_R \right )$\\
\cline{1-7}
\multicolumn{1}{c|}{} & \rule{0pt}{2.5ex}  $\mT_{pq}^P = M_{pq} \left ( \mI_P \right )$ & $\mT_{pq}^D = \emptyset$ & $\mT_{pq} = M_{pq} \left ( \mI \right )$ & $\mS_{pq}^P= M_{pq} \left ( \mJ_P \right )\cup \tilde{0}$ & $\mS_{pq}^D= M_{pq} \left ( \mJ_D \right )$ & $\mS_{pq}=M_{pq} \left ( \mJ \right )\cup \tilde{0}$\\
\cline{2-7}
\end{tabular}
\end{table*}


Using $M_{pq}$ we can map the sets of baseline indices to sets of ghost positions. Interestingly enough, when the array is not redundant then $M_{pq}$ is an isomorphism, i.e., it is structure preserving. If we inspect Table~\ref{tab:tax_per_baseline_vis_index} whilst keeping the aforementioned fact in mind we realize that $\mT_{pq} = M_{pq} \left ( \mI  \right )$ can be divided into the following mutually exclusive subsets: $\mT_{pq}^{PR}$, $\mT_{pq}^{PB}$, $\mT_{pq}^{DR}$  and  $\mT_{pq}^{DB}$. We can use a similar argument to divide the set $\mS_{pq} = M_{pq} \left ( \mJ  \right )\cup\{\bzero\}$ into four mutually exclusive subsets, namely $\mS_{pq}^{PR}$, $\mS_{pq}^{PB}$, $\mS_{pq}^{DR}$  and  $\mS_{pq}^{DB}$. The per-baseline ghosts can therefore, by extension, also be divided into four categories, namely \emph{red proto-ghosts} (associated with $\mT_{pq}^{PR}$ and $\mS_{pq}^{PR}$), \emph{blue proto-ghosts} (associated with $\mT_{pq}^{PB}$ and $\mS_{pq}^{PB}$), \emph{red deutero-ghosts} (associated with $\mT_{pq}^{DR}$ and $\mS_{pq}^{DR}$) and \emph{blue deutero-ghosts} (associated with $\mT_{pq}^{DB}$ and $\mS_{pq}^{DB}$). The augmented per-baseline ghost taxonomy is summarized in Table~\ref{tab:tax_per_baseline_pos_index}.

We can make a number of interesting observations from Table~\ref{tab:tax_per_baseline_pos_index}. First, phase-only calibration does not produce deutero-ghosts, which implies that the flux of all ghosts only decreases in proportion to the number of antennas in the array. Moreover, phase-only calibration produces an equal amount of red (potential positive/amplification) and blue (potential negative/suppressing)\footnote{The sign refers to the sign of the ghost itself and not to the sign of the coefficient associated with it.} ghosts. In the case of full-complex calibration, all red ghosts are deutero-ghosts while all blue ghosts are proto-ghosts. 
As a result, the red ghosts will typically be less bright in the case of full-complex calibration than they will be in the case of phase-only calibration. Moreover, the red ghosts will be less bright than the blue ghosts in the case of full-complex calibration, while they are of similar brightness in the case of phase-only calibration.


Note that our classification of ghosts assumes that the array is not redundant. If the array is redundant, some per-baseline-ghosts would have the same position vector and would therefore be added together; changing their characteristics. This would require a more elaborate classification scheme, which we do not pursue here, since it not necessary to understand the symmetry relationships that exist between pairs of ghosts.


\section{Comparison}
\label{sec:comparison}

In this section we compare the full-complex ghost response of Paper II with the phase-only results we obtained in Sec.~\ref{sec:g_pat_phase}. We first consider the ghost-pattern of an individual baseline in Sec.~\ref{sec:com_bas} before studying the conglomerated patterns in Sec.~\ref{sec:com_con}. All of the results in this section were obtained using the seven dish Karoo Array Telescope (KAT-7) array layout. KAT-7 is a precursor to the MeerKAT array and is located 
in the Karoo (South Africa). The layout of this array was depicted in Paper II. Moreover, we only consider the two-source test case from Sec.~\ref{ssec:two_source} with the unmodeled source being 0.2 Jy at $(1^{\circ},0^{\circ})$, $\delta_0$ = -74.66$^{\circ}$ and $\nu$ = 1.445 GHz. We denote the observational frequency with $\nu$.

\subsection{Per-baseline Results}
\label{sec:com_bas}

\begin{figure*}
\centering
\subfigure[full-complex: $\bb_{45}$ -- Real]
{\includegraphics[width=0.34\textwidth]{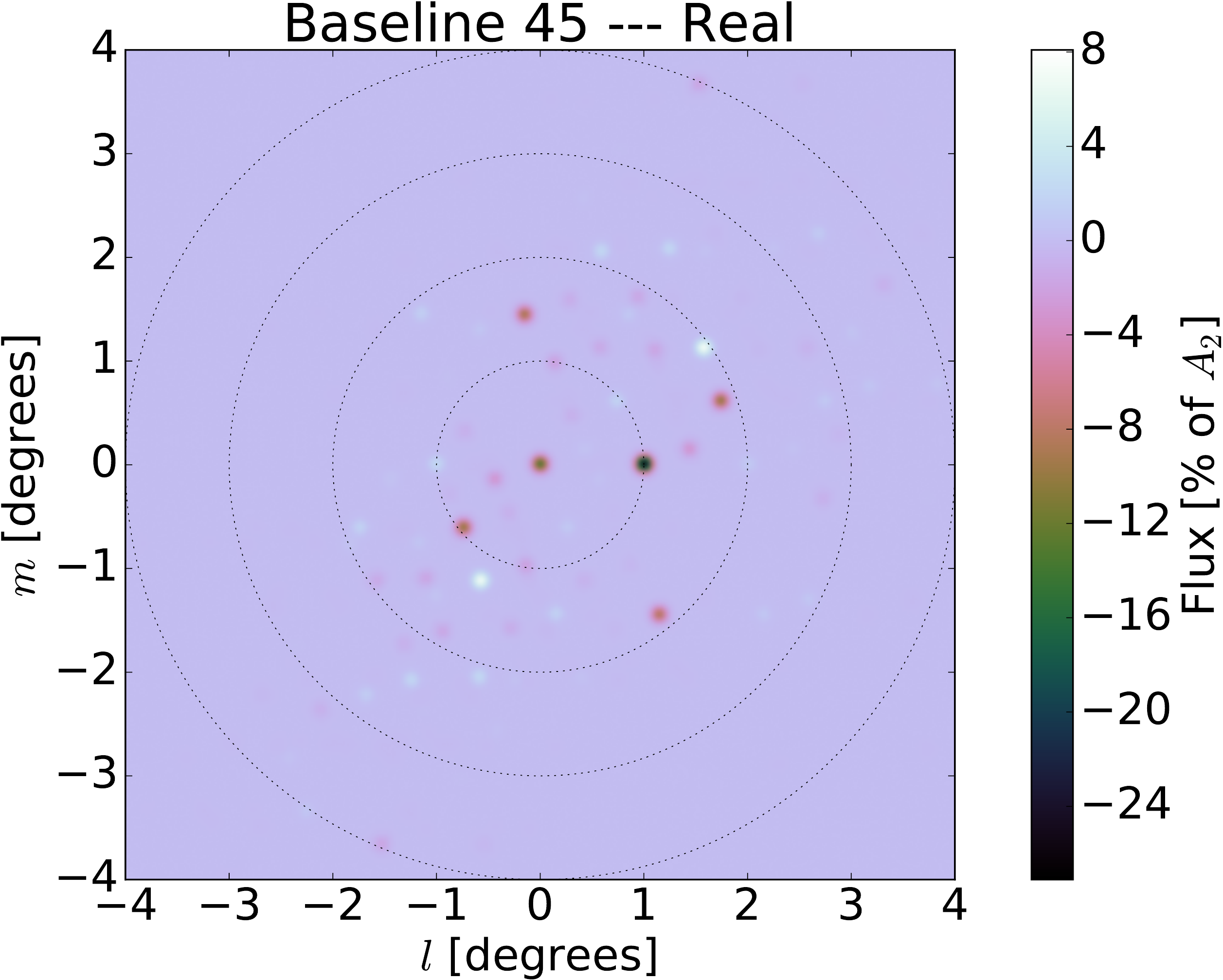}\label{fig:stef_r}}
\subfigure[full-complex: $\bb_{45}$ -- Imag]
{\includegraphics[width=0.34\textwidth]{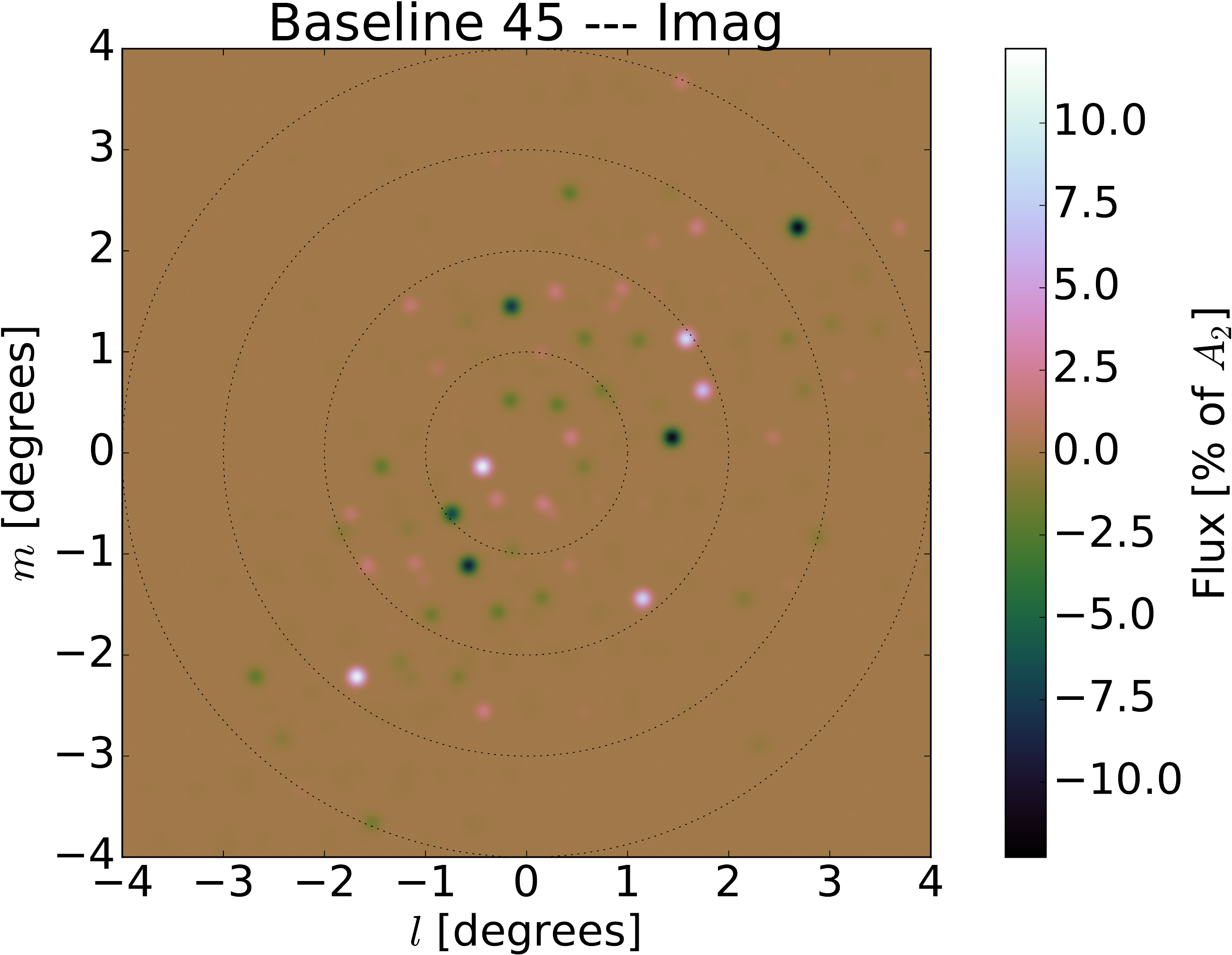}\label{fig:stef_i}}
\subfigure[full-complex: ghost pattern]
{\includegraphics[width=0.257\textwidth]{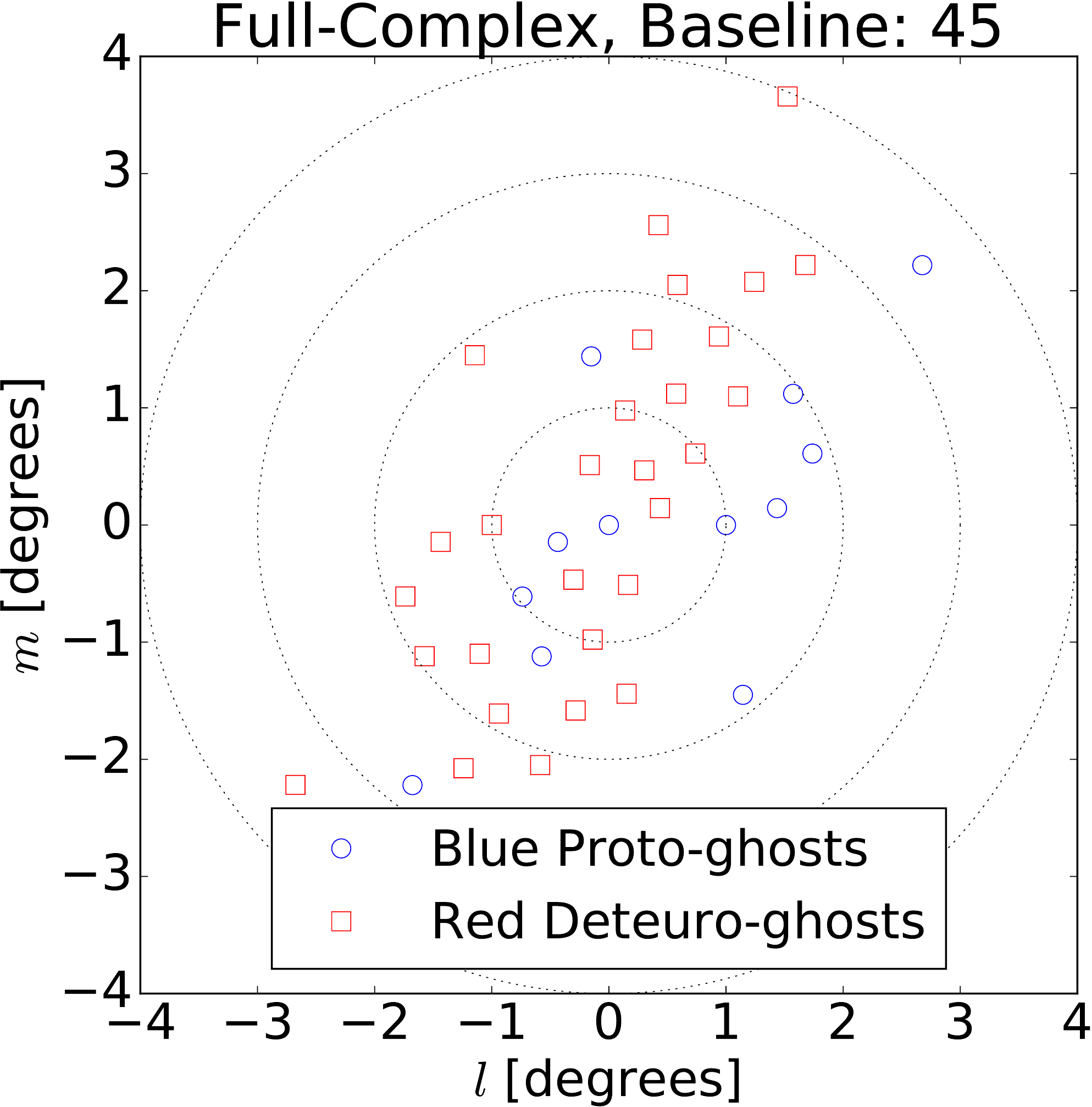}\label{fig:full_pat_45}}

\subfigure[phase-only: $\bb_{45}$ -- Real]
{\includegraphics[width=0.34\textwidth]{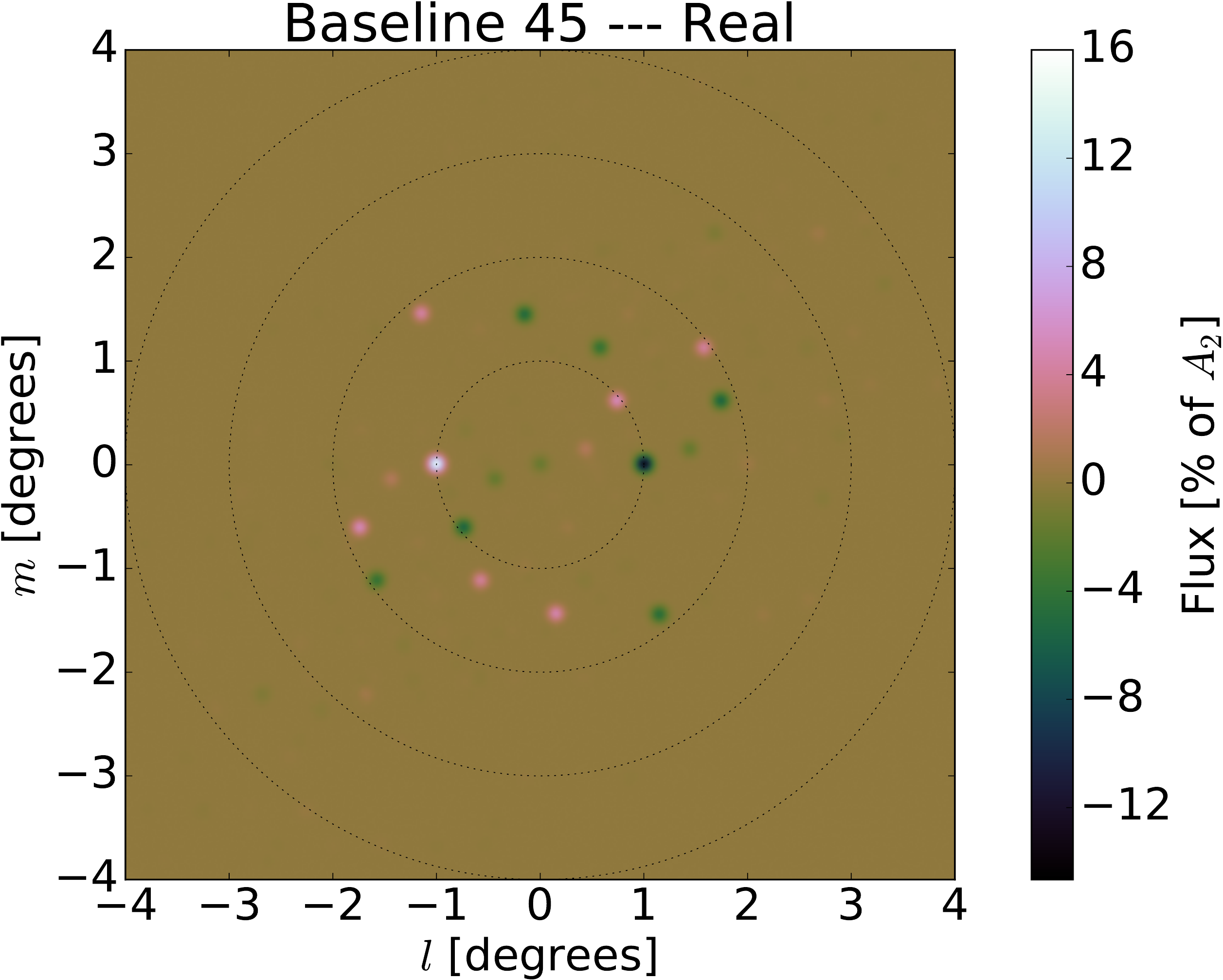}\label{fig:phase_r}}
\subfigure[phase-only: $\bb_{45}$ -- Imag]{\includegraphics[width=0.34\textwidth]{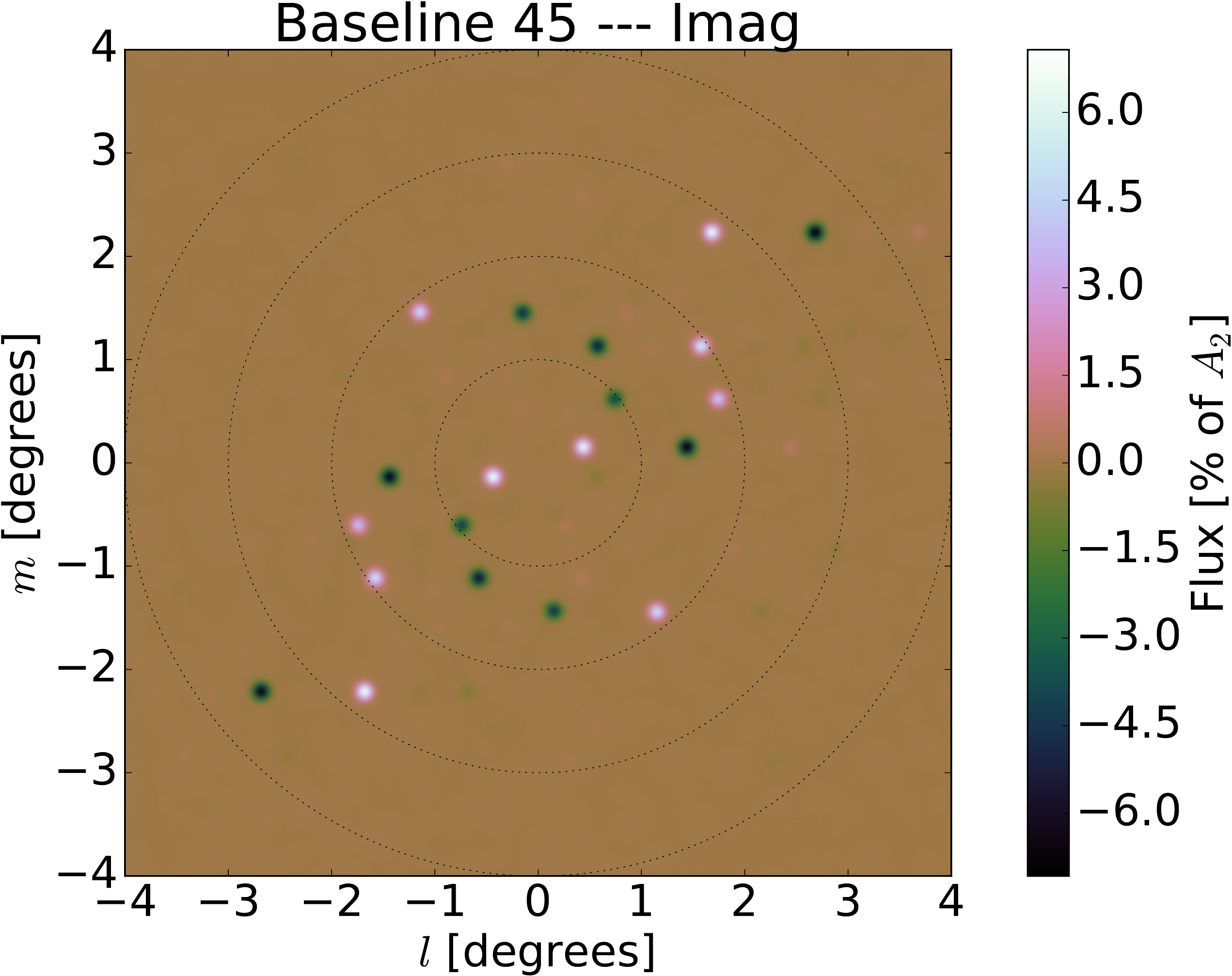}
\label{fig:phase_i}}
\subfigure[phase-only: ghost pattern]
{\includegraphics[width=0.257\textwidth]{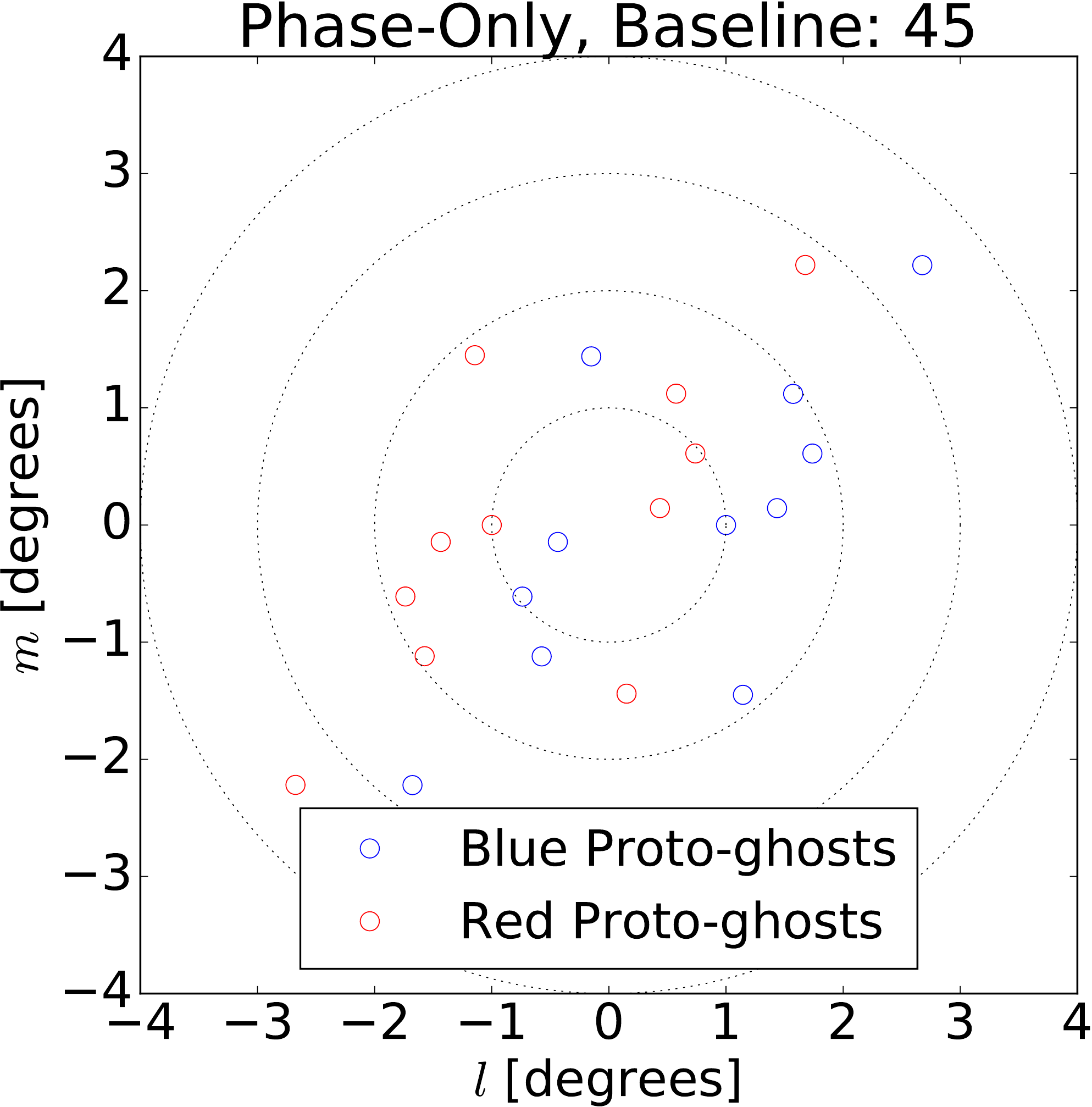}\label{fig:phase_pat_45}}
\caption{The left and middle column show the real (left column) and imaginary (middle column) part of the extrapolated artefact map for baseline $\bb_{45}$ obtained for full-complex calibration (top row) and phase-only calibration (bottom row). These maps were created by using KAT-7. The modelled source was in the center and the unmodelled source (0.2 Jy) was at $(1^{\circ},0^{\circ})$. Moreover, $\delta_0$ = -74.66$^{\circ}$ and $\nu$ = 1.445 Ghz. The two panels in the right column show the theoretical ghost patterns for full-complex (top) and phase-only (bottom) calibration for this scenario. \label{fig:ghost_pattern_bas_45}}
\end{figure*}

In Fig.~\ref{fig:ghost_pattern_bas_45} we present extrapolated artefact maps which were created by using the KAT-7 array layout. More specifically, we have plotted $\mathcal{F}^{-1}\{g_{45}^{-1}(\widehat{X}_{45}^{-1}(\bal))-1\}$ and $\mathcal{F}^{-1}\{\rho_{45}^{-1}(\widehat{X}_{45}^{-1}(\bal))-1\}$ in Figs.~\ref{fig:stef_r}--\ref{fig:stef_i} and Figs.~\ref{fig:phase_r}--\ref{fig:phase_i} respectively. The inverse Fourier transform is denoted by $\mathcal{F}^{-1}\{\cdot\}$. Fig.~\ref{fig:full_pat_45} and Fig.~\ref{fig:phase_pat_45} show the theoretical ghost patterns based on the index sets summarized in Table~\ref{tab:tax_per_baseline_pos_index}. The theoretical ghost positions shown in Figs.~\ref{fig:full_pat_45} and \ref{fig:phase_pat_45} help us visualize the similarities and differences that exist between full-complex calibration and phase-only calibration. The brightest ghosts in the artefact maps match perfectly with the theoretical ghost positions; indicating that the results from raw extrapolation and perturbation align well.   

We can draw the following conclusions from Table~\ref{tab:tax_per_baseline_pos_index} and Fig.~\ref{fig:ghost_pattern_bas_45}:

\begin{enumerate}
\item \emph{Proto-ghosts}: Full-complex calibration only produces blue proto-ghosts. In contrast, phase-only calibration produces an equal number of red and blue proto-ghosts. The blue and red proto-ghosts form symmetrically opposite each other around the modelled source. Moreover, the set of blue proto-ghosts that are created by phase-only calibration and full-complex calibration are almost equal.
\item \emph{Deutero-ghosts}: All the deutero-ghosts that are produced by full-complex calibration are red. No deutero-ghosts are generated by phase-only calibration. The position vector set associated with the red proto-ghosts is a subset of the position vector set that is associated with the full-complex deutero-ghosts, i.e.  $\mT_{pq}^{PR}\subset\mS_{pq}^{DR}$. We therefore produce fewer first order ghosts, i.e. the ghosts associated with the first order perturbation of the visibility gains, with phase-only calibration when compared with full-complex calibration. We do however produce more proto-ghosts.
\end{enumerate}

In Paper II we realized that the following three ghosts were exceptionally bright:
\begin{itemize}
\item The \emph{primary suppressor}. The primary suppressor forms on top of the modelled source, and it can therefore suppress the modelled source. It is important to mention here that, in the full-complex case, the primary suppressor only forms when we incorporate the autocorrelations during calibration, which is rarely done in practice (Paper II). The primary suppressor forms at $\bzero$.
\item The \emph{secondary suppressor}. The secondary suppressor lies on top of the unmodelled source, i.e. at $\bs_0$. This ghost can suppress the unmodelled source.  
\item The \emph{anti-ghost}. The anti-ghost appears symmetrically opposite the secondary suppressor around the primary suppressor, i.e. at $-\bs_0$. 
\end{itemize}
Concentrating on these specific sources in Table~\ref{tab:tax_per_baseline_pos_index} and Fig.~\ref{fig:ghost_pattern_bas_45}, we find:
\begin{enumerate}
\item \emph{Primary suppressor}: The primary suppressor is a blue proto-ghost, but only forms if full-complex calibration is employed. Mathematically speaking $\mS_{pq}^{PB}\backslash\mT_{pq}^{PB} = \{\bzero\}$. We denote set-theoretic difference by ``$\backslash$''. Intuitively, this is in agreement with the fact that the primary suppressor disappears when the autocorrelations are ignored in full-complex calibration, since the autocorrelations do not contain phase information. As we showed in Paper II, the primary suppressor is completely real and since it is blue it will appear as a negative source. This can be confirmed by inspecting Figs.~\ref{fig:stef_r}--\ref{fig:stef_i}.
\item \emph{Secondary suppressor and its anti-ghost}: In the case of full-complex calibration, the secondary suppressor is a blue proto-ghost, and its anti-ghost is a red deutero-ghost. As is the case with full-complex calibration the phase-only secondary suppressor is also a blue proto-ghost, but its anti-ghost is a red proto-ghost. This explains why the phase-only anti-ghost is so much brighter than the full-complex anti-ghost in Fig.~\ref{fig:ghost_pattern_bas_45}. In Paper II we showed that the secondary suppressor and its anti-ghost are real valued. The fact that the secondary suppressor is real valued and blue explains why the secondary suppressor ends up being a negative ghost. The anti-ghost of the secondary suppressor manifests as a positive ghost, because it is real valued and red.  
\end{enumerate}

\subsection{Conglomerated Results}
\label{sec:com_con}

In the previous section we only looked at the per-baseline ghost response but, as we mentioned in Sec.~\ref{ssec:imaging}, the conglomerated ghost pattern and the ghost pattern of each individual baseline are completely different. It turns out that some ghost add up coherently for all baselines while others do not. In Paper II we developed the following taxonomy for ghosts in the conglomerated pattern to address this issue:
\begin{enumerate}
\item \emph{Line ghosts}: All ghosts that form at $\tilde{\bs}\in \bs_0\mathbfss{Z}$ are line ghosts. Line ghosts are present in each baseline and therefore add up coherently when we image.
\item \emph{Scattered ghosts}: If a ghost is not a line ghost, it is a scattered ghost. In the non-redundant case, scattered ghosts are only present in one baseline, which is why they appear dimmer in the conglomerated ghost pattern.
\end{enumerate}

\begin{figure}
\centering
\subfigure[full-complex]
{\includegraphics[width=0.8\columnwidth]{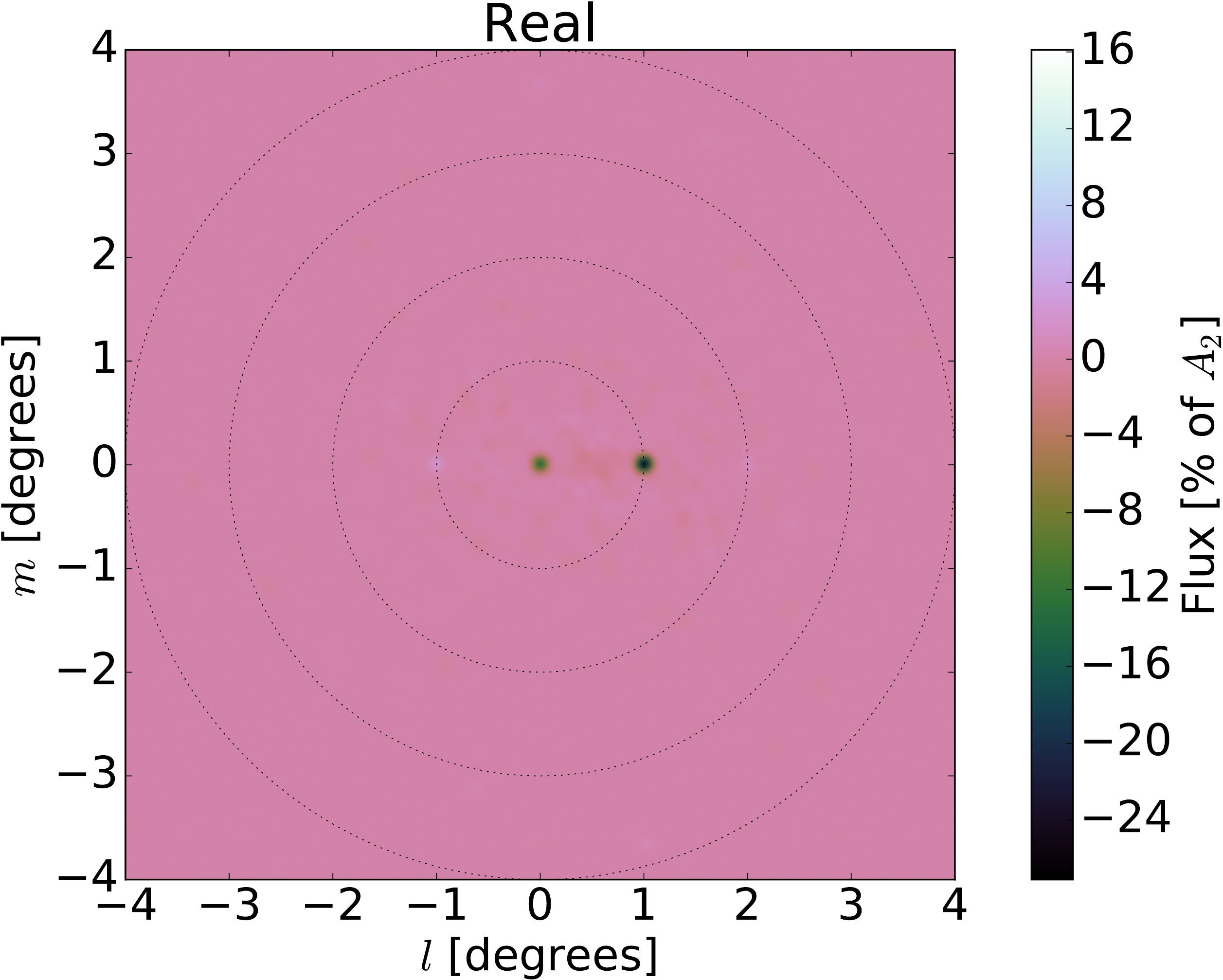}
\label{fig:stef_whole}}
\subfigure[phase-only]
{\includegraphics[width=0.8\columnwidth]{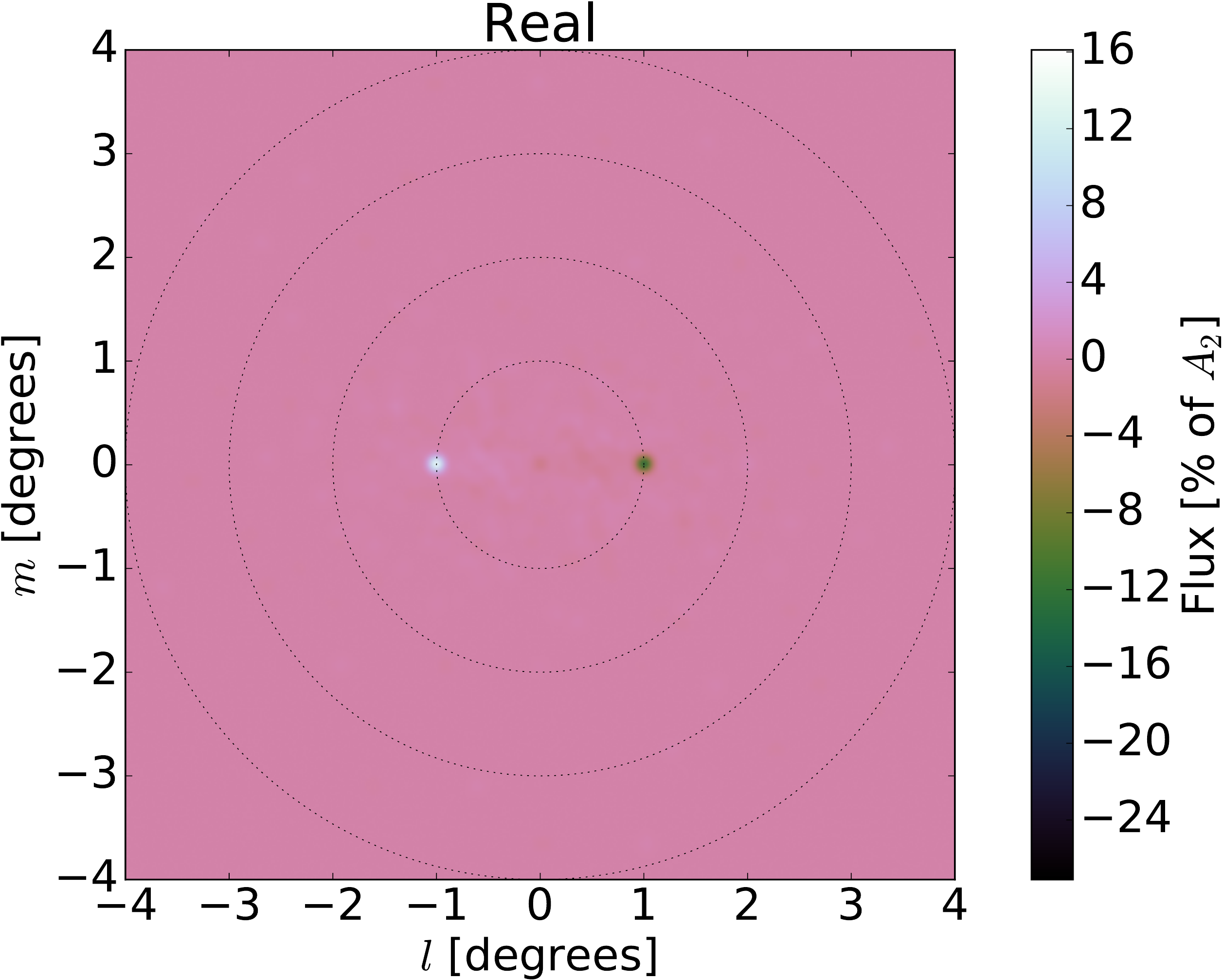} \label{fig:phase_whole}}
\caption{Conglomerated ghost pattern for full-complex calibration (top) and phase-only calibration (bottom). The suppression ghost, the ghost forming on top of the unmodelled source, is brighter in the top panel than in the bottom panel. In the case of the anti-ghost, the ghost forming symmetrically opposite the suppression ghost around the modelled source, we observe the exact opposite. Moreover, fewer bright ghosts are visible in the bottom panel than in the top panel.
\label{fig:con_patterns}} 
\end{figure}

The extrapolated conglomerated ghost patterns for full-complex and phase-only calibration for the full KAT-7 array are shown in Fig.~\ref{fig:stef_whole} and Fig.~\ref{fig:phase_whole} respectively. These maps are clearly different from the per-baseline results shown in Fig.~\ref{fig:ghost_pattern_bas_45}. One of the major differences is that the secondary suppressor and its anti-ghost are much brighter relative to the remaining background ghosts.

\begin{figure*}
\centering
\subfigure[$\Gcal^{\odot-1} - \breve{\bone}$: Line]{\includegraphics[width=0.4\textwidth]{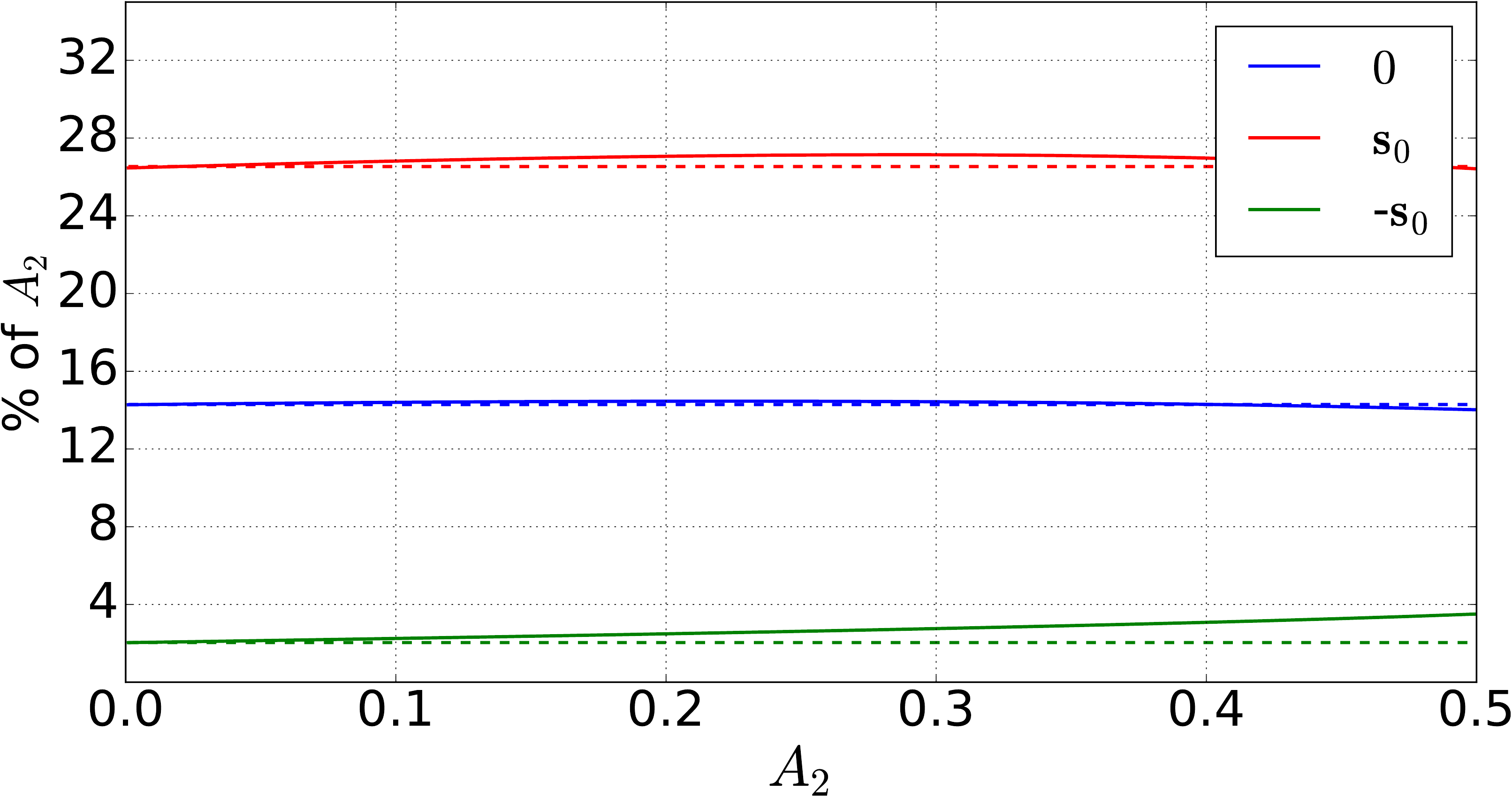}\label{fig:g_line}}
\subfigure[$\Gcal^{\odot-1} - \breve{\bone}$: Scattered]{\includegraphics[width=0.4\textwidth]{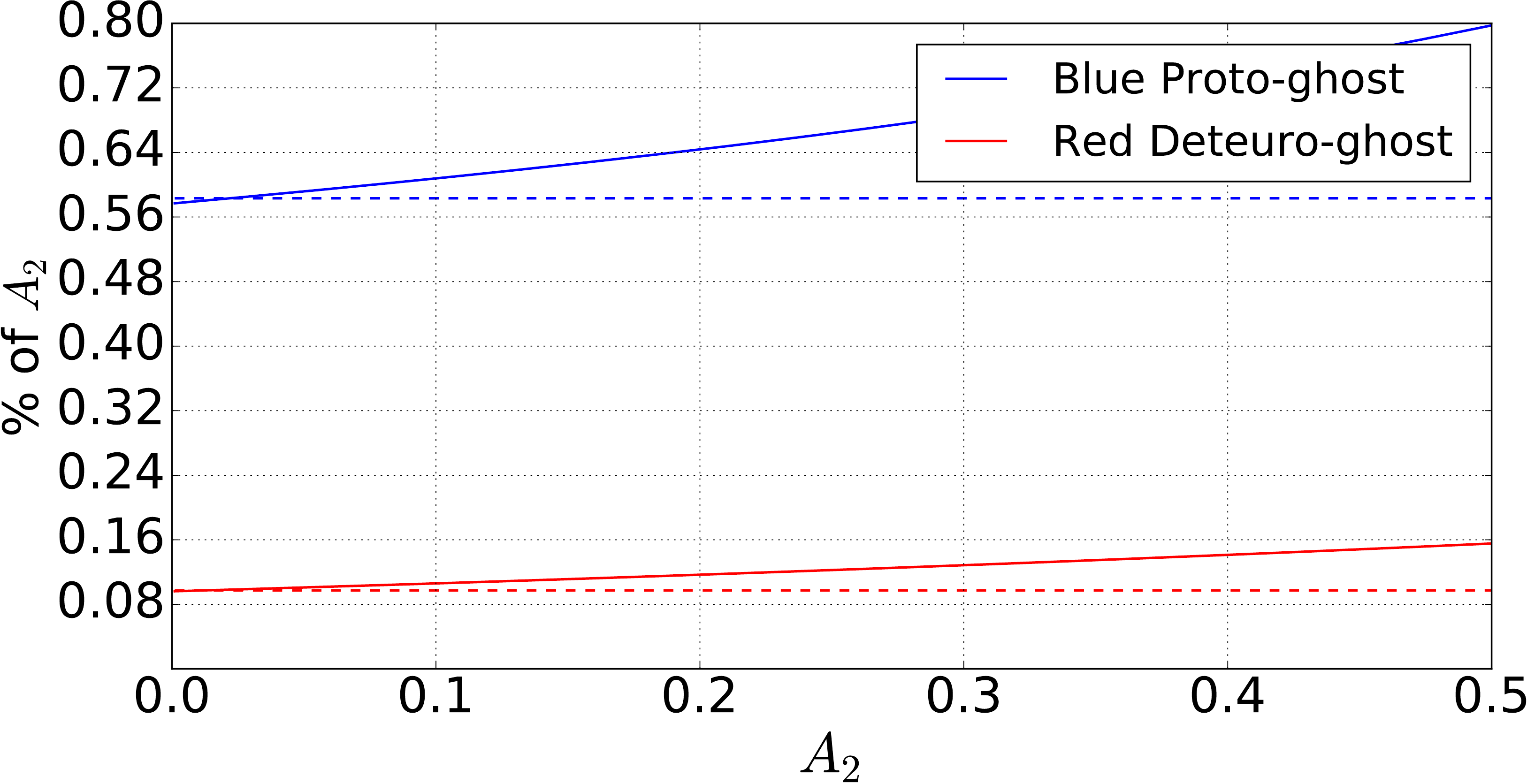}\label{fig:g_sc}}

\subfigure[$\Rcal^{\Delta}_{\Gcal}$: Line]{\includegraphics[width=0.4\textwidth]{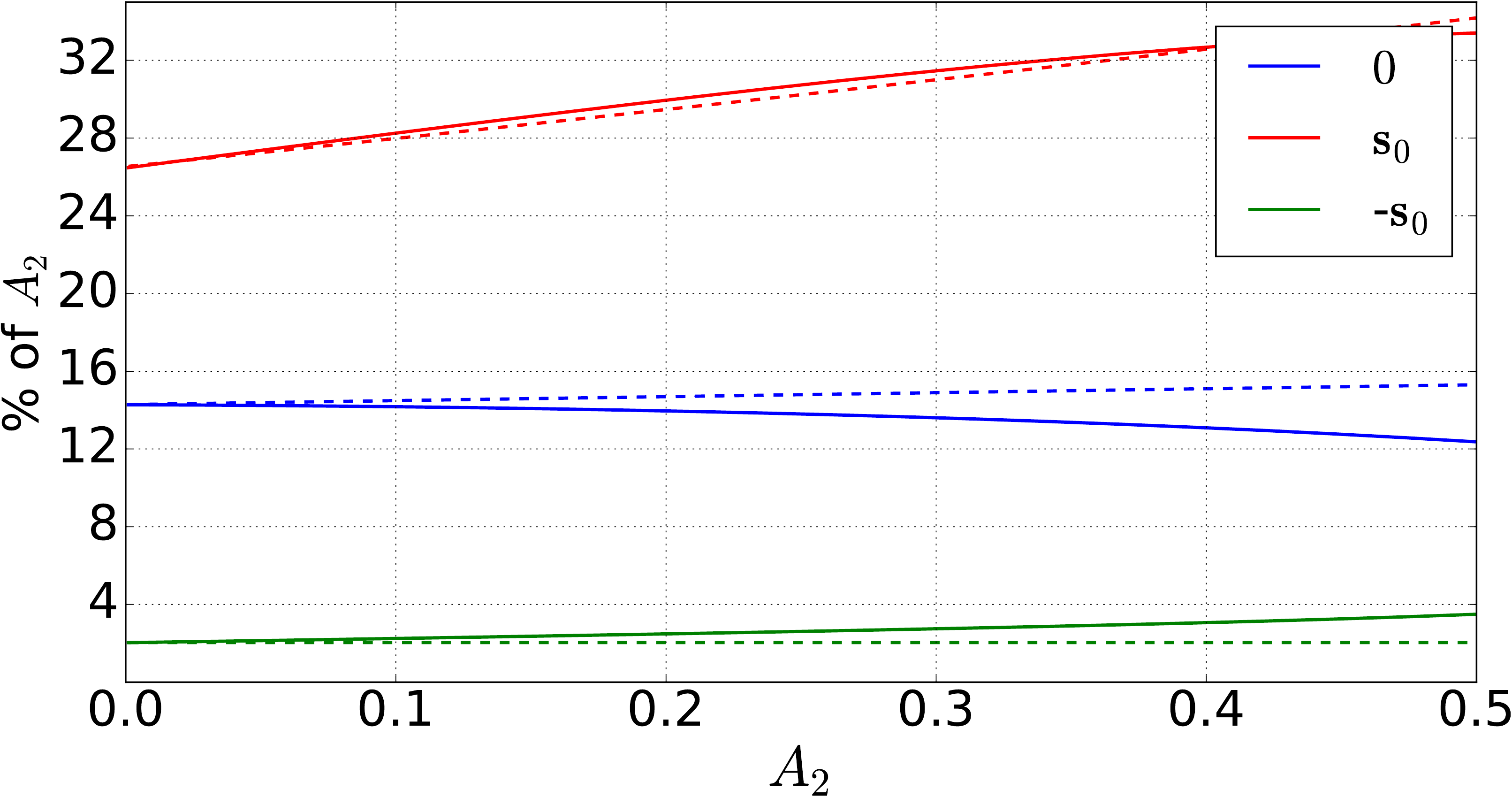}\label{fig:gr_line}}
\subfigure[$\Rcal^{\Delta}_{\Gcal}$: Scattered]{\includegraphics[width=0.4\textwidth]{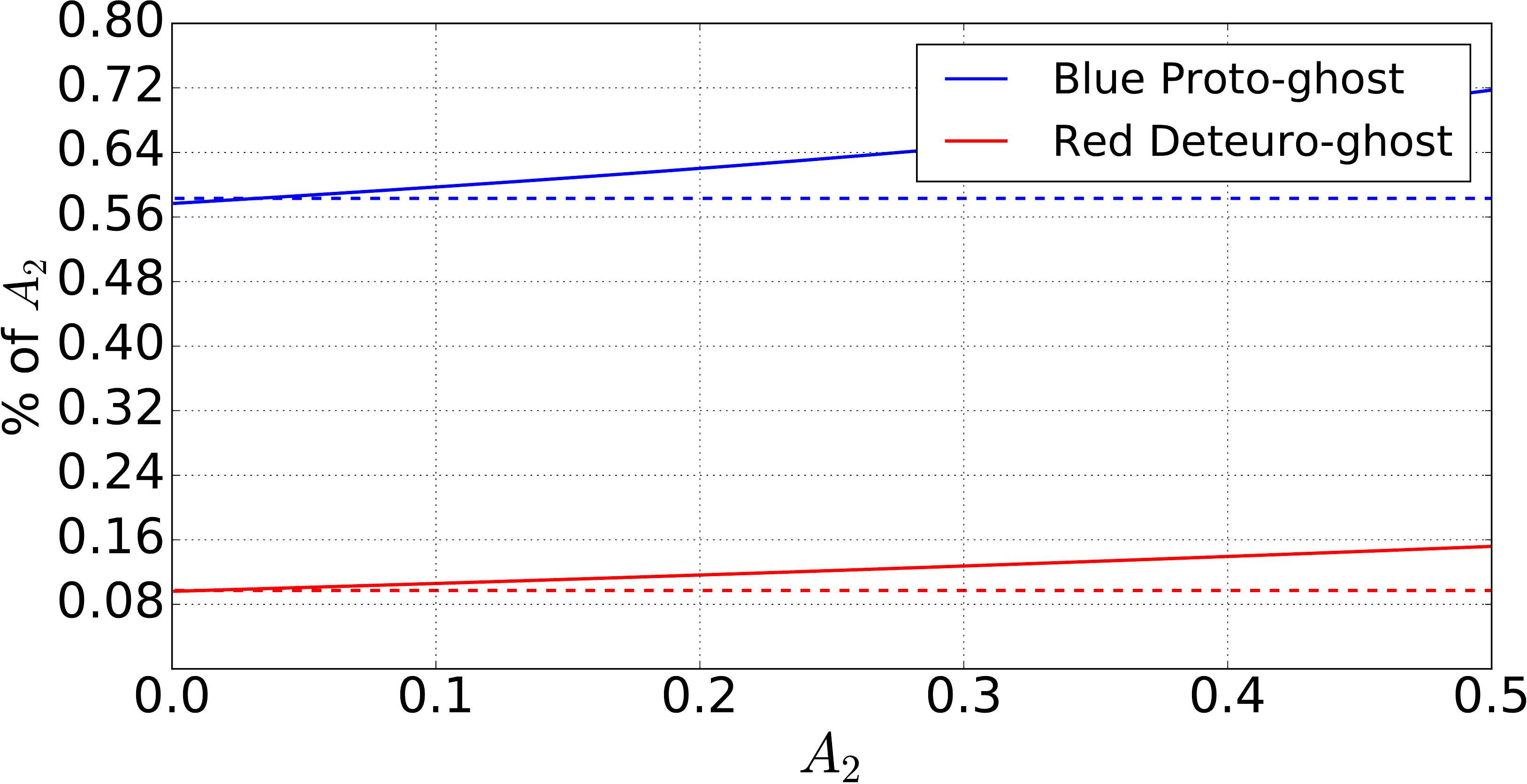}\label{fig:gr_sc}}
\caption{The relative amplitude  of the primary suppressor, secondary suppressor and its anti-ghost on the left and the relative amplitude of a random proto-ghost and its anti-ghost on the right. These curves were created by employing full-complex calibration. The dashed line represents the theoretically derived values that we obtained with perturbation, while the solid line represents a measured value which was obtained from a clean
artefact map that was created via extrapolation. \label{fig:GT_curves}} 
\end{figure*}

\begin{figure*}
\centering
\subfigure[$\Pcal^{\odot-1} - \breve{\bone}$: Line]{\includegraphics[width=0.4\textwidth]{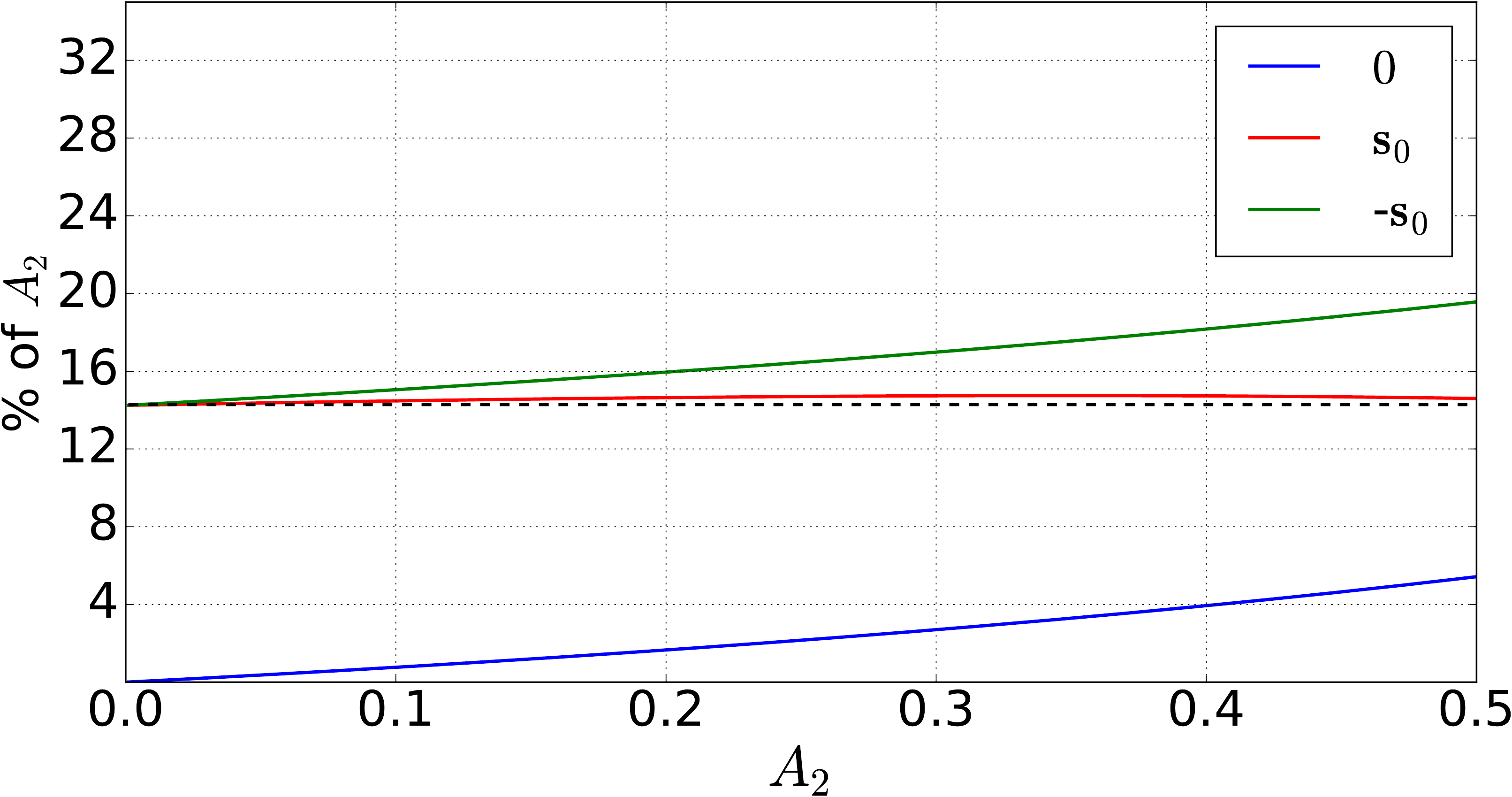}\label{fig:p_line}}
\subfigure[$\Pcal^{\odot-1} - \breve{\bone}$: Scattered]{\includegraphics[width=0.4\textwidth]{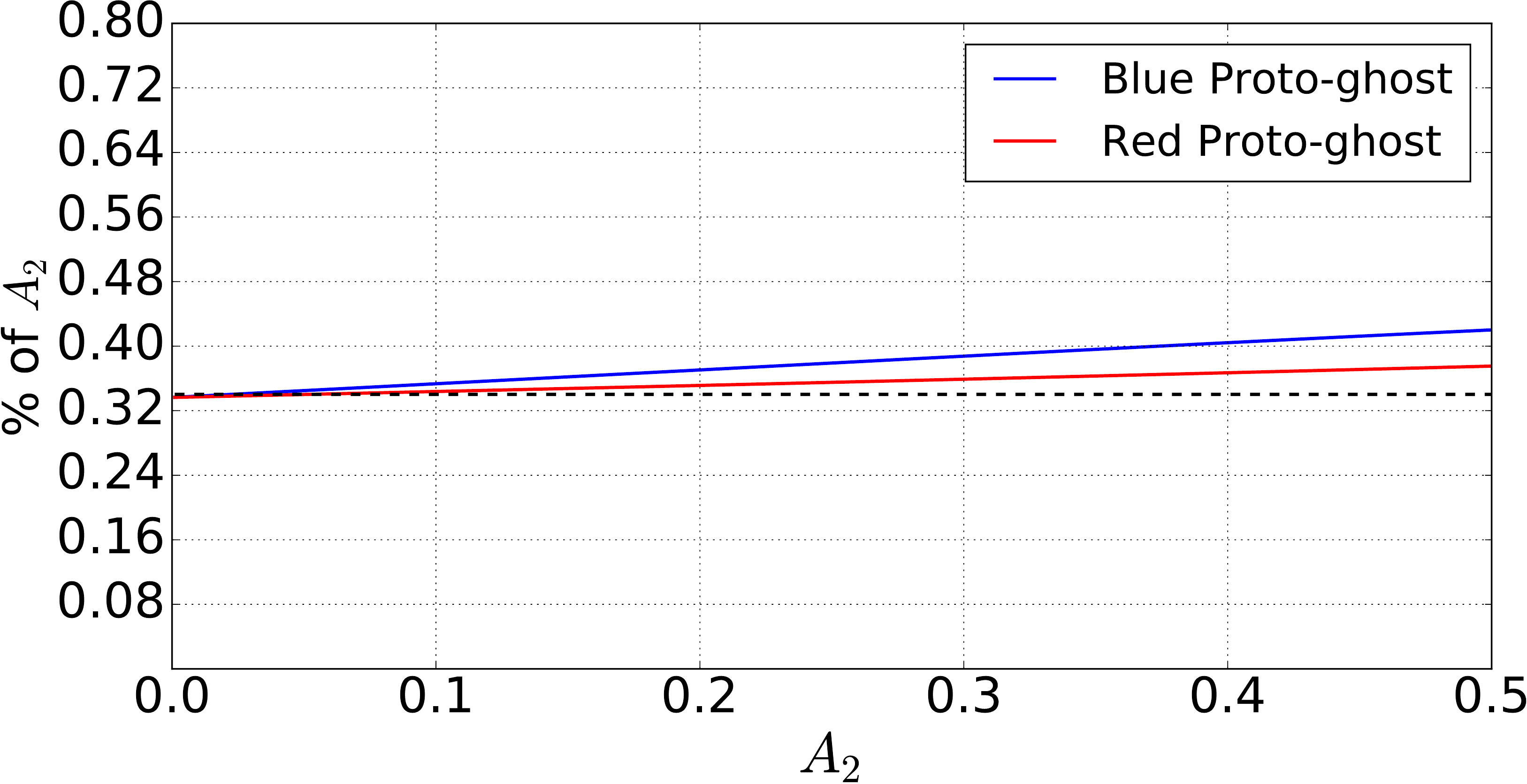}\label{fig:p_sc}}

\subfigure[$\Rcal^{\Delta}_{\Pcal}$: Line]{\includegraphics[width=0.4\textwidth]{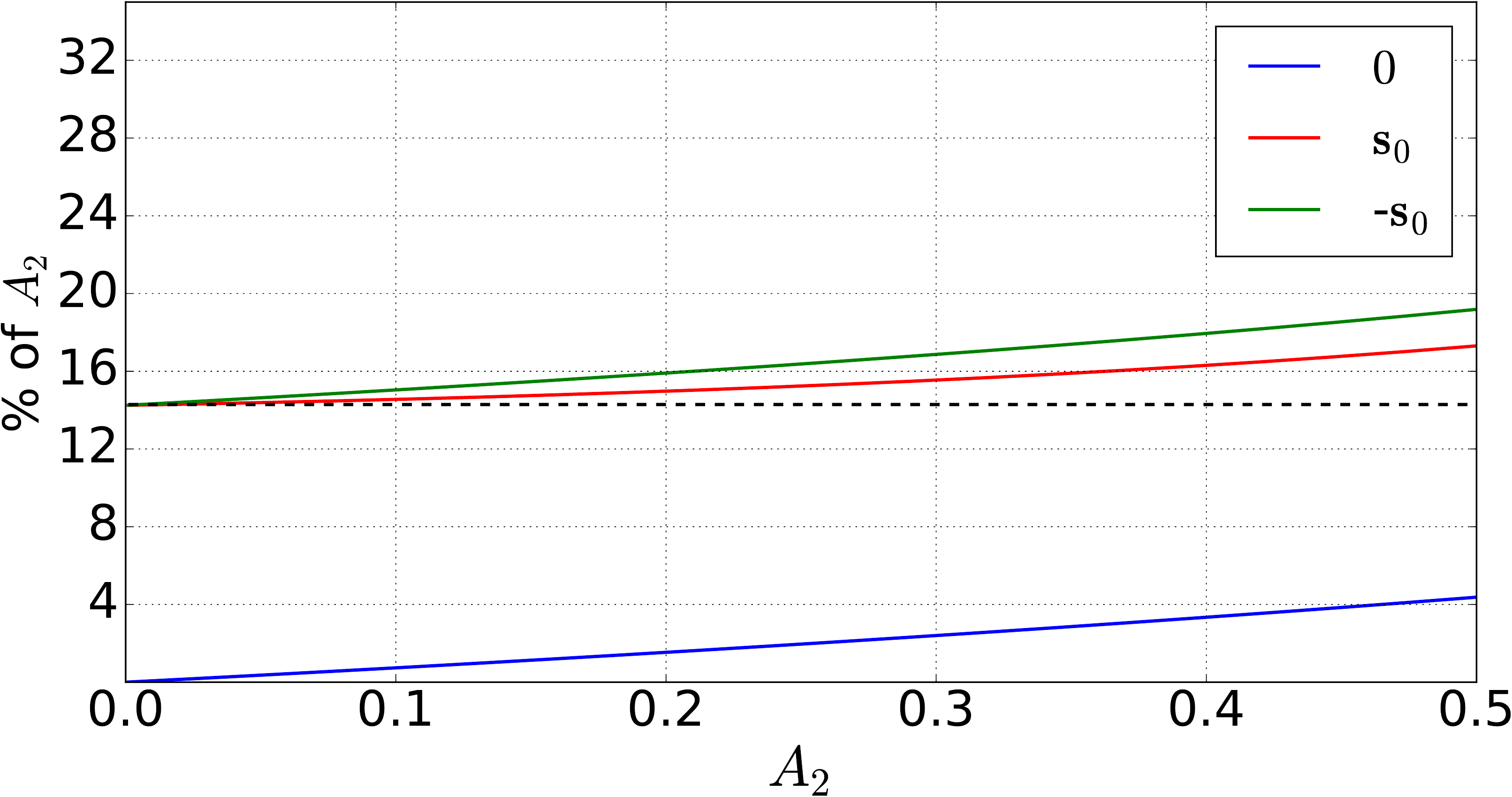}\label{fig:pr_line}}
\subfigure[$\Rcal^{\Delta}_{\Pcal}$: Scattered]{\includegraphics[width=0.4\textwidth]{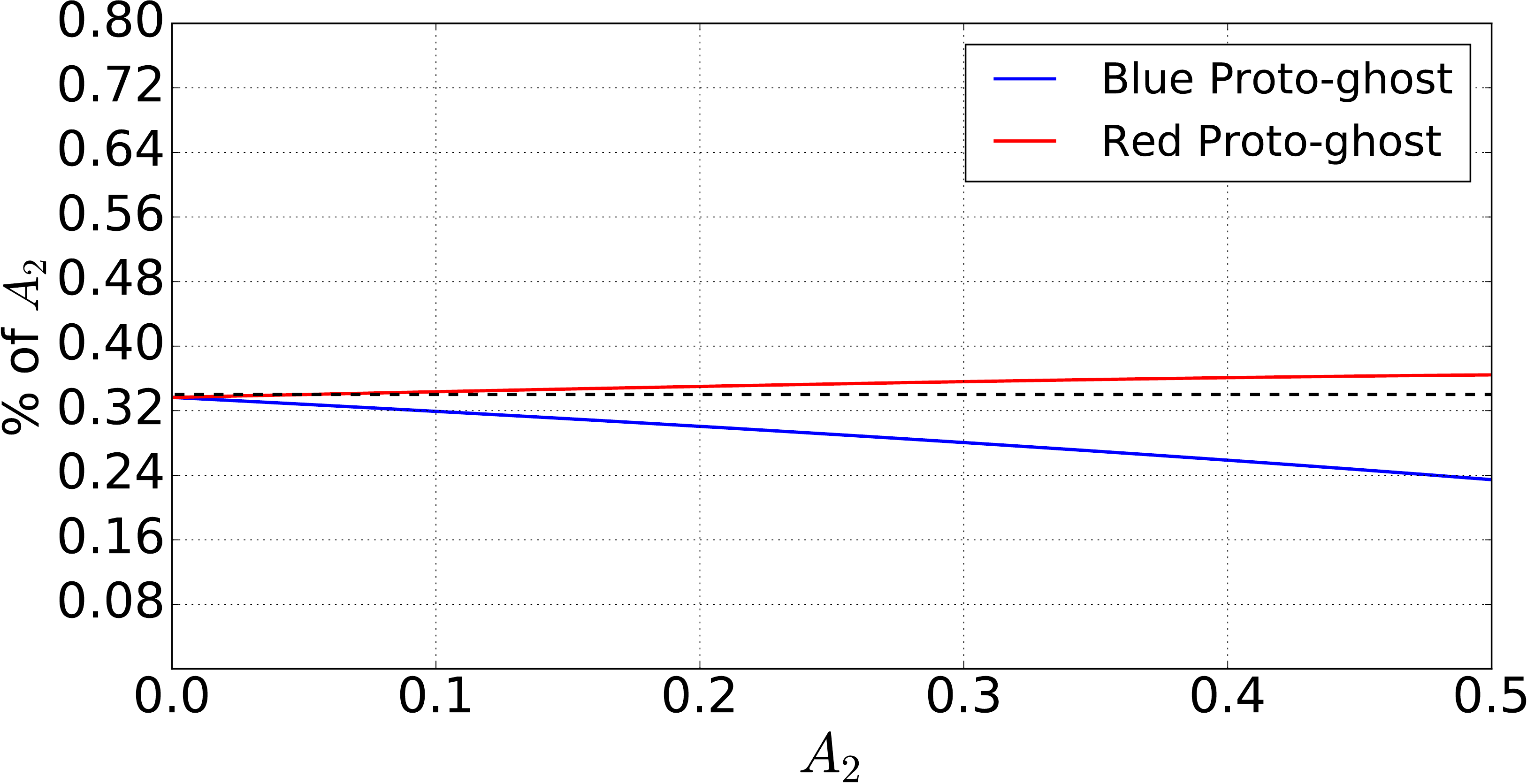}\label{fig:pr_sc}}
\caption{The relative amplitude  of the primary suppressor, secondary suppressor and its anti-ghost on the left and the relative amplitude of a random proto-ghost and its anti-ghost on the right. These curves were created by employing phase-only calibration. The dashed line represents the theoretically derived values that we obtained with perturbation, while the solid line represents a measured value which was obtained from a clean
artefact map that was created via extrapolation. \label{fig:PT_curves}} 
\end{figure*}

The fluxes of the three most important line ghosts in the conglomerated ghost pattern are shown in Fig.~\ref{fig:g_line} and Fig.~\ref{fig:p_line}. The scattered equivalents of Fig.~\ref{fig:g_line} and Fig.~\ref{fig:p_line} are depicted in Fig.~\ref{fig:g_sc} and Fig.~\ref{fig:p_sc}. Fig.~\ref{fig:GT_curves} and Fig.~\ref{fig:PT_curves} were produced by employing full-complex and phase-only calibration respectively. To be more descriptive, Figs.~\ref{fig:g_line}--\ref{fig:g_sc} and Figs.~\ref{fig:p_line}--\ref{fig:p_sc} illustrate the full-complex equivalent of Eq.~\eqref{eq:total_flux} (which can be found in Paper II) and Eq.~\eqref{eq:total_flux} respectively. Similarly, Figs.~\ref{fig:gr_line}--\ref{fig:gr_sc} and Figs.~\ref{fig:pr_line}--\ref{fig:pr_sc} were created by employing the full-complex equivalent of Eq.~\eqref{eq:total_flux2} and Eq.~\eqref{eq:total_flux2} respectively. 


In Fig.~\ref{fig:g_line} and Fig.~\ref{fig:p_line} the blue curve represents the primary suppression ghost, the red curve the secondary suppression ghost and the green curve the anti-ghost of the secondary suppressor. In Fig.~\ref{fig:g_sc} the blue curve represents a random blue full-complex proto-ghost, while its anti-ghost is depicted in red. Its anti-ghost is a red deutero-ghost. In Fig.~\ref{fig:p_sc} the blue curve is associated with a random blue phase-only proto-ghost, the red curve is associated with its anti-ghost, a red proto-ghost.

The dashed curves in Fig.~\ref{fig:GT_curves} and Fig.~\ref{fig:PT_curves} are the perturbation theoretical estimates we derived in Sec.~\ref{ssec:perturbation_phase_only} and in Paper II. In effect, we plot Eq.~\eqref{eq:c_pq_rs_phase} and Eq.~\eqref{eq:cpqrs}. The fact that the results we obtained from perturbation and raw extrapolation agree supports the validity of the first order perturbation derivation in Sec.~\ref{sec:g_pat_phase}. The colors of the dashed lines indicate with which source we can associate each theoretical curve. In Fig.~\ref{fig:PT_curves} we use black dash lines instead, because in the case of phase-only calibration perturbation predicts the same absolute flux value for every ghost and anti-ghost pair.
This can be verified by inspecting Eq.~\eqref{eq:c_pq_rs_phase} and Eq.~\eqref{eq:symmetry}. Perturbation also predicts that the primary suppressor will always have zero flux. This explains why the dashed blue line is absent in Fig.~\ref{fig:PT_curves}.

The primary suppressor is on average 14\% of $A_2$ and 2.5\% of $A_2$ when we employ full-complex and phase-only calibration respectively. Moreover, in the case of full-complex calibration the secondary suppressor and its anti-ghost are on average 30\% of $A_2$ and 2\% of $A_2$ respectively. For the phase-only setup, we find that on average both the secondary suppressor and its anti-ghost are 15\% of $A_2$.

A random blue proto-ghost is on average 0.7\% of $A_2$ and 0.35\% of $A_2$ when we perform full-complex and phase-only calibration respectively. The blue proto-ghost's anti-ghost is on average 0.1\% of $A_2$ and 0.35\% of $A_2$ when we use the full-complex and phase-only calibration algorithms respectively.

Also note that correcting our visibilities does not significantly impact the brightness of our ghosts (compare the top panel of Fig.~\ref{fig:GT_curves} with the bottom panel in Fig.~\ref{fig:GT_curves}). The slight differences between the top and bottom panel Fig.~\ref{fig:GT_curves} are discussed in Paper I and II. The same reasoning can be applied to Fig.~\ref{fig:PT_curves}.

We can reach the following conclusions by comparing Fig.~\ref{fig:stef_whole}, Fig.~\ref{fig:phase_whole}, Fig.~\ref{fig:GT_curves} and  Fig.~\ref{fig:PT_curves} with each other:
\begin{enumerate}
 \item \emph{Line ghosts}: The primary suppressor, the secondary suppressor and its anti-ghost are all line ghosts, which is why they are so much brighter than the other ghosts in Fig~\ref{fig:con_patterns}. According to perturbation theory, the primary suppressor only forms when performing full-complex calibration. However, Fig.~\ref{fig:PT_curves} clearly indicates that a phase-only primary suppressor does form when $A_2$ becomes large enough. The phase-only primary suppressor, however, is much dimmer than its full-complex counterpart. Moreover, the phase-only anti-ghost in  Fig.~\ref{fig:phase_whole} and Fig.~\ref{fig:PT_curves} is much brighter than the anti-ghost in Fig.~\ref{fig:stef_whole} and Fig.~\ref{fig:GT_curves} since the anti-ghost which is produced by phase-only calibration is a proto-ghost, while the anti-ghost produced by full-complex calibration is a deutero-ghost. The full-complex secondary suppressor is also brighter than its phase-only counterpart. This can be explained by comparing Eq.~\eqref{eq:c_pq_rs_phase} and Eq.~\eqref{eq:cpqrs}. These two equations tell us that the secondary suppressor is proportional to $2/N$ and $1/N$ when we employ full-complex and phase-only calibration respectively.
 \item \emph{Scattered ghosts}: We see that we produce less scattered ghosts with phase-only calibration than with full-complex calibration, since we do not produce any deutero-ghosts while performing phase-only calibration. Furthermore, in the case of phase-only calibration, the blue and red symmetric ghost pairs in Fig.~\ref{fig:PT_curves} are of the same order of magnitude (both are proto-ghosts). This is not true when performing full-complex calibration (Fig.~\ref{fig:full_pat_45} clearly shows that at least one of the ghosts in each symmetric pair is a deutero-ghost). This follows trivially from Eq.~\eqref{eq:symmetry}, which is only true for phase-only calibration. Moreover, the full-complex proto-ghosts in Fig.~\ref{fig:GT_curves} are brighter than the phase-only proto-ghosts  in Fig.~\ref{fig:PT_curves}, which is corroborated by Eq.~\eqref{eq:c_pq_rs_phase} and Eq.~\eqref{eq:cpqrs}.  
\end{enumerate}

\section{Primary beam}
\label{sec:p_beam}

In this section we replicate and then try to explain the synthetic anti-ghost experiment in \citet{Stewart2014}, which we discussed in detail in Sec.~\ref{sec:intro} (see also Fig.~\ref{fig:adam_exp}). In Sec.~\ref{sec:spiral_experiment}, we will modify Stewart's experiment slightly to be able to explain his results using just a single graph. Our findings are discussed in detail in Sec.~\ref{sec:spiral_results}. Although the strange behaviour of the anti-ghost appears to contradict some of the results in Paper II and Sec.~\ref{sec:g_pat_phase}, we will see that the primary beam correction causes the apparent dependence of the anti-ghost's brightness on the position of the unmodelled source.
The results in this section were obtained using standard reduction software packages and techniques. We provide the reader with more implementation specifics in Sec.~\ref{sec:spiral_experiment}.

\subsection{Spiral Experiment}
\label{sec:spiral_experiment}

In this section we present an experiment that can shed some light on Stewart's puzzling result (see Sec.~\ref{sec:intro}). We actually discuss a simple, modified version of Stewart's experiment below. Instead of sampling the locations of the unmodelled source from a rectangular grid, we propose that the possible locations of the unmodelled source all lie on a spiral. As will become apparent later, altering Stewart's experiment in this manner allows us to create a one dimensional plot which can help us shed some light on Stewart's findings.


Just as with Stewart's experiment, we decided to conduct our experiment with the LOFAR antenna layout. To mirror Stewart's original experiment as much as possible we used 11min observations. Moreover, the field center and the NCP coincided. The measurement set we used was created assuming 33 LOFAR stations\footnote{LBA CS 1--7, 11, 13, 17, 21, 24, 26, 28, 30--32, 101, 103, 201, 301, 302, 401, 501; LBA RS 106, 205, 208, 306--307, 406, 503, 508--509.}. We also assumed a monochromatic observation at a frequency of $\nu = 58.66~$MHz. 

Furthermore, we only considered a two-source scenario. We place an 88.7 Jy source at (02h22m52.1496s, 86$^{\circ}$18$'$59.184$''$)\footnote{These coordinates are J2000 coordinates.}. This bright source is included in our calibration model. We also introduce a second source of 60 Jy to our simulation. This second source is not included in our calibration model. Moreover, the modelled source is placed at the same location as 3C 61.1.

We start our experiment by placing the unmodelled source at the field center. We then generate the visibilities associated with the modelled and unmodelled source, which we then store in the measurement set. The modelled and unmodelled source were attenuated by the array's primary beam before the visibilities were generated. The effect of beam rotation was ignored. We then perform phase-only calibration. We then image and clean the resulting corrected visibilities and measure the \emph{apparent and intrinsic flux} of the modelled source, the unmodelled source and the anti-ghost.

The intrinsic flux density of a source is the true flux value of a specific source and its apparent flux is the flux it appears to have after it has been attenuated by the observing array's primary beam response. 

It is important to note here that we apply the aforementioned terminology to both the input / a priori models that enabled us to create the visibilities that we used, and the output / a posteriori simulated images that we generated, during the spiral experiment in the discussion that is found in Sec.~\ref{sec:spiral_results}. 


We can now change the position of the unmodelled source and then repeat the aforementioned simulation, calibration and imaging steps to generate another simulated observation. If we continue in this way, varying the position of the unmodelled source followed by the generation of a simulated observation, we will be able to produce multiple snapshot images (the unmodelled source will be in a different location in every snapshot). 

The unmodelled source can be placed at completely random positions or it can follow a deterministic trajectory. We opted for a deterministic trajectory. During the course of the spiral experiment, the unmodelled source follows an Archimedean spiral trajectory, which can be described by the equation
\begin{equation}
\label{eq:spiral}
 r(\theta) = \frac{-\theta}{2\pi},
\end{equation}
where $r$ is the radial distance (in degrees) of the unmodelled source from the field center and $\theta$ is the polar angle of the spiral. In Eq.~\eqref{eq:spiral}, $\theta$ is a positive real number and is measured in radians. The results we obtained by performing this experiment are presented in Sec.~\ref{sec:spiral_results}.

\begin{figure*}
\centering
\subfigure[Source trajectories]
{\includegraphics[width=0.43\textwidth]{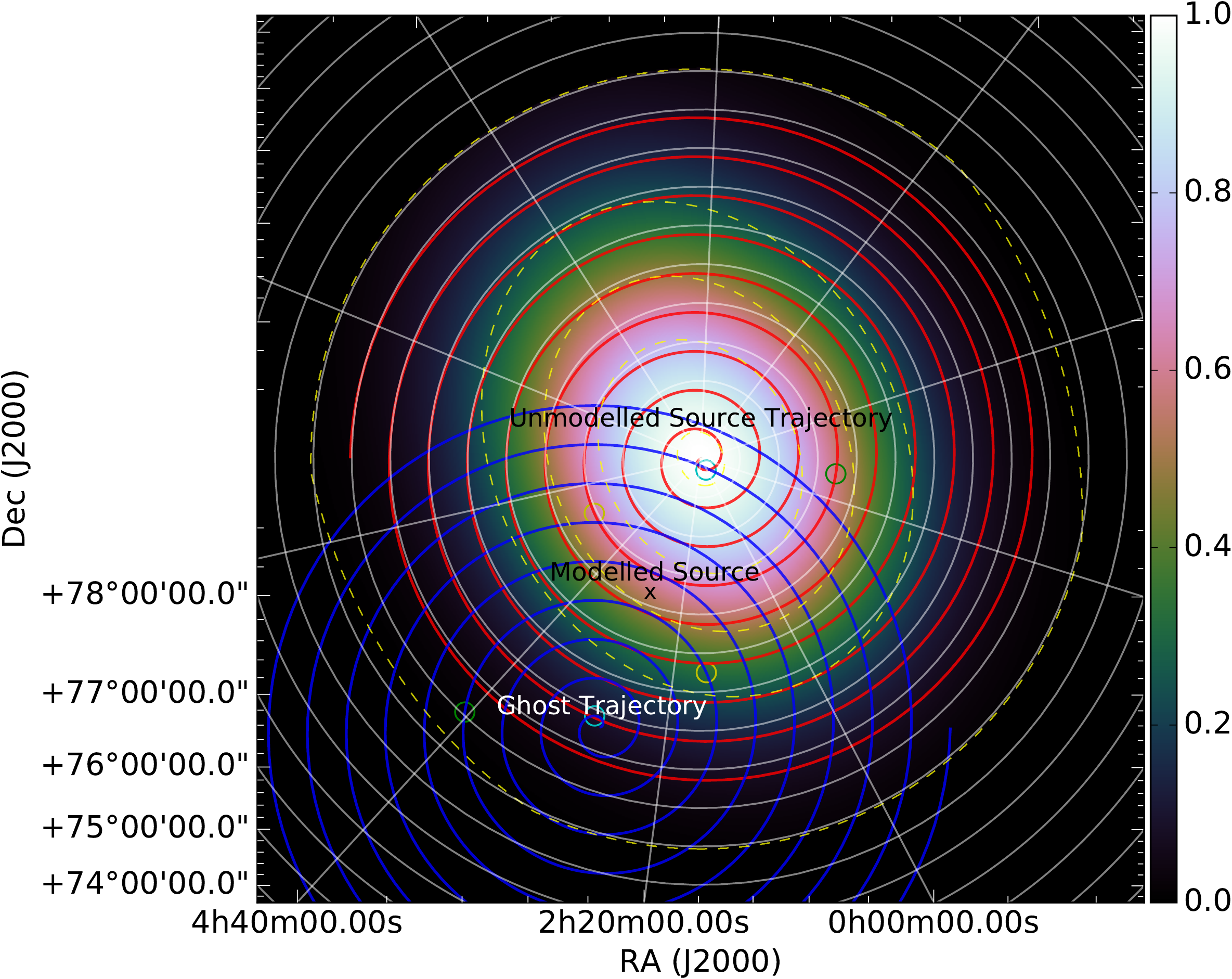}
\label{fig:spiral_trajectory}}
\subfigure[$r=0.32^{\circ}$]
{\includegraphics[width=0.45\textwidth]{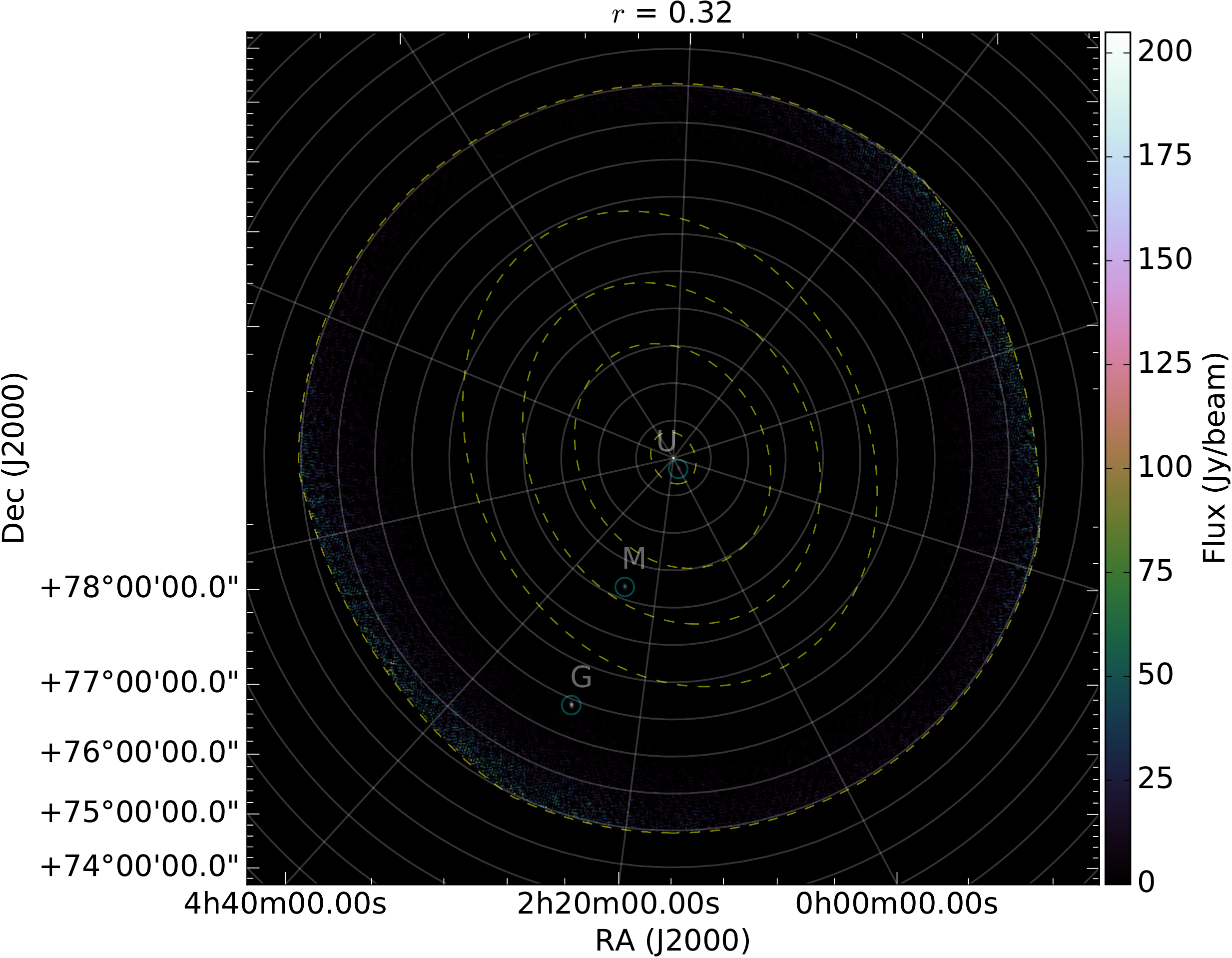}
\label{fig:spiral_1}}
\subfigure[$r=3.08^{\circ}$]
{\includegraphics[width=0.46\textwidth]{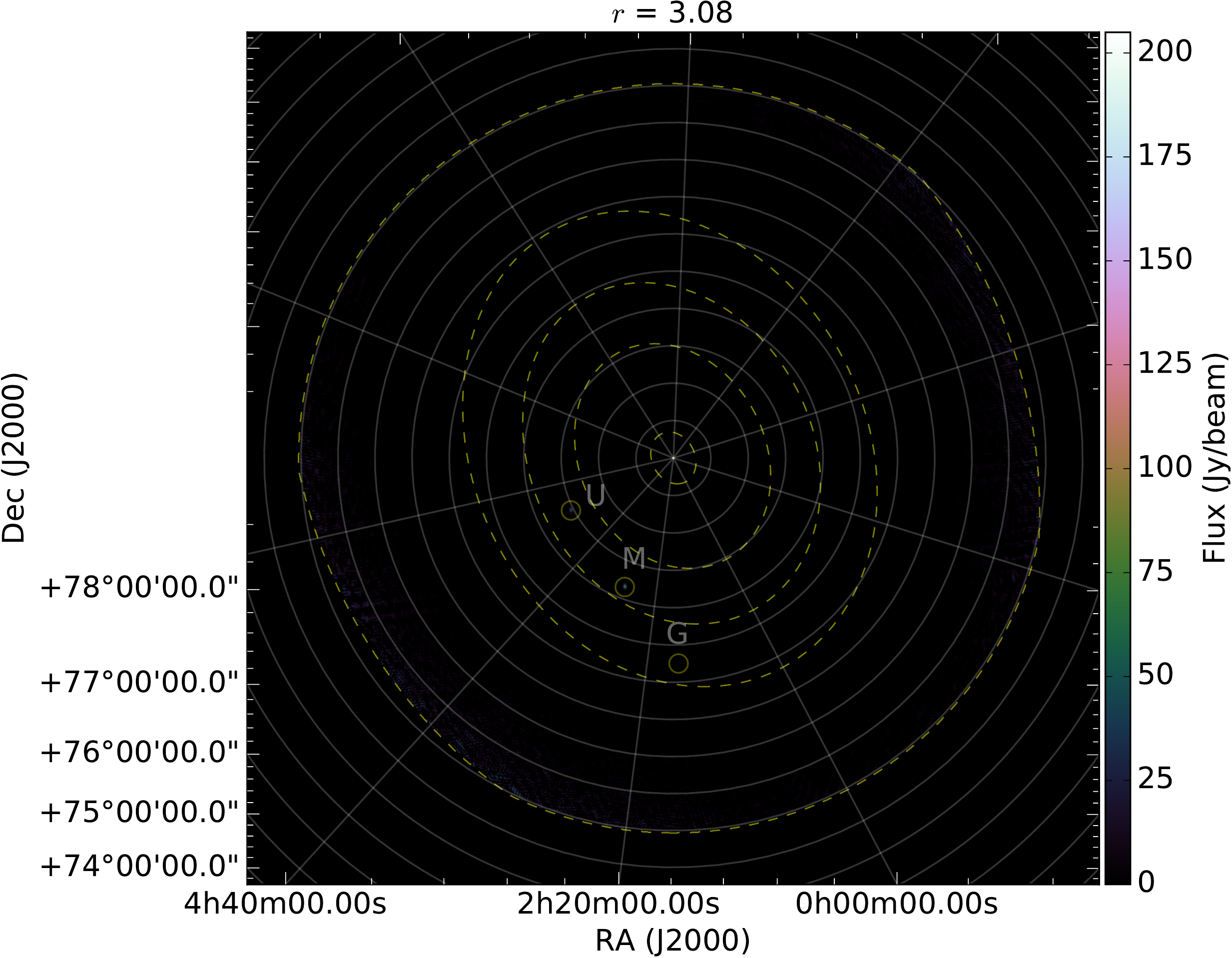}
\label{fig:spiral_2}}
\subfigure[$r=3.48^{\circ}$]
{\includegraphics[width=0.46\textwidth]{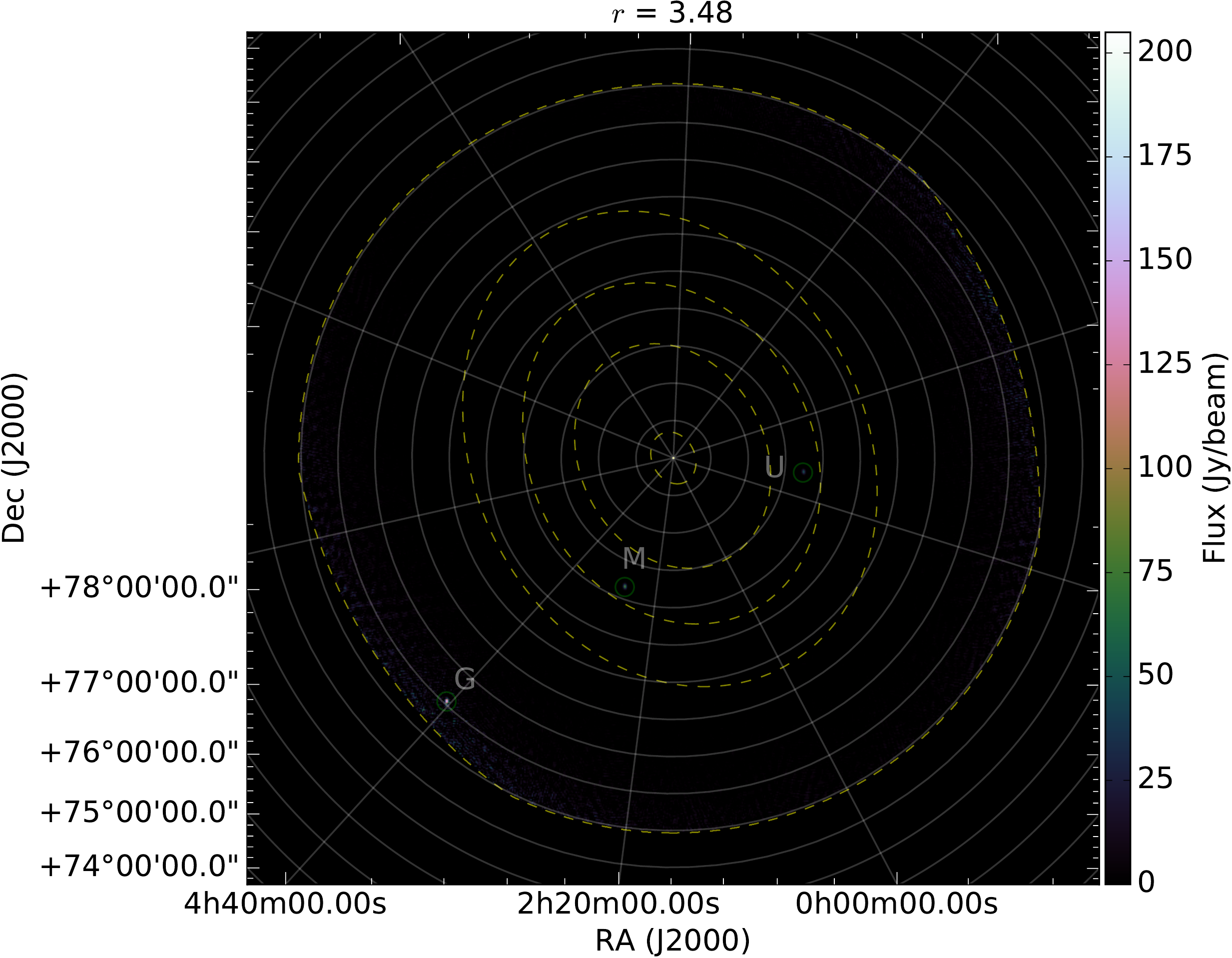}
\label{fig:spiral_3}}
\caption{The source trajectories that the unmodelled source and its anti-ghost follow during the spiral experiment are depicted in Fig~\ref{fig:spiral_trajectory}. The LOFAR primary beam model, which we used in the spiral experiment, is also shown in Fig~\ref{fig:spiral_trajectory}. Figs.~\ref{fig:spiral_1}--\ref{fig:spiral_3} depict three distinct spiral experiment snapshots. The radial distance the unmodelled source is from the field center is given in the individual sub-captions of the three figures. The contours of the LOFAR primary beam model are depicted in Figs.~\ref{fig:spiral_trajectory}--\ref{fig:spiral_3} with dashed yellow lines. The positions of the modelled source, the unmodelled source and the anti-ghost in Figs.~\ref{fig:spiral_1}--\ref{fig:spiral_3} are indicated with labeled circular markers. The labels M, U, and G were used to tag the modelled source, the unmodelled source and the anti-ghost respectively. The positions of the modelled source, the unmodelled source and the anti-ghost in each of the three snapshot can also be found in Fig~\ref{fig:spiral_trajectory} and are indicated with circular markers. The color of the marker should be used as a key to identify to which snapshot each position belongs.\label{fig:fits_files}}
\end{figure*}

The red spiral in Fig.~\ref{fig:spiral_trajectory} depicts the trajectory that the unmodelled source follows during the experiment. The anti-ghost in turn follows the blue trajectory in Fig.~\ref{fig:spiral_trajectory}. Moreover, the LOFAR primary beam which we used during the spiral experiment is presented in Fig.~\ref{fig:spiral_trajectory}.

Fig.~\ref{fig:fits_files} also contains three other images: Fig.~\ref{fig:spiral_1} -- Fig.~\ref{fig:spiral_3}. Each of these figures represents a spiral experiment snapshot. Fig.~\ref{fig:spiral_1}, Fig.~\ref{fig:spiral_2} and Fig.~\ref{fig:spiral_3} were obtained by placing the unmodelled source  $0.32^{\circ}$, $3.08^{\circ}$ and $3.48^{\circ}$ away from the field center respectively. The anti-ghost is clearly visible in Fig.~\ref{fig:spiral_1} and Fig.~\ref{fig:spiral_3}, but is not visible in Fig.~\ref{fig:spiral_2}. This confirms Stewart's observation (see Sec.~\ref{sec:intro}) that the brightness of the anti-ghost is highly dependent on the position of the unmodelled source. The contours of the primary beam that was used in this experiment are depicted in Fig.~\ref{fig:fits_files} with yellow dashed lines. 

Most of the software we used to perform the spiral experiment were created from first principals, i.e. we created a custom built software package which could perform the following
functions:

\begin{enumerate}
\item The \emph{creation of the visibilities} associated with the modelled and the unmodelled source.
\item The \emph{calibration task}. A phase-only variant of the StEFCal calibration algorithm was implemented \citep{Salvini2014}.
\item The incorporation of the \emph{primary beam} into the experiment's pipeline.
\item The \emph{source finding task}. 
\end{enumerate}

We also employed \textsc{lwimager} (an FFT-based imager derived
from the \textsc{CASA} (Common Astronomy Software Applications) \footnote{https://casa.nrao.edu/} libraries and functionally equivalent to the \textsc{CASA} imager). 

\subsection{Spiral Experiment Results}
\label{sec:spiral_results}

\subsubsection{Preliminary observations}
\label{sec:pre_obs}

\begin{figure*}
\centering
\includegraphics[width=\textwidth]{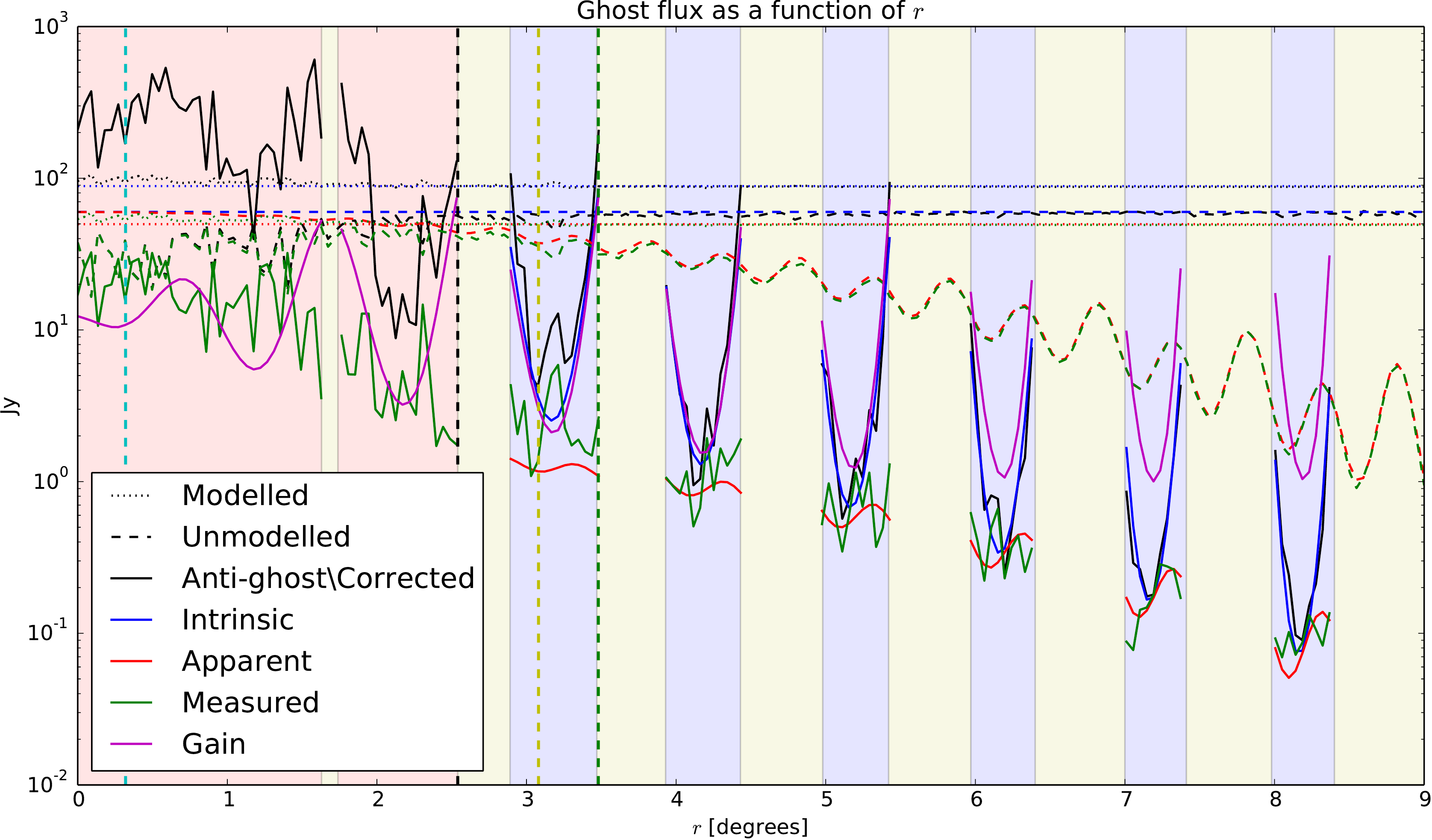}
\caption{The flux values of the spiral experiment for the modelled source, the unmodelled source and the anti-ghost as a function of the radial distance $r$ (the distance the unmodelled source is located from the field center). The curves that are associated with the modelled source, the unmodelled source and the anti-ghost are depicted by dotted, dashed and solid lines respectively. Four curves are associated with each source, each one depicted in a different color. The blue curves represent the true intrinsic flux of a source, the red curves depict the true apparent flux of a source, the green curves represent the measured apparent flux values of a source and the black curves denote the measured intrinsic flux of a source. The solid magenta line is an exception and denotes the beam gain the anti-ghost experiences when the unmodelled source is $r$ degrees from the field center. It is not measured in Jy, but is actually a unitless quantity. The yellow blocks represent regions where the radial distance the ghost is from the field center is larger than $9^{\circ}$. The graph is also divided into two regions, a red region ($r \leq 2.5^{\circ}$) and a blue region ($r > 2.5^{\circ}$). The anti-ghost appears much brighter in the red region than in the blue region on average. The colored vertical dashed lines indicate the values of $r$ at which the snapshots in Fig.~\ref{fig:fits_files} were made.
\label{fig:spiral_results}}
\end{figure*}

The results of the spiral experiment are shown in Fig.~\ref{fig:spiral_results}. We find the different flux values of the modelled source, the unmodelled source and the anti-ghost for the spiral experiment as a function of the radial distance $r$ (see Eq.~\eqref{eq:spiral}). The line-types indicate which curves can be associated with which source: the dotted curves, the dashed curves and the solid curves in Fig.~\ref{fig:spiral_results} can be associated with the modelled source, the unmodelled source and the anti-ghost respectively. The only exception is the solid magenta line, which denotes the beam gain the anti-ghost experiences when the unmodelled source is $r$ degrees from the field center (and is not measured in Jy, but is a unitless gain value).

In Fig.~\ref{fig:spiral_results}, there are four different-colored curves for each of the three aforementioned sources/line-types (i.e. there are twelve curves in total). The blue and red curves denote the true intrinsic and the true apparent flux values of the sources with which they are associated, i.e., true simulated model and true primary beam attenuated simulated model flux values. The values of the red and blue curves are known a priori. The solid red and blue line, however, warrant further explanation. The solid red line represents the theoretical flux density of the anti-ghost, as derived in Sec.~\ref{sec:g_pat_phase}, and the solid blue curve represents the product of the primary beam gain (magenta curve) and the theoretical flux density of the anti-ghost, i.e. the effect that the primary beam had on the theoretical flux density of the anti-ghost once we correct for it. The green and black curves denote the measured intrinsic and the apparent flux values of the sources with which they are associated, i.e., raw measured flux values and primary beam corrected measured flux values. The green and black curve values were obtained from the images produced by the spiral experiment. To summarize, the red and blue curves can be interpreted as theoretical predictions, i.e. as the values that are fed into the spiral experiment, while the green and black curves can be interpreted as observed values, i.e. the quantities that are produced by the spiral experiment.

The yellow bars in Fig.~\ref{fig:spiral_results} indicate regions where the anti-ghost lies outside the boundaries of the images that were created by the spiral experiment. This explains why the yellow regions do not contain any solid curves. 

Note that, since the modelled source remains at the same location during the experiment, its intrinsic flux curves are scalar multiples of its apparent flux curves. This assertion can be verified by comparing the dotted blue curve with the dotted red curve. A similar conclusion can be drawn by comparing the dotted black curve with the dotted green curve. In contrast, the apparent flux of the unmodelled source slowly dims as it spirals outward due to the array's primary beam response. This can be verified by inspecting the dashed red curve.

We have indicated two regions in Fig.~\ref{fig:spiral_results}. In the red region with $r \leq 2.5^{\circ}$, the apparent flux of the unmodelled source is larger than the apparent flux of the modelled source, i.e. the red dashed line is above the red dotted line. In this region, the assumption $A_2 \ll A_1$ of our perturbation analysis is clearly violated. The opposite is true in the blue region with $r > 2.5^{\circ}$. Not unexpectedly, we find that when the apparent flux of the unmodelled source is larger than the modelled source we get a very different ghost response, than the well studied case in which the apparent flux of the unmodelled source is smaller than the apparent flux of the modelled source.

\subsubsection{Model Accuracy}
\label{sec:acc_verus_in}

\begin{figure*}
\centering
\subfigure[Secondary suppressor]{\includegraphics[width=0.45\textwidth]{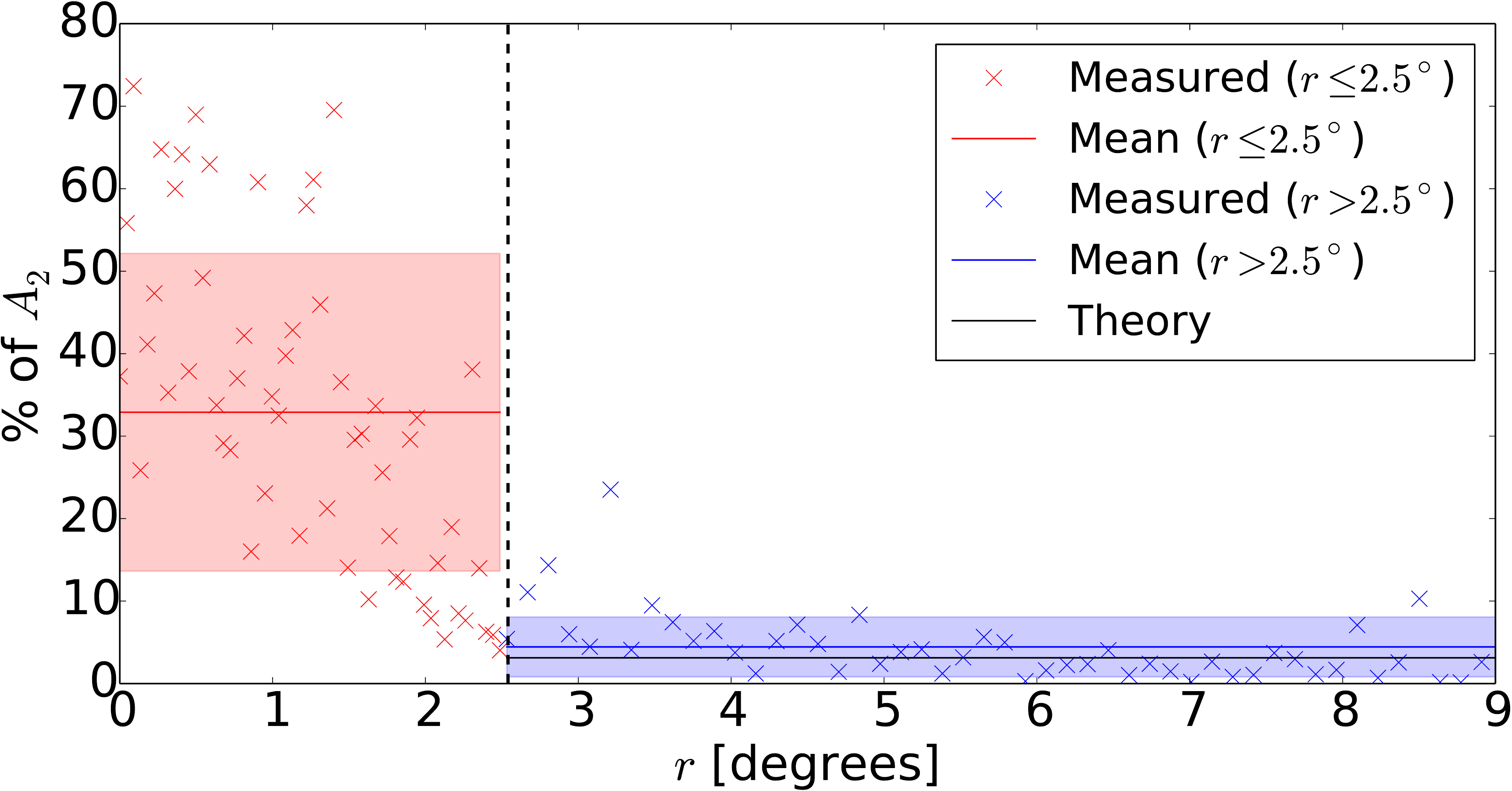}\label{fig:sup_fig}}
\subfigure[Anti-ghost]{\includegraphics[width=0.45\textwidth]{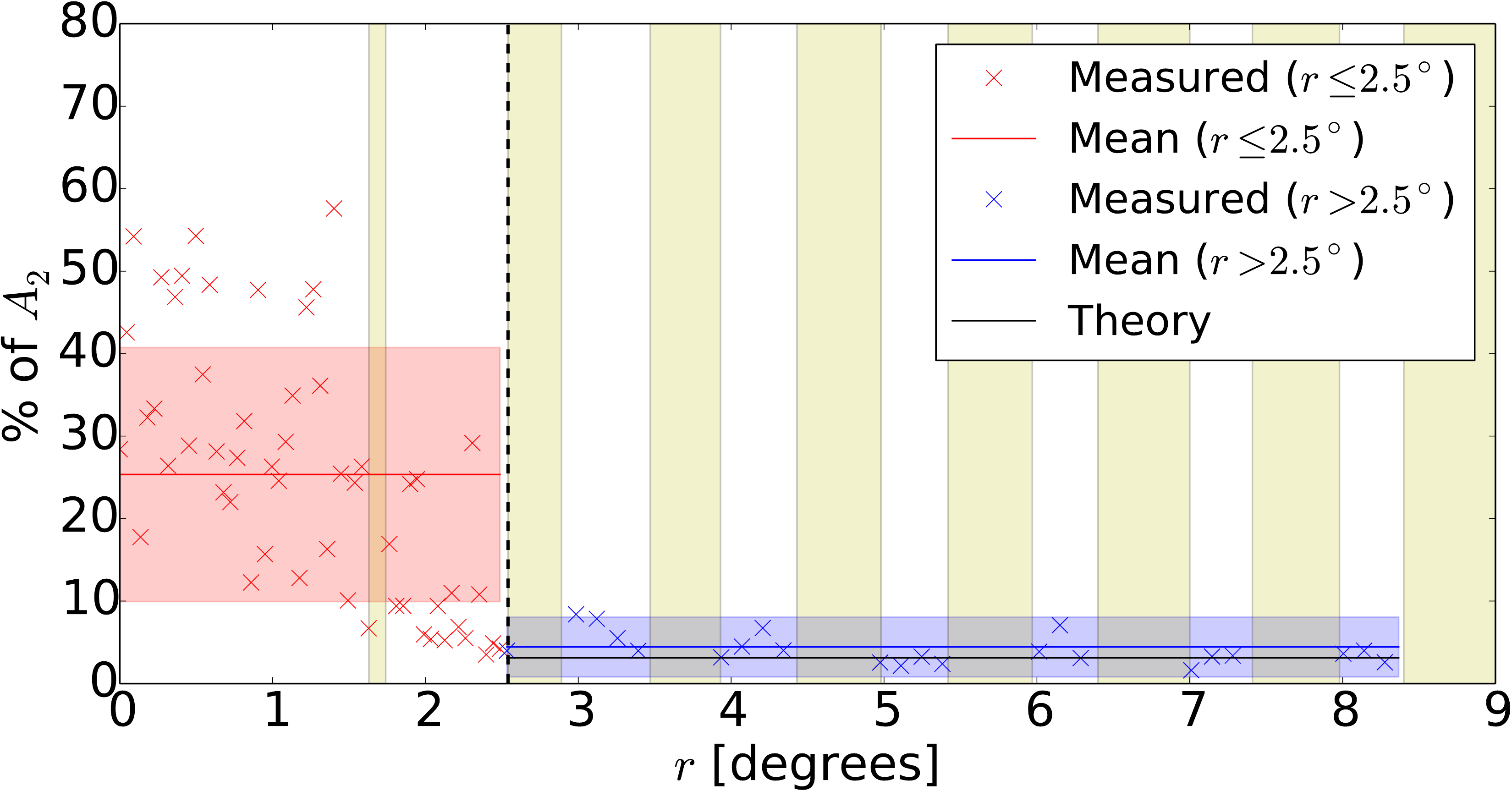}\label{fig:anti_fig}}
\caption{The crosses represent the measured flux values of the secondary suppressor and the anti-ghost (as a \% of $A_2$) as a function of the radial distance $r$. The plots are divided into two regions: a red region and a blue region. The colored horizontal lines represent the mean flux values within each region, while the horizontal filled in blocks represent the standard deviation about the mean within each region. The black line represents the theoretical value of the secondary suppressor and the anti-ghost as predicted in Sec.~\ref{sec:g_pat_phase}. In Fig.~\ref{fig:anti_fig}, the yellow blocks represent regions where the radial distance the ghost is from the field center is larger than $9^{\circ}$. 
\label{fig:sup_anti}} 
\end{figure*}

The fact that the ghost response is different in the red and blue regions of Fig.~\ref{fig:spiral_results} is easier to spot if  Fig.~\ref{fig:sup_anti} is inspected instead. In Fig.~\ref{fig:sup_fig}, we have the measured value of the secondary suppressor (as a \% of $A_2$) as a function of the radial distance the unmodelled source is from the field center. In Fig.~\ref{fig:anti_fig}, we have the measured value of the anti-ghost (as a \% of $A_2$) as a function of the radial distance the unmodelled source is from the field center. The crosses in each region of Fig.~\ref{fig:sup_anti} represent actual measured flux values (as a \% of $A_2$). These values were obtained from the images that were produced during the spiral experiment. The coloured lines in each region represent the mean flux value measured within each region and the filled in block in each region depicts the standard deviation about the mean flux in each region. The black line in both of these figures is the theoretical value of the secondary suppressor and the anti-ghost, as derived in Sec.~\ref{sec:g_pat_phase}. 

Inspection of Fig.~\ref{fig:spiral_results} and Fig.~\ref{fig:sup_anti} reveals the following: when the calibration model is accurate enough, i.e. $A_2 < A_1$ or $r > 2.5^{\circ}$, the measured flux values are close to the theoretical value, which we derived in Sec.~\ref{sec:g_pat_phase}. However, when the model becomes too inaccurate, i.e. $A_2 \geq A_1$ or $r \leq 2.5^{\circ}$, then it deviates completely from the two black lines in Fig.~\ref{fig:sup_anti}. Fig.~\ref{fig:sup_anti}, therefore, shows us that when the unmodelled source is much brighter than the modelled source we produce a much brighter secondary suppressor and anti-ghost than that predicted in Sec.~\ref{sec:g_pat_phase}. This result does not contradict anything in Sec.~\ref{sec:g_pat_phase}, because one of the main assumptions that we make in Sec.~\ref{sec:prelim} is that $A_2 \ll A_1$. A similar result was obtained in \citet{Grobler2015} for the WSRT array. Fig.~\ref{fig:sup_anti} also shows us that the theory we derived in Sec.~\ref{sec:g_pat_phase} holds up if $A_2 < A_1$. We can now use the insight we gained from inspecting Fig.~\ref{fig:sup_anti} to make sense of the remaining curves in Fig.~\ref{fig:spiral_results}.

\subsubsection{Modelled source}

The only interesting thing one can conclude by inspecting the curves of the modelled source is that the modelled source seems to be amplified a bit when we calibrate with a very inaccurate model ($A_2 \geq A_1$), but that this quickly improves when the calibration model becomes more accurate. This conclusion can be reached by comparing either the dotted blue and black curves or the dotted red and green curves with each other.

\subsubsection{Unmodelled source}

The first thing one notices when inspecting the curves associated with the unmodelled source is that the measured apparent flux curve (dashed green) of the unmodelled source is far below the true apparent flux curve (dashed red) when $r \leq 2.5^{\circ}$. We are now able to explain this as, according to Sec.~\ref{sec:acc_verus_in}, the suppression ghost becomes exceptionally bright if the model is too inaccurate ($A_2 \geq A_1$). In contrast, when $r > 2.5^{\circ}$ we see that the dashed green and dashed red curves are virtually indistinguishable, which can also be explained with the aid of Sec.~\ref{sec:acc_verus_in}. Sec.~\ref{sec:acc_verus_in} also tells us that if an accurate enough model ($A_2 < A_1$) is used during calibration the unmodelled source is only suppressed slightly if $N$ is large. Moreover, when $r \leq 2.5^{\circ}$ the black dashed line and the green dashed line are fairly close to each other, as the primary beam gain is close to unity when $r \leq 2.5^{\circ}$. As we mentioned previously, when $r > 2.5^{\circ}$, the degree of suppression is reduced significantly, which also explains why the black dashed curve ends up lying just below the intrinsic flux value of the unmodelled source (the dashed blue curve).

\subsubsection{Anti-ghost}
\label{sec:anti_ghost}

When we inspect the anti-ghost curves we see that the anti-ghost behaves differently when $r \leq 2.5^{\circ}$ than when $r > 2.5^{\circ}$. When $r \leq 2.5^{\circ}$, we are actually using a very inaccurate calibration model ($A_2 \geq A_1$), implying that the anti-ghost will be exceptionally bright in this region of the graph (see Sec.~\ref{sec:acc_verus_in}). We can confirm this by inspecting the solid green curve in the red region of Fig.~\ref{fig:spiral_results}. The anti-ghost can now be made even brighter if we also correct for the primary beam. This can be verified by inspecting the solid black curve in the red region of Fig.~\ref{fig:spiral_results}. 

The combination of these two factors explain why the anti-ghost is clearly visible during the course of the spiral experiment when $r \leq 2.5^{\circ}$. Fig.~\ref{fig:spiral_1} is a snapshot of the spiral experiment. In this snapshot the unmodelled source is $0.34^{\circ}$ from the field center. Fig.~\ref{fig:spiral_1} should be used as a visual aid that shows us just how bright  the anti-ghost can be when $r \leq 2.5^{\circ}$.

When $r > 2.5^{\circ}$ we see that the anti-ghost becomes very dim and becomes close to the theoretically predicted value (the solid red and green lines are close together). This is expected as, when $r > 2.5^{\circ}$, we satisfy the criterion $A_2 < A_1$ (see Sec.~\ref{sec:acc_verus_in}). However, if we now apply the beam gain we see that the anti-ghost can still become exceptionally bright even if the ``apparent'' anti-ghost is fairly weak. This is due to the relative positions of the modelled and unmodelled source. If the relative positions of the modelled and unmodelled source just happens to be of such an orientation that the anti-ghost that forms lies in  a region of the primary beam where the primary beam attenuation is high, then the anti-ghost can be significantly amplified when we do the primary beam correction. If the anti-ghost forms in a region of the primary beam where there is little attenuation then the anti-ghost will remain fairly faint even after we do the primary beam correction. Note that the theoretical ``intrinsic'' flux curve (the solid blue curve which is associated with the anti-ghost) accurately describes the behaviour of the measured intrinsic anti-ghost flux values (solid black curve). 

Fig.~\ref{fig:spiral_2} is a snapshot from the spiral experiment which shows us how hard it can be to spot the anti-ghost when it forms in a region of the primary beam where there is little attenuation. 
In stark contrast, Fig.~\ref{fig:spiral_3} is a snapshot from the spiral experiment that shows that the anti-ghost can be very bright even if the model is quite accurate, 
provided the anti-ghost happens to form in a region of the primary beam where the attenuation is high. The radial distances at which the three snapshots were made are 
depicted in Fig.~\ref{fig:spiral_results} with three vertical dashed lines.

The most important observation that we can make while inspecting Fig.~\ref{fig:spiral_results} is that the solid black line is highly correlated with the solid magenta line. This clearly shows the large impact of the primary beam on the visibility of the anti-ghost. This observation helps explain Stewart's original result. Stewart observed that the brightness of the anti-ghost was highly dependent on the location of the unmodelled source, since the location of the unmodelled source determines the location of its anti-ghost, which may thus end up in a spot where the primary beam has hardly any gain.


\section{Conclusion}

In this paper we showed that spurious symmetrization can be quite bright even if we employ a large number of antennas. We found that two factors working in unison are needed for this to occur: phase-only calibration and the array's primary beam. We showed that phase-only calibration produces a brighter anti-ghost than full-complex calibration. This already bright anti-ghost can in turn be further amplified by the primary beam correction. We studied both of these factors in detail. It is crucial for us to be cognisant of these results while reducing transient data, in order to prevent the misclassification of artefacts as sources.   

We first present the major results we obtained while investigating phase-only calibration. We used extrapolation (see Sec.~\ref{sec:extrap_phase_only}) in conjunction with perturbation analysis (see Sec.~\ref{ssec:perturbation_phase_only}) to derive the two-source phase-only ghost-pattern, which we then compared to the full-complex pattern in Sec.~\ref{sec:comparison}. We summarize the main conclusions we made in Sec.~\ref{sec:comparison} below.

\begin{description}
\item[\textbf{Bright anti-ghost.}] Phase-only calibration produces a brighter anti-ghost if we compare it to the anti-ghost which forms when we perform full-complex calibration. The reason for this is that, in the case of phase-only calibration, the anti-ghost is a proto-ghost, while for full-complex calibration it is a deutero-ghost. This can be confirmed by inspecting Table~\ref{tab:tax_per_baseline_pos_index}.

\item[\textbf{Fewer ghosts.}] Phase only calibration produces fewer scattered ghosts than full-complex calibration as no deutero-ghosts form when we employ phase-only calibration.

\item[\textbf{Color.}] We extended the basic per-baseline taxonomy in Paper II in Sec.~\ref{ssec:tax_baseline}. We realized that we can use the sign of a ghost's real amplitude factor to assign one of the two following false colors to it: red and blue. If a ghost is completely real, its color determines whether it manifests as a positive or negative ghost and, consequently, whether it suppresses or amplifies. Real blue ghosts are suppressors and real red ghosts are amplifiers. The primary suppressor, secondary supressor and its anti-ghost are all usually real valued which in turn means that only their colour is generally important. The primary suppressor and secondary suppressor are blue which explains why they suppress flux. The anti-ghost of the secondary suppressor is red which explains why it manifests as a positive source during imaging. The fact that the anti-ghost is red is important. If this were not the case, spurious symmetrization would not be possible.

\item[\textbf{Symmetric pairs.}] The blue proto-ghosts of phase-only calibration and full-complex calibration in general form a symmetrical pair with a red anti-ghost. However, the red anti-ghosts are proto-ghosts in the case of phase-only calibration and deutero-ghosts when full-complex calibration is employed. This explains why the anti-ghosts of the phase-only blue proto-ghosts are brighter than their full-complex counterparts.
\end{description}

To conclude our phase-only calibration summary, we present ghost flux values as predicted by Eq.~\eqref{eq:c_pq_rs_phase} and Eq.~\eqref{eq:cpqrs} for different values of $N$ in Table \ref{tab:flux_table}. The reader can use this table to directly compare the ghost responses of full-complex and phase-only calibration and by implication enable them to validate most of the aforementioned conclusions.

\begin{table*}
\caption{The flux of the primary suppressor, the secondary suppressor, the anti-ghost of the secondary suppressor, a proto-ghost and a deutero-ghost for a non-redundant interferometer with 7, 14, 27 and 100 antennas as a percentage of $A_2$. In this table $N$ and $B$ denote the number of antennas and baselines in the array respectively. We generated the values in this table by employing two different calibration strategies, namely full-complex and phase-only calibration. We use the abbreviations FC and PO to denote full-complex and phase-only calibration respectively.}
\label{tab:flux_table}
\begin{tabular}{|l|c|l|llll|}
  \cline{1-7}
  \multicolumn{1}{|l|}{\textbf{Ghost}} & \multicolumn{1}{|c|}{\textbf{Approach}} & \multicolumn{1}{|l|}{\textbf{Formula}} & \multicolumn{4}{|c|}{$N$}\\
  \cline{4-7}
   & & fraction of $A_2$ &7&14&27&100\\
  \cline{1-7}
  \multirow{2}{*}{Primary}&FC& $\frac{1}{N}$ & 14\% & 7\% & 4\% & 1\% \\
  &PO&N/A & N/A & N/A & N/A & N/A\\
  \cline{1-7}
  \multirow{2}{*}{Secondary}&FC& $\frac{2}{N}-\frac{1}{N^2}$ & 27\% & 14\% & 7\% & 2\% \\
  &PO&$\frac{1}{N}$ & 14\% & 7\% & 4\% & 1\%\\
  \cline{1-7}
  \multirow{2}{*}{Anti-ghost}&FC& $\frac{1}{N^2}$ & 2\% & 0.5\% & 0.1\% & 0.01\% \\
  &PO&$\frac{1}{N}$ & 14\% & 7\% & 4\% & 1\%\\
  \cline{1-7}
  \multirow{2}{*}{Proto} &FC& $\frac{1}{B}(\frac{1}{N}-\frac{1}{N^2})$ & 0.6\% & 0.07\% & 0.01\% & 0.0002\%\\
  &PO&$\frac{1}{2BN}$&0.3\%&0.04\%&0.005\%&0.0001\%\\
  \cline{1-7}
  \multirow{2}{*}{Deutero} &FC& $\frac{1}{BN^2}$ & 0.1\% & 0.006\% & 0.0004\% & 0.000002\% \\
  &PO&N/A & N/A & N/A & N/A & N/A \\
  \cline{1-7}
\end{tabular}
\end{table*}

We investigated the effect that primary beam correction can have on the anti-ghost in Sec.~\ref{sec:p_beam}. We found that the primary beam can amplify the brightness of the anti-ghost in two main ways.

\begin{description}
\item[\textbf{Model accuracy.}] If the apparent flux of the unmodelled source happens to be larger than the apparent flux of the brightest source in the field then our calibration model becomes very inaccurate. If we calibrate with a very inaccurate model ($A_2 \geq A_1$) then a very bright anti-ghost is created. We discussed this in detail in Sec.~\ref{sec:acc_verus_in} and Sec.~\ref{sec:anti_ghost}.

\item[\textbf{Source positions.}] If the relative positions of the modelled and unmodelled source is such that the anti-ghost forms in a region of the primary beam where the attenuation is high, then the anti-ghost is significantly amplified when we perform our beam correction (see Section~\ref{sec:anti_ghost}).
\end{description}


\section*{Acknowledgements}

This work is based upon research supported by the South African Research Chairs Initiative of the Department of Science and Technology and National Research Foundation. This work is funded by IBM, ASTRON, the Dutch Ministry of Economic Affairs and the Province of Drenthe. It is also part of the SKA-TSM project and supported by The Northern Netherlands Provinces Alliance (SNN), Koers Noord and the Province of Drenthe, and the European Community FP7 programme MIDPREP, Grant Agreement PIRSES-GA-2013-612599.

The authors would like to thank Gianni Bernardi and Sandeep Sirothia for the valuable contributions that they made. We would also like to thank the reviewer, Dr Maxim Voronkov, for the useful feedback he provided; his feedback has definitely improved the quality of the paper.


\bibliographystyle{mn2e}
\bibliography{g_paper}

\begin{thebibliography}{}

\bibitem[\protect\citeauthoryear{Cornwell \& Fomalont}{Cornwell \&
  Fomalont}{1999}]{Taylor1999}
Cornwell T.~J.,  Fomalont E.~B.,  1999, in Taylor G.~B.,  Carilli C.~L.,
  Perley R.~A.,  eds, {ASP Conf. Ser.} Vol.~180, {Synthesis Imaging in Radio
  Astronomy II}.
Astron. Soc. Pac., San Francisco, p.~197

\bibitem[\protect\citeauthoryear{Dewdney, Hall, Schilizzi \& Lazio}{Dewdney
  et~al.}{2009}]{Dewdney2009}
Dewdney P.~E.,  Hall P.~J.,  Schilizzi R.~T.,    Lazio T. J. L.~W.,  2009,
  {Proc. of the IEEE}, 97, 1482

\bibitem[\protect\citeauthoryear{Green}{Green}{2011}]{Green2011}
Green D.,  2011, Bull. Astr. Soc. India, 39, 289

\bibitem[\protect\citeauthoryear{Grobler, Nunhokee, Smirnov, van Zyl \& de
  Bruyn}{Grobler et~al.}{2014}]{Grobler2014}
Grobler T.~L.,  Nunhokee C.~D.,  Smirnov O.~M.,  van Zyl A.~J.,    de Bruyn
  A.~G.,  2014, MNRAS, 439, 4030

\bibitem[\protect\citeauthoryear{Grobler \& Smirnov}{Grobler \&
  Smirnov}{2015}]{Grobler2015}
Grobler T.~L.,  Smirnov O.~M.,  2015, in {Radio Science Conference (URSI
  AT-RASC), 2015 1st URSI Atlantic} {Calibration artefacts: Phase only
  calibration}.
pp~1--1

\bibitem[\protect\citeauthoryear{Heald, Pizzo, Orr{\'u}, Breton, Carbone,
  Ferrari, Hardcastle, Jurusik, Macario, Mulcahy et~al.,}{Heald
  et~al.}{2015}]{Heald2015}
Heald G.,  Pizzo R.,  Orr{\'u} E.,  Breton R.,  Carbone D.,  Ferrari C.,
  Hardcastle M.,  Jurusik W.,  Macario G.,  Mulcahy D.,    et~al., 2015, A\&A,
  582, A123

\bibitem[\protect\citeauthoryear{Jonas}{Jonas}{2009}]{Jonas2009}
Jonas J.~L.,  2009, {Proc. of the IEEE}, 97, 1522

\bibitem[\protect\citeauthoryear{Linfield}{Linfield}{1986}]{Linfield1986}
Linfield R.~P.,  1986, AJ, 92, 213

\bibitem[\protect\citeauthoryear{Rau, Bhatnagar, Voronkov \& Cornwell}{Rau
  et~al.}{2009}]{Rau2009}
Rau U.,  Bhatnagar S.,  Voronkov M.~A.,    Cornwell T.~J.,  2009, Proc. of the
  IEEE, 97, 1472

\bibitem[\protect\citeauthoryear{Salvini \& Wijnholds}{Salvini \&
  Wijnholds}{2014}]{Salvini2014}
Salvini S.,  Wijnholds S.~J.,  2014, A\&A, 571, A97

\bibitem[\protect\citeauthoryear{Smirnov \& Tasse}{Smirnov \&
  Tasse}{2015}]{Smirnov2015}
Smirnov O.,  Tasse C.,  2015, MNRAS, 449, 2668

\bibitem[\protect\citeauthoryear{Smirnov}{Smirnov}{2010}]{Smirnov2010ghosts}
Smirnov O.~M.,  2010, {Ghostbusters: The Unknown Unknowns Of Selfcal},
  presentation at CALIM2010 conference, Dwingeloo,
  \\http://www.astron.nl/calim2010/presentations/\\14\_Ghostbusters\_Smirnov.pdf

\bibitem[\protect\citeauthoryear{{Smirnov}}{{Smirnov}}{2011}]{RRIME1}
{Smirnov} O.~M.,  2011, \aap, 527, A106

\bibitem[\protect\citeauthoryear{Stewart, Fender, Broderick, Hassall,
  Mu{\~n}oz-Darias, Rowlinson, Swinbank, Staley, Molenaar, Scheers
  et~al.,}{Stewart et~al.}{2016}]{Stewart2016}
Stewart A.,  Fender R.,  Broderick J.,  Hassall T.,  Mu{\~n}oz-Darias T.,
  Rowlinson A.,  Swinbank J.,  Staley T.,  Molenaar G.,  Scheers B.,    et~al.,
  2016, MNRAS, 456, 2321

\bibitem[\protect\citeauthoryear{Stewart}{Stewart}{2014}]{Stewart2014}
Stewart A.~J.,  2014, {LOFAR Image Plane Transient Candidate \#3}, presentation
  at LOFAR TKP meeting, Amsterdam,
  \\https://speakerdeck.com/transientskp/lofar-image-plane-transient-candidate-number-3

\bibitem[\protect\citeauthoryear{Swinbank, Staley, Molenaar, Rol, Rowlinson,
  Scheers, Spreeuw, Bell, Broderick, Carbone et~al.,}{Swinbank
  et~al.}{2015}]{Swinbank2015}
Swinbank J.~D.,  Staley T.~D.,  Molenaar G.~J.,  Rol E.,  Rowlinson A.,
  Scheers B.,  Spreeuw H.,  Bell M.~E.,  Broderick J.~W.,  Carbone D.,
  et~al., 2015, Astr. and Comp., 11, 25

\bibitem[\protect\citeauthoryear{van~der Veen \& Wijnholds}{van~der Veen \&
  Wijnholds}{2013}]{vanderVeen2013}
van~der Veen A.-J.,  Wijnholds S.~J.,  2013, in Bhattacharyya S.~S.,
  Deprettere E.~F.,  Leupers R.,   Takala J.,  eds, {Signal Processing Tools
  for Radio Astronomy}, {Handbook of Signal Processing Systems}.
Springer, New York, p.~421

\bibitem[\protect\citeauthoryear{van Haarlem et~al.,}{van Haarlem
  et~al.}{2013}]{Haarlem2013}
van Haarlem M.~P.,  et~al., 2013, A\&A, 556, 1

\bibitem[\protect\citeauthoryear{Wijnholds, Grobler \& Smirnov}{Wijnholds
  et~al.}{2016}]{Wijnholds2016}
Wijnholds S.~J.,  Grobler T.~L.,    Smirnov O.~M.,  2016, MNRAS, 457, 2331

\bibitem[\protect\citeauthoryear{Wijnholds, van~der Tol, Nijboer \& van~der
  Veen}{Wijnholds et~al.}{2010}]{Wijnholds2010}
Wijnholds S.~J.,  van~der Tol S.,  Nijboer R.,    van~der Veen A.-J.,  2010,
  IEEE Sig. Process. Mag., 27, 30

\bibitem[\protect\citeauthoryear{Wilkinson, Conway \& Biretta}{Wilkinson
  et~al.}{1988}]{Wilkinson1988}
Wilkinson P.~N.,  Conway J.,    Biretta J.,  1988, in Reid M.~J.,  Moran J.~M.,
   eds, {Proc. IAU Symp. 129} Vol.~129, {The impact of VLBI on Astrophysics and
  Geophysics}.
Springer, Netherlands, p.~509

\bibitem[\protect\citeauthoryear{Yatawatta, Kazemi \& Zaroubi}{Yatawatta
  et~al.}{2012}]{Yatawatta2012}
Yatawatta S.,  Kazemi S.,    Zaroubi S.,  2012, in {Innovative Parallel
  Computing (InPar), 2012} {GPU accelerated nonlinear optimization in radio
  interferometric calibration}.
pp~1--6

\end{thebibliography}
 
\appendix

\section{Derivation of Eq.~\refpinv}
\label{sec:app:deriv_pinv}

In this appendix we derive a closed form expression for the pseudo-inverse of Eq.~\eqref{eq:mat_to_invert}. We do this by employing eigenvalue decomposition, i.e.
we first find
\begin{equation}
\label{eq:EVD}
2(N\bI - \bone\bone^H) = \bV \bLambda \bV^H,
\end{equation}
which enables us to compute
\begin{equation}
\left [ 2(N\bI - \bone\bone^H) \right ]^\dagger = \left ( \bV \bLambda \bV^H \right )^\dagger = \bV \bLambda^\dagger \bV^H,
\end{equation}
where $\bLambda^\dagger$ is defined such that all non-zero eigenvalues are replaced by their reciprocals while the zero eigenvalues remain zero.

It is rather easy to show that
\begin{equation}
\label{eq:V}
\bV = \left [ \frac{1}{\sqrt{N}}\bone, \bV_0 \right ],
\end{equation}
where the vector of ones has length $N$ and $\bV_0$ provides an orthonormal basis for the null space of the explicitly stated eigenvector. Substituting Eq.~\eqref{eq:V} into \eqref{eq:EVD} shows that
\begin{equation}
\bLambda = \diag{\left [ 0, 2N, \cdots, 2N \right ]}.
\end{equation}
We therefore have
\begin{eqnarray}
\bLambda^\dagger & = & \diag{\left [ 0, \frac{1}{2N}, \cdots, \frac{1}{2N} \right ]}\nonumber\\
& = & \frac{1}{2N} \bI - \frac{1}{2N} \diag{\left [ 1, 0, \cdots, 0 \right ]}.
\end{eqnarray}
Hence
\begin{eqnarray}
\bV \bLambda^\dagger \bV^H & = & \frac{1}{2N} \bV \left ( \bI - \diag{\left [  1, 0, \cdots, 0 \right ]} \right ) \bV^H\nonumber\\
& = & \frac{1}{2N} \bI - \frac{1}{2 N^2}\bone\bone^H,
\end{eqnarray}
which completes our derivation of Eq.~\eqref{eq:inv_matrix}.

\end{document}